\documentclass[a4paper]{article}
\usepackage{eurosym}
\usepackage{amsmath, amsfonts, amsthm,amssymb}
\usepackage{mathrsfs}
\usepackage[margin=2cm]{geometry}
\usepackage{xcolor}
\usepackage{colortbl}
\usepackage{graphicx}
\usepackage{caption}
\usepackage{subfig}
\usepackage{multirow,makecell,multicol}
\usepackage{hhline,array,rotating}
\usepackage{stackengine}
\usepackage{lscape}
\usepackage[T1]{fontenc}

\newcolumntype{M}[1]{>{\centering\arraybackslash}m{#1}}
\newcommand\xrowht[2][0]{\addstackgap[.5\dimexpr#2\relax]{\vphantom{#1}}}
\allowdisplaybreaks
\makeatletter
\renewcommand*{\@fnsymbol}[1]{
  \ifcase#1
  \or 1
  \or *
  \or 2
  \or 3
  \else *
  \fi}
\makeatother
\input{tcilatex}
\begin{document}

\title{Formulation of stress-gradient models describing \\
three-dimensional non-local medium}
\author{Sla\dj an Jeli\'{c}\thanks{
Department of Physics, Faculty of Sciences, University of Novi Sad, Trg D.
Obradovi\'{c}a 4, 21000 Novi Sad, Serbia, sladjan.jelic@df.uns.ac.rs} \ 
\thanks{%
Corresponding author}\quad Du\v{s}an Zorica\thanks{
Department of Physics, Faculty of Sciences, University of Novi Sad, Trg D.
Obradovi\'{c}a 4, 21000 Novi Sad, Serbia, dusan.zorica@df.uns.ac.rs} \ 
\thanks{%
Mathematical Institute, Serbian Academy of Arts and Sciences, Kneza Mihaila
36, 11000 Belgrade, Serbia, dusan\textunderscore zorica@mi.sanu.ac.rs} }
\maketitle

\begin{abstract}
\noindent Based on one-dimensional Eringen stress-gradient non-local model,
by considering non-locality vector and nabla operator instead of
non-locality scalar parameter and second order derivative, eight
three-dimensional Eringen non-local models are formulated and classified
into two groups: three scalar- and five tensor-type non-local models,
according to the type of used non-locality operator which is obtained via
various vector products of non-locality vector and nabla operator. The
compatibility conditions ensuring symmetricity of Cauchy stress tensor in
the case of the tensor-type model are derived. Furthermore, using the
Fourier integral transform with respect to spatial coordinates, non-locality
kernels (Green's functions), reflecting non-locality character of the
material, are derived for each of the proposed models. Except for the one
scalar-type model, all other models account for both local and non-local
contributions to Cauchy stress tensor. Additionally, the isotropy of
proposed models, as well as their non-local isotropy properties, both
depending on non-locality kernel, are examined. All scalar-type models are
isotropic, such that one of them is non-locally isotropic and two of them
correspond to a non-locally anisotropic body, while all tensor-type models
are anisotropic, such that there are two models that do not prefer direction
of non-locality, thus corresponding to a non-locally isotropic body, whereas
three models correspond to a body exhibiting non-locality along a specific
direction(s), thus corresponding to a non-locally anisotropic body.

\end{abstract}

\section{Introduction}

During the second half of the twentieth century, the non-local theory of
elasticity was developed, with Kr\"{o}ner and Eringen as pioneers of this
theory, see \cite{ERINGEN19721,eringen1,eringen2,ERINGEN1977339,Kroner},
postulating that the mechanical state, described in one-dimensional case by
stress $\sigma $, at one point can depend not only on local deformations $%
\varepsilon $, in accordance with the classical theory of elasticity, i.e.,
in accordance with Hooke's law%
\begin{equation*}
\sigma \left( x\right) =E\varepsilon \left( x\right) ,
\end{equation*}%
with $E$ as Young modulus, but also on deformations in the surrounding of
the point, i.e., according to%
\begin{equation*}
\sigma \left( x\right) =\int_{%
\mathbb{R}
}K\left( x,x^{\prime }\right) \varepsilon \left( x^{\prime }\right) \mathrm{d%
}x^{\prime },
\end{equation*}%
where function $K$ is the non-locality kernel, accounting for the measure of
influence of deformation at points $x^{\prime }$ surrounding the point $x$.
Non-local constitutive relations can be written in both integral and
differential form, so that the non-local influence in integral form of
constitutive relation is accounted for through the appropriate kernel
function $K$, also referred as Green's function, which accounts for the
non-local character explicitly, while in the case of differential form of
constitutive relation the non-local character is accounted for implicitly
through the form of equation and the corresponding model parameter.

Similarly as in the case of the one-dimensional stress-gradient Eringen model%
\begin{equation}
\left( 1-l^{2}\frac{\mathrm{d}^{2}}{\mathrm{d}x^{2}}\right) \sigma \left(
x\right) =\sigma ^{local}\left( x\right) ,  \label{1D-Eringen}
\end{equation}%
which becomes%
\begin{equation*}
\left( 1-l^{2}\frac{\mathrm{d}^{2}}{\mathrm{d}x^{2}}\right) \sigma \left(
x\right) =\sigma ^{local}\left( x\right) =E\varepsilon \left( x\right)
\end{equation*}%
in the case of non-local elasticity, by introducing the vector quantity $%
\boldsymbol{L}$ instead of the scalar quantity $l$ as a non-locality
parameter, as well as by replacing the spatial derivative with respect to $x$%
-coordinate $\frac{\mathrm{d}}{\mathrm{d}x}$ with the nabla operator $\nabla$%
, which contains derivatives with respect to three independent coordinates,
by using the scalar product $\cdot $, vector product $\times $, dyadic
product $\otimes $, and tensor contraction $:$, three scalar operators $%
\mathrm{S}_{i}=\mathrm{S}_{i}\left( \nabla \right) $, $i\in \left\{
1,2,3\right\} $, constructed in Section \ref{konstraksn} and listed in Table %
\ref{Modeli-new}, can be distinguished as counterparts of the operator $l^{2}%
\frac{\mathrm{d}^{2}}{\mathrm{d}x^{2}}$ in Eringen stress-gradient model (%
\ref{1D-Eringen}), so that these scalar operators constitute the scalar-type
stress-gradient Eringen three-dimensional models%
\begin{equation}
\left( 1-\mathrm{S}_{i}\left( \nabla \right) \right) \boldsymbol{\hat{\sigma}%
}\left( \boldsymbol{r}\right) =\boldsymbol{\hat{\sigma}}^{local}\left( 
\boldsymbol{r}\right), \quad i\in \left\{ 1,2,3\right\} ,
\label{non-loc-scalar-type}
\end{equation}%
generalizing one-dimensional Eringen model (\ref{1D-Eringen}), where $%
\boldsymbol{\hat{\sigma}}$ is the Cauchy stress tensor depending on radius
vector $\boldsymbol{r}\in 
\mathbb{R}
^{3}$, whose intensity is denoted by $r=|\boldsymbol{r}|$. Although scalar
operators $\mathrm{S}_{i},$ appearing in non-local model (\ref%
{non-loc-scalar-type}), retain the scalar character of the non-locality
operator of the one-dimensional Eringen model (\ref{1D-Eringen}), in the
case of modeling three-dimensional non-local body, one may also consider the
tensor type non-local operators, $\mathbf{\hat{T}}_{i}=\mathbf{\hat{T}}%
_{i}\left( \nabla \right) $, $i\in \left\{ 1,2,3,4,5\right\} $, constructed
in Section \ref{konstraksn} and listed in Table \ref{Modeli-new}, obtained
by using scalar, vector, and dyadic product of non-locality vector $%
\boldsymbol{L}$ and nabla operator $\nabla ,$ implying the tensor-type
stress-gradient Eringen three-dimensional models%
\begin{equation}
\left( \boldsymbol{\hat{I}}-\mathbf{\hat{T}}_{i}\left( \nabla \right)
\right) \boldsymbol{\hat{\sigma}}\left( \boldsymbol{r}\right) =\boldsymbol{%
\hat{\sigma}}^{local}\left( \boldsymbol{r}\right), \quad i\in \left\{
1,2,3,4,5\right\} ,  \label{non-loc-tensor-type}
\end{equation}%
also generalizing one-dimensional Eringen model (\ref{1D-Eringen}), where $%
\boldsymbol{\hat{I}}$ is the identity tensor.

Taking into account that local Cauchy stress tensor $\boldsymbol{\hat{\sigma}%
}^{local}$ needs to be symmetric, the compatibility conditions ensuring its
symmetricity are derived in Section \ref{konstraksn}, assuming that
non-local Cauchy stress tensor $\boldsymbol{\hat{\sigma}}$ is symmetric as
well. Furthermore, using the Fourier integral transform with respect to
spatial coordinates, non-locality kernels $\mathcal{K}_{i},$ $i\in \left\{
1,2,3\right\} $, and $\boldsymbol{\hat{\mathcal{K}}}_{i}$, $i\in \left\{
1,2,3,4,5\right\} ,$ reflecting the non-local character of the material, are
derived in Section \ref{Three-dimensional-cases} for each of the proposed
models and presented in Table \ref{Modeli-new}. While only one scalar-type
model assumes non-locality kernel being functional, another seven models
assume kernel that additionally contains Dirac delta distribution $\delta $
as the distributional part, thus accounting for both local and non-local
contributions to the Cauchy stress tensor. Additionally, the isotropy of
proposed models, understood in the sense that there is no shear-extensional
coupling and there are two independent model parameters, as well as whether
models correspond to a non-locally isotropic material, understood in the
sense that there are no preferred direction(s), are examined in Section \ref%
{Non-locality of the Cauchy stress tensor} by analyzing the character of
non-locality kernel. It is found that all of the scalar-type models are
isotropic, contrary to the tensor-type models which are all anisotropic. On
the other hand, in the case of scalar-type models, one corresponds to a
non-locally isotropic three dimensional body and two correspond to a
non-locally anisotropic body, while in the case of tensor-type models, there
are two models that do not prefer direction of non-locality determined by
the non-locality vector $\boldsymbol{L}$, thus corresponding to a
non-locally isotropic body, whereas each of three models correspond to a
body exhibiting non-locality along a specific direction(s), thus
corresponding to a non-locally anisotropic three-dimensional body.

In the case of three-dimensional non-local elasticity, the Hooke's law%
\begin{equation}
\boldsymbol{\hat{\sigma}}^{local}\left( \boldsymbol{r}\right) =\lambda 
\limfunc{tr}\boldsymbol{\hat{\varepsilon}}\left( \boldsymbol{r}\right) 
\boldsymbol{\hat{I}}+2\mu \boldsymbol{\hat{\varepsilon}}\left( \boldsymbol{r}%
\right)   \label{Hooke's-law}
\end{equation}%
transforms into%
\begin{gather}
\left( 1-\mathrm{S}_{i}\left( \nabla \right) \right) \boldsymbol{\hat{\sigma}%
}\left( \boldsymbol{r}\right) =\boldsymbol{\hat{\sigma}}^{local}\left( 
\boldsymbol{r}\right) =\lambda \limfunc{tr}\boldsymbol{\hat{\varepsilon}}%
\left( \boldsymbol{r}\right) \boldsymbol{\hat{I}}+2\mu \boldsymbol{\hat{%
\varepsilon}}\left( \boldsymbol{r}\right) \quad \text{and}
\label{scalar-non-local-elasticity} \\
\left( \boldsymbol{\hat{I}}-\mathbf{\hat{T}}_{i}\left( \nabla \right)
\right) \boldsymbol{\hat{\sigma}}\left( \boldsymbol{r}\right) =\boldsymbol{%
\hat{\sigma}}^{local}\left( \boldsymbol{r}\right) =\lambda \limfunc{tr}%
\boldsymbol{\hat{\varepsilon}}\left( \boldsymbol{r}\right) \boldsymbol{\hat{I%
}}+2\mu \boldsymbol{\hat{\varepsilon}}\left( \boldsymbol{r}\right) 
\label{tensor-non-local-elasticity}
\end{gather}%
for scalar- and tensor-type stress-gradient Eringen three-dimensional
models, where $\lambda $ and $\mu $ are Lam\'{e} coefficients and $%
\boldsymbol{\hat{\varepsilon}}$ is the infinitesimal strain tensor. Local
Cauchy stress tensor $\boldsymbol{\hat{\sigma}}^{local}$ can be assumed
differently if one considers non-local generalization of non-isotropic
elasticity, or even coupling non-locality with viscoelasticity.

\begin{sidewaystable}
\begin{center}
\begin{tabular}{|c|c|c|c|c|c|} \hline \xrowht{25pt}
 & Model & \multicolumn{2}{c|}{Non-locality operator}  & \multicolumn{2}{c|}{Non-locality kernels}  \\ \cline{1-6} \xrowht{35pt}

\multirow{3}{*}{\multirowcell{1}{\rotatebox{90}{Scalar-type}}} & $S_1$ & \multirow{3}{*}{\multirowcell{4}{ ${\rm{S}}_{i}(\nabla)$ }} & $\left( \boldsymbol{L}\cdot \boldsymbol{L}\right) \left( \nabla \cdot \nabla \right)$  & \multirow{3}{*}{\multirowcell{4}{  ${\mathcal{K}}_{i}  \left( \boldsymbol{r}\right) $ }} & $\frac{1}{4\pi rL^{2}}\mathrm{e}^{-\frac{r}{L}}$ \\ \cline{2-2} \cline{4-4} \cline{6-6} \xrowht{35pt}

 & $S_2$  & & $\left( \boldsymbol{L}\cdot \nabla \right) ^{2} $ &  & $\frac{1}{2L}\delta \left( \boldsymbol{r}_{\perp }\right) \mathrm{e}^{-\frac{\left\vert r_{\parallel
}\right\vert }{L}}$ \\  \cline{2-2} \cline{4-4} \cline{6-6} \xrowht{35pt}

 & $S_3$ & & $\left(\boldsymbol{L}\times \nabla \right) \cdot \left( \boldsymbol{L}\times \nabla\right) $   &  & $\frac{1}{2\pi L^{2}}\delta \left( r_{\parallel }\right) K_{0}\left( \frac{r_{\perp }}{L}\right) $\\ \cline{1-6}  \xrowht{35pt}

\multirow{11.5}{*}{\multirowcell{3}{\rotatebox{90}{Tensor-type}}} & $T_{1}$ & \multirow{5}{*}{\multirowcell{10.75}{ $\boldsymbol{\hat{{\rm{T}}}}_{i}(\nabla)$ }} &$\left( \boldsymbol{L}\otimes \nabla \right) ^{\mathrm{T}}\left( \boldsymbol{%
L}\otimes \nabla \right) $ &  \multirow{5}{*}{\multirowcell{10.75}{  $\boldsymbol{\hat{{\mathcal{K}}}}_{i}  \left( \boldsymbol{r}\right) $ }} & $2\delta
\left( \boldsymbol{r}\right) \left( \boldsymbol{\hat{I}}-2\frac{\boldsymbol{r%
}\otimes \boldsymbol{r}}{r^{2}}\right) -\frac{1}{4\pi r^{3}}\mathrm{e}^{-%
\frac{r}{L}}\left( \left( 1+\frac{r}{L}\right) \boldsymbol{\hat{I}}+\left(
3\left( 1+\frac{r}{L}\right) +\frac{r^{2}}{L^{2}}\right) \frac{\boldsymbol{r}%
\otimes \boldsymbol{r}}{r^{2}}\right)$ \\  \cline{2-2} \cline{4-4} \cline{6-6} \xrowht{35pt}

 & $T_{2}$  & & $\left( \boldsymbol{L}\otimes \nabla \right) \left( \boldsymbol{L}\otimes
\nabla \right) ^{\mathrm{T}}$ &  &  $\delta
\left( \boldsymbol{r}\right) \left( \boldsymbol{\hat{I}}-\frac{\boldsymbol{L}%
\otimes \boldsymbol{L}}{L^{2}}\right) +\frac{1}{4\pi L^{2}r}\mathrm{e}^{-%
\frac{r}{L}}\frac{\boldsymbol{L}\otimes \boldsymbol{L}}{L^{2}}$ \\  \cline{2-2} \cline{4-4} \cline{6-6} \xrowht{35pt}

 & $T_3$ & & \makecell[l]{$\left( \boldsymbol{L}\otimes \nabla \right) \left( \boldsymbol{L}\otimes
\nabla \right) $ \smallskip \\ $ \quad  =\left( \boldsymbol{L}\otimes \boldsymbol{L}\right) \left(
\nabla \otimes \nabla \right) $}  & & $\delta
\left( \boldsymbol{r}\right) \left( \boldsymbol{\hat{I}}-\frac{\boldsymbol{L}%
\otimes \boldsymbol{L}}{L^{2}}\right) +\frac{1}{4L}\delta \left( \boldsymbol{%
r}_{\perp }\right) \mathrm{e}^{-\frac{\left\vert r_{\parallel }\right\vert }{%
L}}\frac{\boldsymbol{L}\otimes \boldsymbol{L}}{L^{2}}+\frac{\delta \left( 
\boldsymbol{r}_{\perp }\right) }{r_{\perp }}\mathrm{e}^{-\frac{\left\vert
r_{\parallel }\right\vert }{L}}\left( \frac{\boldsymbol{L}}{L}\otimes \frac{%
\boldsymbol{r}_{\perp }}{r_{\perp }}\right) $ \\ \cline{2-2} \cline{4-4} \cline{6-6} \xrowht{35pt}

 & $T_4$ & & \makecell[l]{$\left( \boldsymbol{L}\otimes \nabla \right) ^{\mathrm{T}}\left( \boldsymbol{%
L}\otimes \nabla \right) ^{\mathrm{T}} $ \smallskip \\   $ \quad =\left( \nabla \otimes \nabla \right)
\left( \boldsymbol{L}\otimes \boldsymbol{L}\right) $  } & & $\delta
\left( \boldsymbol{r}\right) \left( \boldsymbol{\hat{I}}-\frac{\boldsymbol{L}%
\otimes \boldsymbol{L}}{L^{2}}\right) +\frac{1}{4L}\delta \left( \boldsymbol{%
r}_{\perp }\right) \mathrm{e}^{-\frac{\left\vert r_{\parallel }\right\vert }{%
L}}\frac{\boldsymbol{L}\otimes \boldsymbol{L}}{L^{2}}+\frac{\delta \left( 
\boldsymbol{r}_{\perp }\right) }{r_{\perp }}\mathrm{e}^{-\frac{\left\vert
r_{\parallel }\right\vert }{L}}\left( \frac{%
\boldsymbol{r}_{\perp }}{r_{\perp }} \otimes  \frac{\boldsymbol{L}}{L}\right) $ \\ \cline{2-2} \cline{4-4} \cline{6-6} \xrowht{35pt}

 & $T_5$ & &  $\left( \boldsymbol{L}\times \nabla \right)
\otimes \left( \boldsymbol{L}\times \nabla \right)$ &  & 
\makecell{ $\delta \left( \boldsymbol{r}\right) \left( \boldsymbol{\hat{I}}-\left( 
\frac{\boldsymbol{L}}{L}\times \frac{\boldsymbol{r}_{\perp }}{r_{\perp }}%
\right) \otimes \left( \frac{\boldsymbol{L}}{L}\times \frac{\boldsymbol{r}%
_{\perp }}{r_{\perp }}\right) \right) -\frac{1}{2\pi L}\delta \left(
r_{\parallel }\right) \frac{1}{r_{\perp }}K_{1}\left( \frac{r_{\perp }}{L}%
\right) \frac{\boldsymbol{r}_{\perp }\otimes \boldsymbol{r}_{\perp }}{%
r_{\perp }^{2}} $\\
$+\frac{1}{2\pi L}\delta \left( r_{\parallel }\right) \left( \frac{1}{L}%
K_{0}\left( \frac{r_{\perp }}{L}\right) +\frac{1}{r_{\perp }}K_{1}\left( 
\frac{r_{\perp }}{L}\right) \right) \left( \left( \frac{\boldsymbol{L}}{L}%
\times \frac{\boldsymbol{r}_{\perp }}{r_{\perp }}\right) \otimes \left( 
\frac{\boldsymbol{L}}{L}\times \frac{\boldsymbol{r}_{\perp }}{r_{\perp }}%
\right) \right) $} \\ \cline{1-6}
\end{tabular}
\end{center}
\caption{Scalar and tensor operators, as well as non-locality kernels, corresponding to non-local models, where $\delta$ denotes Dirac delta distribution, while $K_0$ and $K_1$ are modified Bessel functions of the second kind (Macdonald functions) of the zeroth and first orders, as well as where $r_{\parallel }= \boldsymbol{r} \cdot \frac{\boldsymbol{L} }{L}$, $\boldsymbol{r}_{\perp }= \boldsymbol{r}-\boldsymbol{r}_{\parallel }$ with $\boldsymbol{r}_{\parallel }=r_{\parallel } \frac{\boldsymbol{L} }{L}$, and $r_{\perp }= \left\vert \boldsymbol{r}_{\perp } \right\vert $.}
\label{Modeli-new}
\end{sidewaystable}

Eringen's monograph \textit{Nonlocal Continuum Field Theories} \cite{eringen}
presents a unified theoretical framework for continuum field theories
incorporating spatial and temporal non-local effects, where a
straightforward generalization of Eringen model (\ref{1D-Eringen}) can be
found, taking the form of model $S_{1}$, see (\ref{non-loc-scalar-type}) and
Table \ref{Modeli-new}. The generalized three-dimensional Eringen non-local
model has been employed for modeling non-local material behavior in various
theoretical frameworks, including wave propagation theory \cite%
{LazarAgiSofi1}, stability theory \cite{Wu2020}, thermo-elasticity \cite%
{Dastjerdi}, as well as magneto-electro-elasticity \cite{Arefi}.

Solving the three-dimensional inhomogeneous Helmholtz-type constitutive
equation, which accounts for anisotropic non-locality through a symmetric
length scale tensor with six independent components, Lazar and Po found the
non-locality kernel in the general form (B.12) in Appendix B of \cite%
{LazarPo}, which reduces to the non-locality kernel of model $S_{1}$ when
diagonal terms of the tensor are equal and off-diagonal terms are equal to
zero. Model $S_{1}$ and corresponding non-locality kernel $\mathcal{K}%
_{1}\left( r\right) \mathcal{=}\frac{1}{4\pi rL^{2}}\mathrm{e}^{-\frac{r}{L}}
$, see Table \ref{Modeli-new}, are originally proposed by Eringen \cite%
{eringen1} and later employed in studies on the three-dimensional non-local
problems, see \cite{LazarAgiSofi1,LazarPo1}. The mentioned kernel represents
Green's function corresponding to the isotropic Helmholtz operator \cite%
{LazarPo}, and it is of the isotropic Yukawa-type potential form, associated
with materials of cubic crystal symmetry \cite{LazarAgPo}.

On the other hand, according to authors' knowledge, except for model $S_{1}$%
, all other proposed models and their corresponding non-locality kernels,
listed in Table \ref{Modeli-new}, represent new constitutive relations.
Namely, there are articles \cite%
{Parisa,Lanzoni,Shaat2021,TunaKirca,TunaKirca1} in which the non-locality
kernel is postulated as a direct generalization of the non-locality kernel $%
\mathcal{K}\left( x\right) \mathcal{=}\frac{1}{2l}\mathrm{e}^{-\frac{%
\left\vert x\right\vert }{l}}$ for the one-dimensional Eringen model, see (%
\ref{1D-Eringen-kernel}) below, so that, being of the form $\frac{1}{2L}%
\mathrm{e}^{-\frac{\left\vert r_{\parallel }\right\vert }{L}}$, it resembles
to the kernel $\mathcal{K}_{2}\left( r\right) \mathcal{=}\frac{1}{2L}\delta
\left( \boldsymbol{r}_{\perp }\right) \mathrm{e}^{-\frac{\left\vert
r_{\parallel }\right\vert }{L}}$ which corresponds to model $S_{2}$, however
with a slight difference since it does not contain the Dirac delta
distribution $\delta \left( \boldsymbol{r}_{\perp }\right) $. Similarly,
Eringen proposed the non-locality kernel for the case of two-dimensional
materials, see \cite{eringen1,eringen3,eringen}, used in \cite{Farajpour}
and derived in \cite{LazarAgiSofi} starting from the Helmholtz-type
constitutive equation corresponding to two-dimensional non-local material,
which accounts for non-locality through a symmetric length scale tensor with
three independent tensor components. Eringen's two-dimensional kernel in the
more general case, for a three-dimensional non-local material, takes the
form similar to the kernel $\mathcal{K}_{3}\left( r\right) \mathcal{=}\frac{1%
}{2\pi L^{2}}\delta \left( r_{\parallel }\right) K_{0}\left( \frac{r_{\perp }%
}{L}\right) $ corresponding to model $S_{3}$, see Table \ref{Modeli-new},
such that the difference between these two kernels is the Dirac delta
distribution $\delta \left( r_{\parallel }\right) $.

Finally, several expressions for the tensor-type non-local kernels
associated with an anisotropic operator and formulated by the means of an
arbitrary matrix, together with their derivation and a discussion of
specific cases corresponding to particular structural configurations of
materials, are presented in \cite{LazarPo1,LazarPo}, while in \cite{LazarMA}
the tensor-type non-local kernel for the two-dimensional case is presented,
with the tensor defined by the Burgers vector and direction of the
dislocation line.

Besides non-locality kernels characterized by exponential or Bessel-type
attenuation, in the literature, various types of non-locality kernels, such
as power-law error-function-based, bell-shaped, and conical kernels, have
been proposed in \cite{eringen1,Patnaik,Polizzotto,Schwartz}. 
However, analytical derivations in which the constitutive relations yield a
specific form of the non-locality kernel are still lacking.

\section{Construction of non-local operators\label{konstraksn}}

In order to generalize the operator $l^{2}\frac{\mathrm{d}^{2}}{\mathrm{d}%
x^{2}}$ in one-dimensional Eringen stress-gradient model (\ref{1D-Eringen})
to scalar- and tensor-type non-local operators $\mathrm{S}_{i}=\mathrm{S}%
_{i}\left( \nabla \right) $, $i\in\{1,2,3\}$, and $\mathbf{\hat{T}}_{i}=%
\mathbf{\hat{T}}_{i}\left( \nabla \right)$, $i\in\{1,2,3,4,5\}$, appearing
in three-dimensional Eringen stress-gradient models (\ref%
{non-loc-scalar-type}) and (\ref{non-loc-tensor-type}), one uses
non-locality vector $\boldsymbol{L} $ and nabla operator $\nabla $, along
with different vector products, namely scalar product $\cdot $, vector
product $\times $, dyadic product $\otimes $, as well as tensor contraction $%
:$. The expressions representing both scalar- and tensor-type non-locality
operators are summarized in Table \ref{Modeli-new}, while their construction
and their possibly equivalent forms are discussed in the sequel. Also, the
compatibility conditions, guaranteeing symmetricity of the Cauchy stress
tensor $\boldsymbol{\hat{\sigma}}$, are derived.


Scalar-type non-locality operators require scalar product between
non-locality vector $\boldsymbol{L}$ and nabla operator $\nabla $, such that
one has%
\begin{equation}
\mathrm{S}_{1}\left( \nabla \right) =\left( \boldsymbol{L}\cdot \boldsymbol{L%
}\right) \left( \nabla \cdot \nabla \right) \quad \text{and\quad }\mathrm{S}%
_{2}\left( \nabla \right) =\left( \boldsymbol{L}\cdot \nabla \right) ^{2}
\label{S1-i-S2}
\end{equation}%
Note, due to the property of tensor contraction $\left( \boldsymbol{a}%
\otimes \boldsymbol{b}\right) :\left( \boldsymbol{c}\otimes \boldsymbol{d}%
\right) =\left( \boldsymbol{a}\cdot \boldsymbol{c}\right) \left( \boldsymbol{%
b}\cdot \boldsymbol{d}\right) $, where tensors are constructed through the
dyadic product of vectors, one has equivalent forms of scalar-type
non-locality operators $\mathrm{S}_{1}$ and $\mathrm{S}_{2}$, namely%
\begin{equation*}
\mathrm{S}_{1}\left( \nabla \right) =\left( \boldsymbol{L}\otimes \nabla
\right) :\left( \boldsymbol{L}\otimes \nabla \right) =\left( \boldsymbol{L}%
\otimes \nabla \right) ^{\mathrm{T}}:\left( \boldsymbol{L}\otimes \nabla
\right) ^{\mathrm{T}}=\left( \boldsymbol{L}\cdot \boldsymbol{L}\right)
\left( \nabla \cdot \nabla \right),
\end{equation*}%
and%
\begin{equation*}
\mathrm{S}_{2}\left( \nabla \right) =\left( \boldsymbol{L}\otimes \nabla
\right) :\left( \boldsymbol{L}\otimes \nabla \right) ^{\mathrm{T}}=\left( 
\boldsymbol{L}\otimes \nabla \right) ^{\mathrm{T}}:\left( \boldsymbol{L}%
\otimes \nabla \right) =\left( \boldsymbol{L}\otimes \boldsymbol{L}\right)
:\left( \nabla \otimes \nabla \right) =\left( \nabla \otimes \nabla \right)
:\left( \boldsymbol{L}\otimes \boldsymbol{L}\right)=\left( \boldsymbol{L}%
\cdot \nabla \right) ^{2} .
\end{equation*}%
Moreover, there is one additional form of scalar-type non-locality operator $%
\mathrm{S}_{1}$ and two additional forms of scalar-type non-locality
operator $\mathrm{S}_{2}$, namely 
\begin{equation*}
\mathrm{S}_{1}\left( \nabla \right) =\boldsymbol{L}\left( \boldsymbol{L}%
\otimes \nabla \right) \nabla =\left( \boldsymbol{L}\cdot \boldsymbol{L}%
\right) \left( \nabla \cdot \nabla \right)
\end{equation*}%
and%
\begin{equation*}
\mathrm{S}_{2}\left( \nabla \right) =\nabla \left( \boldsymbol{L}\otimes 
\boldsymbol{L}\right) \nabla =\boldsymbol{L}\left( \nabla \otimes \nabla
\right) \boldsymbol{L}=\left( \boldsymbol{L}\cdot \nabla \right) ^{2},
\end{equation*}%
due to $\boldsymbol{a}\left( \boldsymbol{b}\otimes \boldsymbol{c}\right) 
\boldsymbol{d}=\boldsymbol{a}\cdot \left( \left( \boldsymbol{b}\otimes 
\boldsymbol{c}\right) \boldsymbol{d}\right) =\left( \boldsymbol{a}\cdot 
\boldsymbol{b}\right) \left( \boldsymbol{c}\cdot \boldsymbol{d}\right) $,
since $\left( \boldsymbol{a}\otimes \boldsymbol{b}\right) \boldsymbol{c}=%
\boldsymbol{a}\left( \boldsymbol{b} \cdot \boldsymbol{c}\right) $. Contrary
to the case of scalar-type non-locality operators $\mathrm{S}_{1}$ and $%
\mathrm{S}_{2}$, which are not uniquely defined, the operator%
\begin{equation*}
\mathrm{S}_{3}\left( \nabla \right) =\left( \boldsymbol{L}\times \nabla
\right) \cdot \left( \boldsymbol{L}\times \nabla \right)
\end{equation*}%
does not have another equivalent form. Operators $\mathrm{S}_{i}$, $%
i\in\{1,2,3\}$, do not require compatibility conditions in order to ensure
symmetrical form of the Cauchy stress tensor, since they are scalar-type
operators.

Tensor-type non-locality operators require dyadic product between
non-locality vector $\boldsymbol{L}$ and nabla operator $\nabla $, such that
one has%
\begin{gather*}
\mathbf{\hat{T}}_{1}\left( \nabla \right) =\left( \boldsymbol{L}\otimes
\nabla \right) ^{\mathrm{T}}\left( \boldsymbol{L}\otimes \nabla \right) ,%
\text{\quad }\mathbf{\hat{T}}_{2}\left( \nabla \right) =\left( \boldsymbol{L}%
\otimes \nabla \right) \left( \boldsymbol{L}\otimes \nabla \right) ^{\mathrm{%
T}}, \\
\mathbf{\hat{T}}_{3}\left( \nabla \right) =\left( \boldsymbol{L}\otimes
\nabla \right) \left( \boldsymbol{L}\otimes \nabla \right) ,\quad \text{%
and\quad }\mathbf{\hat{T}}_{4}\left( \nabla \right) =\left( \boldsymbol{L}%
\otimes \nabla \right) ^{\mathrm{T}}\left( \boldsymbol{L}\otimes \nabla
\right) ^{\mathrm{T}}.
\end{gather*}%
Note, due to the property $\left( \boldsymbol{a}\otimes \boldsymbol{b}%
\right) \left( \boldsymbol{c}\otimes \boldsymbol{d}\right) =\left( \left( 
\boldsymbol{a}\otimes \boldsymbol{b}\right) \boldsymbol{c}\right) \otimes 
\boldsymbol{d}=\left( \boldsymbol{b}\cdot \boldsymbol{c}\right) \left( 
\boldsymbol{a}\otimes \boldsymbol{d}\right) $, since $\left( \boldsymbol{a}%
\otimes \boldsymbol{b}\right) \boldsymbol{c}=\boldsymbol{a}\left( 
\boldsymbol{b}\cdot \boldsymbol{c}\right) $ one has equivalent forms of
tensor-type non-locality operators $\mathbf{\hat{T}}_{1}$ - $\mathbf{\hat{T}}%
_{4}$, namely%
\begin{gather*}
\mathbf{\hat{T}}_{1}\left( \nabla \right) =\left( \boldsymbol{L}\cdot 
\boldsymbol{L}\right) \left( \nabla \otimes \nabla \right) ,\text{\quad }%
\mathbf{\hat{T}}_{2}\left( \nabla \right) =\left( \boldsymbol{L}\otimes 
\boldsymbol{L}\right) \left( \nabla \cdot \nabla \right) , \\
\mathbf{\hat{T}}_{3}\left( \nabla \right) =\left( \boldsymbol{L}\cdot \nabla
\right) \left( \boldsymbol{L}\otimes \nabla \right) ,\quad \text{and\quad }%
\mathbf{\hat{T}}_{4}\left( \nabla \right) =\left( \boldsymbol{L}\cdot \nabla
\right) \left( \boldsymbol{L}\otimes \nabla \right) ^{\mathrm{T}}.
\end{gather*}%
Moreover, each of tensor-type non-locality operators $\mathbf{\hat{T}}_{3}$
and $\mathbf{\hat{T}}_{4}$ has one more additional form, namely%
\begin{equation*}
\mathbf{\hat{T}}_{3}\left( \nabla \right) =\left( \boldsymbol{L}\otimes 
\boldsymbol{L}\right) \left( \nabla \otimes \nabla \right) \quad \text{%
and\quad }\mathbf{\hat{T}}_{4}\left( \nabla \right) =\left( \nabla \otimes
\nabla \right) \left( \boldsymbol{L}\otimes \boldsymbol{L}\right) .
\end{equation*}%
Similarly as in the case of scalar-type non-locality operator $\mathrm{S}%
_{3} $, there is a uniquely constructed tensor-type non-locality operator%
\begin{equation*}
\mathbf{\hat{T}}_{5}\left( \nabla \right) =\left( \boldsymbol{L}\times
\nabla \right) \otimes \left( \boldsymbol{L}\times \nabla \right) ,
\end{equation*}%
obtained through the vector product of non-locality vector $\boldsymbol{L}$
and nabla operator $\nabla $.

Assuming symmetricity of the Cauchy stress tensor $\boldsymbol{\hat{\sigma}}$%
, tensor-type operators $\mathbf{\hat{T}}_{i},$ $i\in \{1,2,3,4,5\},$
require the compatibility conditions, guaranteeing symmetrical form of the
local Cauchy stress tensor $\boldsymbol{\hat{\sigma}}^{local}$, since one of
its constituents is expressed in terms of the action of tensor-type operator
on the Cauchy stress tensor $\boldsymbol{\hat{\sigma}}$, see (\ref%
{non-loc-tensor-type}), and therefore the compatibility condition requests
that the off-diagonal tensor components satisfy%
\begin{equation}
\left[ \mathbf{\hat{T}}_{i}\left( \nabla \right) \boldsymbol{\hat{\sigma}}%
\left( \boldsymbol{r}\right) \right] _{\alpha \beta }=\left[ \mathbf{\hat{T}}%
_{i}\left( \nabla \right) \boldsymbol{\hat{\sigma}}\left( \boldsymbol{r}%
\right) \right] _{\beta \alpha },\quad \alpha ,\beta \in \left\{
1,2,3\right\} ,\quad \alpha \neq \beta .  \label{compatibility-condition}
\end{equation}

In the case of model $T_{1},$ the compatibility condition (\ref%
{compatibility-condition}) reads%
\begin{equation}
L^{2}\func{div}\left( \func{curl}\boldsymbol{\hat{\sigma}}\left( \boldsymbol{%
r}\right) \right) ^{\mathrm{T}}=\boldsymbol{L}\left( \boldsymbol{L}\otimes
\nabla \right) \left( \func{curl}\boldsymbol{\hat{\sigma}}\left( \boldsymbol{%
r}\right) \right) ^{\mathrm{T}}=\boldsymbol{0},  \label{cc-T1}
\end{equation}%
since in the Cartesian coordinate system one has%
\begin{align*}
\mathbf{\hat{T}}_{1}\left( \nabla \right) \boldsymbol{\hat{\sigma}}\left( 
\boldsymbol{r}\right) & =L^{2}\left( \nabla \otimes \nabla \right) 
\boldsymbol{\hat{\sigma}}\left( \boldsymbol{r}\right)  \\
& =L^{2}\left( \partial _{x_{\alpha }x_{\mu }}^{2}\sigma _{\nu \beta
}\right) \left( \boldsymbol{e}_{\alpha }\otimes \boldsymbol{e}_{\mu }\right)
\left( \boldsymbol{e}_{\nu }\otimes \boldsymbol{e}_{\beta }\right)  \\
& =L^{2}\left( \partial _{x_{\alpha }x_{\mu }}^{2}\sigma _{\nu \beta
}\right) \delta _{\mu \nu }\left( \boldsymbol{e}_{\alpha }\otimes 
\boldsymbol{e}_{\beta }\right)  \\
& =L^{2}\left( \partial _{x_{\alpha }x_{\mu }}^{2}\sigma _{\mu \beta
}\right) \left( \boldsymbol{e}_{\alpha }\otimes \boldsymbol{e}_{\beta
}\right) ,
\end{align*}%
where the summation convention is assumed, implying%
\begin{eqnarray*}
&&L^{2}\partial _{x_{\alpha }x_{\mu }}^{2}\sigma _{\mu \beta }=L^{2}\partial
_{x_{\beta }x_{\mu }}^{2}\sigma _{\mu \alpha },\quad \alpha \neq \beta \quad 
\text{i.e.,} \\
&&L^{2}\partial _{x_{\mu }}\left( \partial _{x_{\alpha }}\sigma _{\mu \beta
}-\partial _{x_{\beta }}\sigma _{\mu \alpha }\right) =L^{2}\epsilon _{\alpha
\beta \nu }\partial _{x_{\mu }}\partial _{x_{\alpha }}\sigma _{\mu \beta }=0,
\end{eqnarray*}%
by the compatibility condition (\ref{compatibility-condition}), with $\alpha
\neq \beta \neq \nu $ and values of $\alpha ,\beta ,\nu $ being of even
parity, where $\epsilon _{\alpha \beta \nu }$ is the Levi-Civita symbol, so
that the previous expression reduces to (\ref{cc-T1}), due to%
\begin{align}
\boldsymbol{a}\left( \boldsymbol{b}\times \boldsymbol{\hat{\sigma}}\left( 
\boldsymbol{r}\right) \right) ^{\mathrm{T}}& =\left( a_{\mu }\boldsymbol{e}%
_{\mu }\right) \left( \left( b_{\alpha }\boldsymbol{e}_{\alpha }\right)
\times \left( \sigma _{\beta \eta }\left( \boldsymbol{e}_{\beta }\otimes 
\boldsymbol{e}_{\eta }\right) \right) \right) ^{\mathrm{T}}  \notag \\
& =\left( a_{\mu }\boldsymbol{e}_{\mu }\right) \left( b_{\alpha }\sigma
_{\beta \eta }\left( \left( \boldsymbol{e}_{\alpha }\times \boldsymbol{e}%
_{\beta }\right) \otimes \boldsymbol{e}_{\eta }\right) \right) ^{\mathrm{T}}
\notag \\
& =\left( a_{\mu }\boldsymbol{e}_{\mu }\right) \left( \epsilon _{\alpha
\beta \nu }b_{\alpha }\sigma _{\beta \eta }\left( \boldsymbol{e}_{\nu
}\otimes \boldsymbol{e}_{\eta }\right) \right) ^{\mathrm{T}}  \notag \\
& =\epsilon _{\alpha \beta \nu }a_{\mu }b_{\alpha }\sigma _{\beta \eta
}\left( \boldsymbol{e}_{\mu }\left( \boldsymbol{e}_{\eta }\otimes 
\boldsymbol{e}_{\nu }\right) \right)   \notag \\
& =\epsilon _{\alpha \beta \nu }a_{\mu }b_{\alpha }\sigma _{\beta \eta
}\delta _{\mu \eta }\boldsymbol{e}_{\nu }  \notag \\
& =\epsilon _{\alpha \beta \nu }a_{\mu }b_{\alpha }\sigma _{\mu \beta }%
\boldsymbol{e}_{\nu }.  \label{cc-rotor}
\end{align}

In the case of model $T_{2},$ the compatibility condition (\ref%
{compatibility-condition}) reads%
\begin{equation}
\boldsymbol{L}\left( \boldsymbol{L}\times \left( \left( \nabla \cdot \nabla
\right) \boldsymbol{\hat{\sigma}}\left( \boldsymbol{r}\right) \right)
\right) ^{\mathrm{T}}=\func{div}\left( \left( \boldsymbol{L}\otimes \nabla
\right) ^{\mathrm{T}}\left( \boldsymbol{L}\times \boldsymbol{\hat{\sigma}}%
\left( \boldsymbol{r}\right) \right) ^{\mathrm{T}}\right) =\boldsymbol{0},
\label{cc-T2}
\end{equation}%
since in the Cartesian coordinate system one has%
\begin{align*}
\mathbf{\hat{T}}_{2}\left( \nabla \right) \boldsymbol{\hat{\sigma}}\left( 
\boldsymbol{r}\right) &=\left( \boldsymbol{L}\otimes \nabla \right) \left( 
\boldsymbol{L}\otimes \nabla \right) ^{\mathrm{T}}\boldsymbol{\hat{\sigma}}%
\left( \boldsymbol{r}\right) =\left( \boldsymbol{L}\otimes \boldsymbol{L}%
\right) \left( \nabla ^{2}\boldsymbol{\hat{\sigma}}\left( \boldsymbol{r}%
\right) \right) \\
&=L_{\alpha }L_{\mu }\left( \nabla ^{2}\sigma _{\nu \beta }\right) \left( 
\boldsymbol{e}_{\alpha }\otimes \boldsymbol{e}_{\mu }\right) \left( 
\boldsymbol{e}_{\nu }\otimes \boldsymbol{e}_{\beta }\right) \\
&=L_{\alpha }L_{\mu }\left( \nabla ^{2}\sigma _{\nu \beta }\right) \delta
_{\mu \nu }\left( \boldsymbol{e}_{\alpha }\otimes \boldsymbol{e}_{\beta
}\right) \\
&=L_{\alpha }L_{\mu }\left( \nabla ^{2}\sigma _{\mu \beta }\right) \left( 
\boldsymbol{e}_{\alpha }\otimes \boldsymbol{e}_{\beta }\right) ,
\end{align*}%
implying the following expression%
\begin{eqnarray*}
&&L_{\alpha }L_{\mu }\nabla ^{2}\sigma _{\mu \beta }=L_{\beta }L_{\mu
}\nabla ^{2}\sigma _{\mu \alpha },\quad \alpha \neq \beta \quad \text{i.e.,}
\\
&&L_{\mu }\left( L_{\alpha }\nabla ^{2}\sigma _{\mu \beta }-L_{\beta }\nabla
^{2}\sigma _{\mu \alpha }\right) =\epsilon _{\alpha \beta \nu }L_{\mu
}L_{\alpha }\nabla ^{2}\sigma _{\mu \beta }=0,
\end{eqnarray*}%
by the compatibility condition (\ref{compatibility-condition}), so that the
previous expression reduces to (\ref{cc-T2}), due to (\ref{cc-rotor}).

In the case of model $T_{3},$ the compatibility condition (\ref%
{compatibility-condition}) reads%
\begin{equation}
\left( \boldsymbol{L}\cdot \nabla \right) \left( \func{div}\left( 
\boldsymbol{L}\times \boldsymbol{\hat{\sigma}}\left( \boldsymbol{r}\right)
\right) ^{\mathrm{T}}\right) =\boldsymbol{L}\left( \nabla \otimes \nabla
\right) \left( \boldsymbol{L}\times \boldsymbol{\hat{\sigma}}\left( 
\boldsymbol{r}\right) \right) ^{\mathrm{T}}=\boldsymbol{0},  \label{cc-T3}
\end{equation}%
since in the Cartesian coordinate system one has%
\begin{align*}
\mathbf{\hat{T}}_{3}\left( \nabla \right) \boldsymbol{\hat{\sigma}}\left( 
\boldsymbol{r}\right) &=\left( \boldsymbol{L}\otimes \boldsymbol{L}\right)
\left( \nabla \otimes \nabla \right) \boldsymbol{\hat{\sigma}}\left( 
\boldsymbol{r}\right) \\
&=L_{\alpha }L_{\eta }\left( \partial _{x_{\xi }x_{\mu }}^{2}\sigma _{\zeta
\beta }\right) \left( \boldsymbol{e}_{\alpha }\otimes \boldsymbol{e}_{\eta
}\right) \left( \boldsymbol{e}_{\xi }\otimes \boldsymbol{e}_{\mu }\right)
\left( \boldsymbol{e}_{\zeta }\otimes \boldsymbol{e}_{\beta }\right) \\
&=L_{\alpha }L_{\eta }\left( \partial _{x_{\xi }x_{\mu }}^{2}\sigma _{\zeta
\beta }\right) \delta _{\eta \xi }\delta _{\mu \zeta }\left( \boldsymbol{e}%
_{\alpha }\otimes \boldsymbol{e}_{\beta }\right) \\
&=L_{\alpha }L_{\xi }\left( \partial _{x_{\xi }x_{\mu }}^{2}\sigma _{\mu
\beta }\right) \left( \boldsymbol{e}_{\alpha }\otimes \boldsymbol{e}_{\beta
}\right) \\
&=\left( \boldsymbol{L}\cdot \nabla \right) L_{\alpha }\left( \partial
_{x_{\mu }}\sigma _{\mu \beta }\right) \left( \boldsymbol{e}_{\alpha
}\otimes \boldsymbol{e}_{\beta }\right)
\end{align*}%
implying the following expression%
\begin{eqnarray*}
&&\left( \boldsymbol{L}\cdot \nabla \right) L_{\alpha }\partial _{x_{\mu
}}\sigma _{\mu \beta }=\left( \boldsymbol{L}\cdot \nabla \right) L_{\beta
}\partial _{x_{\mu }}\sigma _{\mu \alpha },\quad \alpha \neq \beta \quad 
\text{i.e.,} \\
&&\left( \boldsymbol{L}\cdot \nabla \right) \partial _{x_{\mu }}\left(
L_{\alpha }\sigma _{\mu \beta }-L_{\beta }\sigma _{\mu \alpha }\right)
=\left( \boldsymbol{L}\cdot \nabla \right) \epsilon _{\alpha \beta \nu
}\partial _{x_{\mu }}\left( L_{\alpha }\sigma _{\mu \beta }\right) =0,
\end{eqnarray*}%
by the compatibility condition (\ref{compatibility-condition}), so that the
previous expression reduces to (\ref{cc-T3}), due to (\ref{cc-rotor}).

In the case of model $T_{4},$ the compatibility condition (\ref%
{compatibility-condition}) reads%
\begin{equation}
\boldsymbol{L}\left( \boldsymbol{L}\cdot \nabla \right) \left( \func{curl}%
\boldsymbol{\hat{\sigma}}\left( \boldsymbol{r}\right) \right) ^{\mathrm{T}%
}=\left( \boldsymbol{L}\otimes \boldsymbol{L}\right) \left( \func{div}\left( 
\func{curl}\boldsymbol{\hat{\sigma}}\left( \boldsymbol{r}\right) \right) ^{%
\mathrm{T}}\right) =\boldsymbol{0},  \label{cc-T4}
\end{equation}%
since in the Cartesian coordinate system one has%
\begin{align*}
\mathbf{\hat{T}}_{4}\left( \nabla \right) \boldsymbol{\hat{\sigma}}\left( 
\boldsymbol{r}\right) &=\left( \nabla \otimes \nabla \right) \left( 
\boldsymbol{L}\otimes \boldsymbol{L}\right) \boldsymbol{\hat{\sigma}}\left( 
\boldsymbol{r}\right) \\
&=L_{\xi }L_{\mu }\left( \partial _{x_{\alpha }x_{\eta }}^{2}\sigma _{\zeta
\beta }\right) \left( \boldsymbol{e}_{\alpha }\otimes \boldsymbol{e}_{\eta
}\right) \left( \boldsymbol{e}_{\xi }\otimes \boldsymbol{e}_{\mu }\right)
\left( \boldsymbol{e}_{\zeta }\otimes \boldsymbol{e}_{\beta }\right) ) \\
&=L_{\xi }L_{\mu }\left( \partial _{x_{\alpha }x_{\eta }}^{2}\sigma _{\zeta
\beta }\right) \delta _{\eta \xi }\delta _{\mu \zeta }\left( \boldsymbol{e}%
_{\alpha }\otimes \boldsymbol{e}_{\beta }\right) \\
&=L_{\xi }L_{\mu }\left( \partial _{x_{\alpha }x_{\xi }}^{2}\sigma _{\mu
\beta }\right) \left( \boldsymbol{e}_{\alpha }\otimes \boldsymbol{e}_{\beta
}\right) , \\
&=\left( \boldsymbol{L}\cdot \nabla \right) L_{\mu }\left( \partial
_{x_{\alpha }}\sigma _{\mu \beta }\right) \left( \boldsymbol{e}_{\alpha
}\otimes \boldsymbol{e}_{\beta }\right)
\end{align*}%
implying the following expression%
\begin{eqnarray*}
&&\left( \boldsymbol{L}\cdot \nabla \right) L_{\mu }\partial _{x_{\alpha
}}\sigma _{\mu \beta }=\left( \boldsymbol{L}\cdot \nabla \right) L_{\mu
}\partial _{x_{\beta }}\sigma _{\mu \alpha },\quad \alpha \neq \beta \quad 
\text{i.e.,} \\
&&\left( \boldsymbol{L}\cdot \nabla \right) L_{\mu }\left( \partial
_{x_{\alpha }}\sigma _{\mu \beta }-\partial _{x_{\beta }}\sigma _{\mu \alpha
}\right) =\left( \boldsymbol{L}\cdot \nabla \right) \epsilon _{\alpha \beta
\nu }L_{\mu }\partial _{x_{\alpha }}\sigma _{\mu \beta }=0,
\end{eqnarray*}%
by the compatibility condition (\ref{compatibility-condition}), so that the
previous expression reduces to (\ref{cc-T4}), due to (\ref{cc-rotor}).

In the case of model $T_{5},$ the compatibility condition (\ref%
{compatibility-condition}) reads%
\begin{equation}
\boldsymbol{L}\left( \left( \left( \nabla \otimes \nabla \right) - 
\boldsymbol{\hat{I}} \left( \nabla \cdot \nabla \right) \right) \left( 
\boldsymbol{L}\times \boldsymbol{\hat{\sigma}}\left( \boldsymbol{r}\right)
\right) ^{\mathrm{T}}-\left( \left( \boldsymbol{L}\otimes \nabla \right)
-\left( \boldsymbol{L}\otimes \nabla \right) ^{\mathrm{T}}\right) \left( 
\func{curl}\boldsymbol{\hat{\sigma}}\left( \boldsymbol{r}\right) \right) ^{%
\mathrm{T}}\right) =\boldsymbol{0},  \label{cc-T5}
\end{equation}%
since in the Cartesian coordinate system one has%
\begin{align*}
\mathbf{\hat{T}}_{5}&\left( \nabla \right) \boldsymbol{\hat{\sigma}}\left( 
\boldsymbol{r}\right) =\left( \left( \boldsymbol{L}\times \nabla \right)
\otimes \left( \boldsymbol{L}\times \nabla \right) \right) \boldsymbol{\hat{%
\sigma}}\left( \boldsymbol{r}\right) \\
&=\left( \left( \epsilon _{ijk}\boldsymbol{e}_{i}L_{j}\partial
_{x_{k}}\right) \otimes \left( \epsilon _{pqr}\boldsymbol{e}%
_{p}L_{q}\partial _{x_{r}}\right) \right) \left( \sigma _{\mu \nu }\left( 
\boldsymbol{e}_{\mu }\otimes \boldsymbol{e}_{\nu }\right) \right) \\
&=\epsilon _{ijk}\epsilon _{pqr}L_{j}L_{q}\left( \partial
_{x_{k}x_{r}}^{2}\sigma _{\mu \nu }\right) \left( \boldsymbol{e}_{i}\otimes 
\boldsymbol{e}_{p}\right) \left( \boldsymbol{e}_{\mu }\otimes \boldsymbol{e}%
_{\nu }\right) \\
&=\epsilon _{ijk}\epsilon _{pqr}L_{j}L_{q}\left( \partial
_{x_{k}x_{r}}^{2}\sigma _{\mu \nu }\right) \delta _{p\mu }\left( \boldsymbol{%
e}_{i}\otimes \boldsymbol{e}_{\nu }\right) \\
&=\epsilon _{ijk}\epsilon _{pqr}L_{j}L_{q}\left( \partial
_{x_{k}x_{r}}^{2}\sigma _{p\nu }\right) \left( \boldsymbol{e}_{i}\otimes 
\boldsymbol{e}_{\nu }\right) \\
&=\left( \delta _{ip}\delta _{jq}\delta _{kr}-\delta _{ip}\delta _{jr}\delta
_{kq}+\delta _{iq}\delta _{jr}\delta _{kp}-\delta _{iq}\delta _{jp}\delta
_{kr}+\delta _{ir}\delta _{jp}\delta _{kq}-\delta _{ir}\delta _{jq}\delta
_{kp}\right) L_{j}L_{q}\partial _{x_{k}x_{r}}^{2}\sigma _{p\nu }\left( 
\boldsymbol{e}_{i}\otimes \boldsymbol{e}_{\nu }\right) \\
&=L_{j}\partial _{x_{k}}\left( L_{j}\partial _{x_{k}}\sigma _{i\nu
}-L_{k}\partial _{x_{j}}\sigma _{i\nu }+L_{i}\partial _{x_{j}}\sigma _{k\nu
}-L_{j}\partial _{x_{i}}\sigma _{k\nu }+L_{k}\partial _{x_{i}}\sigma _{j\nu
}-L_{i}\partial _{x_{k}}\sigma _{j\nu }\right) \left( \boldsymbol{e}%
_{i}\otimes \boldsymbol{e}_{\nu }\right) \\
&=\left( L^{2}\nabla ^{2}\sigma _{i\nu } \!-\! \left( \boldsymbol{L}\cdot
\nabla \right) ^{2}\sigma _{i\nu } \! + \! \left( \boldsymbol{L}\cdot \nabla
\right) L_{i}\partial _{x_{k}}\sigma _{k\nu } \! - \! L^{2}\partial
_{x_{i}}\partial _{x_{k}}\sigma _{k\nu } \! + \! \left( \boldsymbol{L}\cdot
\nabla \right) \partial _{x_{i}}\left( L_{j}\sigma _{j\nu }\right) \! - \!
\nabla ^{2}L_{i}L_{j}\sigma _{j\nu }\right) \left( \boldsymbol{e}_{i}\otimes 
\boldsymbol{e}_{\nu }\right)
\end{align*}%
since%
\begin{equation*}
\epsilon _{ijk}\epsilon _{pqr}=\left\vert 
\begin{array}{ccc}
\delta _{ip} & \delta _{iq} & \delta _{ir} \\ 
\delta _{jp} & \delta _{jq} & \delta _{jr} \\ 
\delta _{kp} & \delta _{kq} & \delta _{kr}%
\end{array}%
\right\vert ,
\end{equation*}%
implying the following expression%
\begin{align*}
&\left( \boldsymbol{L}\cdot \nabla \right) L_{i}\partial _{x_{k}}\sigma
_{k\nu }-L^{2}\partial _{x_{i}}\partial _{x_{k}}\sigma _{k\nu }+\left( 
\boldsymbol{L}\cdot \nabla \right) \partial _{x_{i}}\left( L_{j}\sigma
_{j\nu }\right) -\nabla ^{2}L_{i}L_{j}\sigma _{j\nu } \\
& \quad =\left( \boldsymbol{L}\cdot \nabla \right) L_{\nu }\partial
_{x_{k}}\sigma _{ki}-L^{2}\partial _{x_{\nu }}\partial _{x_{k}}\sigma
_{ki}+\left( \boldsymbol{L}\cdot \nabla \right) \partial _{x_{\nu }}\left(
L_{j}\sigma _{ji}\right) -\nabla ^{2}L_{\nu }L_{j}\sigma _{ji},\quad i\neq
\nu \quad \text{i.e.,} \\
& \left( \boldsymbol{L}\cdot \nabla \right) \partial _{x_{k}}\left(
L_{i}\sigma _{k\nu } \!- \! L_{\nu }\sigma _{ki}\right) \! - \!
L^{2}\partial _{x_{k}}\left( \partial _{x_{i}}\sigma _{k\nu } \! - \!
\partial _{x_{\nu }}\sigma _{ki}\right) \! +\! \left( \boldsymbol{L}\cdot
\nabla \right) L_{j}\left( \partial _{x_{i}}\sigma _{j\nu } \!- \!\partial
_{x_{\nu }}\sigma _{ji}\right) \! - \! \nabla ^{2}L_{j}\left( L_{i}\sigma
_{j\nu } \! - \! L_{\nu }\sigma _{ji}\right) =0 \\
& \left( \boldsymbol{L}\cdot \nabla \right) \epsilon _{i\nu \pi }\partial
_{x_{k}}L_{i}\sigma _{k\nu }-L^{2}\epsilon _{i\nu \pi }\partial
_{x_{k}}\partial _{x_{i}}\sigma _{k\nu }+\left( \boldsymbol{L}\cdot \nabla
\right) \epsilon _{i\nu \pi }L_{j}\partial _{x_{i}}\sigma _{j\nu }-\nabla
^{2}\epsilon _{i\nu \pi }L_{j}L_{i}\sigma _{j\nu }=0
\end{align*}%
by the compatibility condition (\ref{compatibility-condition}), since $%
\sigma _{i\nu }=\sigma _{\nu i}$, so that%
\begin{align*}
&\left( \boldsymbol{L}\cdot \nabla \right) \func{div}\left( \boldsymbol{L}%
\times \boldsymbol{\hat{\sigma}}\left( \boldsymbol{r}\right) \right) ^{%
\mathrm{T}}-L^{2}\func{div}\left( \func{curl}\boldsymbol{\hat{\sigma}}\left( 
\boldsymbol{r}\right) \right) ^{\mathrm{T}}+\left( \boldsymbol{L}\cdot
\nabla \right) \boldsymbol{L}\left( \func{curl}\boldsymbol{\hat{\sigma}}%
\left( \boldsymbol{r}\right) \right) ^{\mathrm{T}}-\nabla ^{2}\boldsymbol{L}%
\left( \boldsymbol{L}\times \boldsymbol{\hat{\sigma}}\left( \boldsymbol{r}%
\right) \right) ^{\mathrm{T}} =\boldsymbol{0} \\
& \left( \left( \boldsymbol{L}\cdot \nabla \right) \nabla -\nabla ^{2}%
\boldsymbol{L}\right) \left( \boldsymbol{L}\times \boldsymbol{\hat{\sigma}}%
\left( \boldsymbol{r}\right) \right) ^{\mathrm{T}}-\left( \left( \boldsymbol{%
L}\cdot \boldsymbol{L}\right) \nabla -\left( \boldsymbol{L}\cdot \nabla
\right) \boldsymbol{L}\right) \left( \func{curl}\boldsymbol{\hat{\sigma}}%
\left( \boldsymbol{r}\right) \right) ^{\mathrm{T}} =\boldsymbol{0}
\end{align*}%
due to (\ref{cc-rotor}), which reduces to (\ref{cc-T5}).

\section{Non-locality kernels\label{Non-locality kernels}}

In the case of one-dimensional Eringen model, given by (\ref{1D-Eringen}),
the non-locality kernel is of the exponential type and takes the following
form%
\begin{equation}
\mathcal{K}\left( x\right) \mathcal{=}\frac{1}{2l}\mathrm{e}^{-\frac{%
\left\vert x\right\vert }{l}}.  \label{1D-Eringen-kernel}
\end{equation}%
Non-locality kernels corresponding to three-dimensional generalization of
Eringen model, along with proposed constitutive models, are listed in Table %
\ref{Modeli-new}.

\subsection{One-dimensional case}

In order to obtain non-locality kernel (\ref{1D-Eringen-kernel}), the
Fourier integral transform method corresponding to one-dimensional case can
be used, such that the one-dimensional Fourier transform, defined by%
\begin{equation*}
\bar{f}\left( k\right) =\mathcal{F}\left[ f\left( x\right) \right] \left(
k\right) =\int_{%
\mathbb{R}
}f\left( x\right) \mathrm{e}^{-\mathrm{i}kx}\mathrm{d}x,\quad k\in 
\mathbb{R}
,
\end{equation*}%
applied to the expression (\ref{1D-Eringen}) yields%
\begin{align*}
\left( 1+l^{2}k^{2}\right) \bar{\sigma}\left( k\right) & =\bar{\sigma}%
^{local}\left( k\right) ,\quad \text{i.e.,} \\
\bar{\sigma}\left( k\right) & =\mathcal{\bar{K}}\left( k\right) \,\bar{\sigma%
}^{local}\left( k\right) ,\quad \text{with\quad }\mathcal{\bar{K}}\left(
k\right) =\frac{1}{l^{2}}\frac{1}{\frac{1}{l^{2}}+k^{2}},
\end{align*}%
since $\mathcal{F}\left[ \frac{\mathrm{d}}{\mathrm{d}x}\right] \left(
k\right) =\mathrm{i}k$, which, after the inverse Fourier transform, defined
by%
\begin{equation*}
g\left( x\right) =\mathcal{F}^{-1}\left[ \bar{g}\left( k\right) \right]
\left( x\right) =\frac{1}{2\pi }\int_{%
\mathbb{R}
}\bar{g}\left( k\right) \mathrm{e}^{\mathrm{i}kx}\mathrm{d}k,
\end{equation*}%
becomes%
\begin{equation*}
\sigma \left( x\right) =\mathcal{K}\left( x\right) \ast _{x}\sigma
^{local}\left( x\right) =\frac{1}{2l}\mathrm{e}^{-\frac{\left\vert
x\right\vert }{l}}\ast _{x}\sigma ^{local}\left( x\right) ,
\end{equation*}%
see (\ref{1D-Eringen-kernel}), since 
\begin{equation}
\mathcal{F}^{-1}\left[ \frac{1}{k^{2}+\lambda ^{2}}\right] \left( x\right) =%
\frac{1}{2\pi }\int_{-\infty }^{\infty }\frac{1}{k^{2}+\lambda ^{2}}\mathrm{e%
}^{\mathrm{i}kx}\mathrm{d}k=\frac{1}{2\lambda }\mathrm{e}^{-\lambda
\left\vert x\right\vert },\quad \text{for}\quad \lambda >0,
\label{IFT-sa-lam}
\end{equation}%
and 
\begin{equation*}
\mathcal{F}^{-1}\left[ \bar{\sigma}\left( k\right) \right] \left( x\right) =%
\mathcal{F}^{-1}\left[ \mathcal{\bar{K}}\left( k\right) \right] \left(
x\right) \ast _{x}\mathcal{F}^{-1}\left[ \bar{\sigma}^{local}\left( k\right) %
\right] \left( x\right) ,
\end{equation*}%
where $\ast _{x}$ denotes one-dimensional convolution%
\begin{equation*}
f\left( x\right) \ast _{x}g\left( x\right) =\int_{%
\mathbb{R}
}f\left( x-x^{\prime }\right) g\left( x^{\prime }\right) \mathrm{d}x^{\prime
}.
\end{equation*}

\subsection{Three-dimensional cases\label{Three-dimensional-cases}}

Using the Fourier integral transform with respect to spatial coordinates,
defined by%
\begin{equation}
\bar{f}\left( \boldsymbol{k}\right) =\mathcal{F}\left[ f\left( \boldsymbol{r}%
\right) \right] \left( \boldsymbol{k}\right) =\int_{%
\mathbb{R}
^{3}}f\left( \boldsymbol{r}\right) \mathrm{e}^{-\mathrm{i}\boldsymbol{k\cdot
r}}\mathrm{d}_{\boldsymbol{r}}V,\quad \boldsymbol{k}\in 
\mathbb{R}
^{3},  \label{FT}
\end{equation}%
one transforms the scalar- and tensor-type stress-gradient Eringen
three-dimensional models, given by (\ref{non-loc-scalar-type}) and (\ref%
{non-loc-tensor-type}), which respectively contain scalar and tensor
non-locality operators $\mathrm{S}_{i}$ and $\mathbf{\hat{T}}_{i}$, given in
Table \ref{Modeli-new}, so that in the Fourier domain one has either%
\begin{equation}
\left( 1+\bar{S}_{i}\left( \boldsymbol{k}\right) \right) \boldsymbol{\bar{%
\hat{\sigma}}}\left( \boldsymbol{k}\right) =\boldsymbol{\bar{\hat{\sigma}}}%
^{local}\left( \boldsymbol{k}\right) ,\quad \text{i.e.,}\quad \boldsymbol{%
\bar{\hat{\sigma}}}\left( \boldsymbol{k}\right) =\mathcal{\bar{K}}_{i}\left( 
\boldsymbol{k}\right) \boldsymbol{\bar{\hat{\sigma}}}^{local}\left( 
\boldsymbol{k}\right) ,\quad \text{with\quad }\mathcal{\bar{K}}_{i}\left( 
\boldsymbol{k}\right) =\frac{1}{1+\bar{S}_{i}\left( \boldsymbol{k}\right) },
\label{Ki-bar-scalar}
\end{equation}%
in the case of scalar-type models, or%
\begin{equation}
\left( \boldsymbol{\hat{I}}+\boldsymbol{\bar{\hat{T}}}_{i}\left( \boldsymbol{%
k}\right) \right) \boldsymbol{\bar{\hat{\sigma}}}\left( \boldsymbol{k}%
\right) =\boldsymbol{\bar{\hat{\sigma}}}^{local}\left( \boldsymbol{k}\right)
,\quad \text{i.e.,}\quad \boldsymbol{\bar{\hat{\sigma}}}\left( \boldsymbol{k}%
\right) =\boldsymbol{\bar{\hat{\mathcal{K}}}}_{i}\left( \boldsymbol{k}%
\right) \boldsymbol{\bar{\hat{\sigma}}}^{local}\left( \boldsymbol{k}\right)
,\quad \text{with\quad }\boldsymbol{\bar{\hat{\mathcal{K}}}}_{i}\left( 
\boldsymbol{k}\right) =\left( \boldsymbol{\hat{I}}+\boldsymbol{\bar{\hat{T}}}%
_{i}\left( \boldsymbol{k}\right) \right) ^{-1},  \label{Ki-bar-tensor}
\end{equation}%
in the case of tensor-type models, where the non-locality scalar and tensor
functions in Fourier domain $\bar{S}_{i}$ and $\boldsymbol{\bar{\hat{T}}}%
_{i} $, as well as non-locality kernels in Fourier domain $\bar{\mathcal{K}}%
_{i}$ and $\boldsymbol{\bar{\hat{\mathcal{K}}}}_{i}$, are given in Table \ref%
{Modeli}.

Fourier transform inversion in (\ref{Ki-bar-scalar}) and (\ref{Ki-bar-tensor}%
) yields the Cauchy stress tensor as the convolution of local stress tensor
with either scalar or tensor non-locality kernel, either as%
\begin{equation*}
\boldsymbol{\hat{\sigma}}\left( \boldsymbol{r}\right) =\mathcal{K}_{i}\left( 
\boldsymbol{r}\right) \ast _{\boldsymbol{r}}\boldsymbol{\hat{\sigma}}%
^{local}\left( \boldsymbol{r}\right) \quad \text{or as\quad }\boldsymbol{%
\hat{\sigma}}\left( \boldsymbol{r}\right) =\boldsymbol{\hat{\mathcal{K}}}%
_{i}\left( \boldsymbol{r}\right) \ast _{\boldsymbol{r}}\boldsymbol{\hat{%
\sigma}}^{local}\left( \boldsymbol{r}\right) ,
\end{equation*}%
with non-locality kernels listed in Table \ref{Modeli-new} and calculated in
the sequel, where the convolution is understood component-wise and given by%
\begin{equation*}
f\left( \boldsymbol{r}\right) \ast _{\boldsymbol{r}}g\left( \boldsymbol{r}%
\right) =\int_{%
\mathbb{R}
^{3}}f\left( \boldsymbol{r}-\boldsymbol{r}^{\prime }\right) g\left( 
\boldsymbol{r}^{\prime }\right) \mathrm{d}_{\boldsymbol{r}^{\prime }}V.
\end{equation*}%
Moreover, the non-locality character of Cauchy stress tensor is considered
in Section \ref{Non-locality of the Cauchy stress tensor}.

\begin{table}[h]
\begin{center}
\begin{tabular}{|c|c|c|c|c|c|c|c|} \hline \xrowht{20pt}
 & Model & \multicolumn{2}{c|}{Non-locality operator}  & \multicolumn{2}{c|}{\makecell{Non-locality operator \\ in Fourier domain}}    &    \multicolumn{2}{c|}{\makecell{Non-locality kernel \\ in Fourier domain}} \\ \cline{1-8} \xrowht{20pt}

\multirow{2.5}{*}{\multirowcell{3}{\rotatebox{90}{Scalar-type}}} & $S_1$ & \multirow{3}{*}{\multirowcell{3.25}{ ${\rm{S}}_{i}(\nabla)$ }} & $\left( \boldsymbol{L}\cdot \boldsymbol{L}\right) \left( \nabla \cdot \nabla \right)$  & \multirow{3}{*}{\multirowcell{3.25}{ $\bar{S}_{i}(\boldsymbol{k})$ }} & $L^{2}k^{2}$ & \multirow{3}{*}{\multirowcell{3.25}{  ${\mathcal{\bar{K}}}_{i}  \left( \boldsymbol{k}\right) $}} & $\frac{1}{1+L^{2}k^{2}}$ \\ \cline{2-2} \cline{4-4} \cline{6-6} \cline{8-8} \xrowht{20pt}

 & $S_2$  & & $\left( \boldsymbol{L}\cdot \nabla \right) ^{2} $ &  & $\left( \boldsymbol{L}\cdot 
\boldsymbol{k}\right) ^{2}$ & & $\frac{1}{1+\left( \boldsymbol{L}\cdot 
\boldsymbol{k}\right) ^{2}}$ \\  \cline{2-2} \cline{4-4} \cline{6-6} \cline{8-8} \xrowht{20pt}

 & $S_3$ & & $\left(\boldsymbol{L}\times \nabla \right) \cdot \left( \boldsymbol{L}\times \nabla\right) $   &  & $L^{2}k^{2}+\left( \boldsymbol{L}\cdot 
\boldsymbol{k}\right) ^{2}$  & & $\frac{1}{1+L^{2}k^{2}-\left( \boldsymbol{L}%
\cdot \boldsymbol{k}\right) ^{2}}$ \\ \cline{1-8}  \xrowht{20pt}

\multirow{6.7}{*}{\multirowcell{4}{\rotatebox{90}{Tensor-type}}} & $T_{1}$ & \multirow{4}{*}{\multirowcell{6.8}{ $\boldsymbol{\hat{{\rm{T}}}}_{i}(\nabla)$ }} & $\left( \boldsymbol{L}\cdot\boldsymbol{L}\right) \left( \nabla \otimes \nabla \right) $ &  \multirow{4}{*}{\multirowcell{6.8}{ $\boldsymbol{\bar{\hat{{T}}}}_{i}(\boldsymbol{k})$ }} & $ L^{2} (\boldsymbol{k} \otimes \boldsymbol{k}) $ & \multirow{4}{*}{\multirowcell{6.8}{  $\boldsymbol{\bar{\hat{{\mathcal{K}}}}}_{i}  \left( \boldsymbol{k}\right) $}} & $\boldsymbol{\hat{I}}-\frac{\boldsymbol{k}\otimes \boldsymbol{k}}{\frac{1}{%
L^{2}}+k^{2}}$\\  \cline{2-2} \cline{4-4} \cline{6-6} \cline{8-8} \xrowht{20pt}

 & $T_{2}$  & & $\left( \boldsymbol{L}\otimes \boldsymbol{L}\right) \left( \nabla \cdot
\nabla \right) $ &  &  $ k^2 (\boldsymbol{L} \otimes \boldsymbol{L}) $ & & $\boldsymbol{\hat{I}}-\frac{k^{2}}{\frac{1}{L^{2}}+k^{2}}\frac{\boldsymbol{L}%
\otimes \boldsymbol{L}}{L^{2}}$ \\  \cline{2-2} \cline{4-4} \cline{6-6} \cline{8-8} \xrowht{20pt}

 & $T_3$ & &$\left( \boldsymbol{L}\cdot \nabla \right) \left( \boldsymbol{L}\otimes
\nabla \right) $  & & $ (\boldsymbol{L} \cdot \boldsymbol{k}) (\boldsymbol{L} \otimes \boldsymbol{k}) $ & & $\boldsymbol{\hat{I}}-\frac{\left( \boldsymbol{L}\cdot \boldsymbol{k}\right)
\left( \boldsymbol{L}\otimes \boldsymbol{k}\right) }{1+\left( \boldsymbol{L}%
\cdot \boldsymbol{k}\right) ^{2}}$ \\ \cline{2-2} \cline{4-4} \cline{6-6} \cline{8-8} \xrowht{20pt}

 & $T_4$ & &$\left( \boldsymbol{L}\cdot \nabla \right)\left( \boldsymbol{L}\otimes \nabla \right) ^{\mathrm{T}} $ & & $ (\boldsymbol{L} \cdot \boldsymbol{k}) ( \boldsymbol{k}  \otimes \boldsymbol{L}) $ & & $\boldsymbol{\hat{I}}-\frac{\left( \boldsymbol{L}\cdot \boldsymbol{k}\right)
\left(\boldsymbol{k} \otimes \boldsymbol{L}\right) }{1+\left( \boldsymbol{L}%
\cdot \boldsymbol{k}\right) ^{2}}$ \\ \cline{2-2} \cline{4-4} \cline{6-6} \cline{8-8} \xrowht{20pt}

 & $T_5$ & &  $\left( \boldsymbol{L}\times \nabla \right)
\otimes \left( \boldsymbol{L}\times \nabla \right)$ &  & $ (\boldsymbol{L} \times \boldsymbol{k}) \otimes (\boldsymbol{L} \times \boldsymbol{k}) $ & & $\boldsymbol{\hat{I}}-\frac{\left( \boldsymbol{L}\times \boldsymbol{k}\right)
\otimes \left( \boldsymbol{L}\times \boldsymbol{k}\right) }{%
1+L^{2}k^{2}-\left( \boldsymbol{L\cdot r}\right) ^{2}}$ \\ \cline{1-8}
\end{tabular}
\end{center}
\caption{Scalar and tensor operators in both spatial and Fourier domains, as well as non-locality kernels in Fourier domain, corresponding to non-local models, where $\bar{S}_{i}(\boldsymbol{k})=-\mathcal{F}\left[ {\rm{S}}_{i}(\nabla)\right] \left(\boldsymbol{ k}\right) $ and $\boldsymbol{\bar{\hat{{T}}}}_{i}(\boldsymbol{k})=-\mathcal{F}\left[ \boldsymbol{\hat{{\rm{T}}}}_{i}(\nabla)\right] \left(\boldsymbol{ k}\right) $.}
\label{Modeli}
\end{table}

\subsubsection{Non-locality kernel of model $S_{1}$}

Model $S_{1}$, see the scalar-type stress-gradient Eringen three-dimensional
model (\ref{non-loc-scalar-type}) and Table \ref{Modeli-new}, constituted as
the straightforward three-dimensional generalization of Eringen model (\ref%
{1D-Eringen}), for the non-locality kernel gives the following expression%
\begin{equation}
\mathcal{K}_{1}\left( r\right) \mathcal{=}\frac{1}{4\pi rL^{2}}\mathrm{e}^{-%
\frac{r}{L}},  \label{K1}
\end{equation}%
see also Table \ref{Modeli-new}.

Namely, in the case of model $S_{1}$, the expression (\ref{Ki-bar-scalar})
for local stress tensor and non-locality kernel, both in Fourier domain,
becomes%
\begin{align}
\left( 1+L^{2}k^{2}\right) \boldsymbol{\bar{\hat{\sigma}}}\left( \boldsymbol{%
k}\right) & =\boldsymbol{\bar{\hat{\sigma}}}^{local}\left( \boldsymbol{k}%
\right) ,\quad \text{i.e.,}  \notag \\
\boldsymbol{\bar{\hat{\sigma}}}\left( \boldsymbol{k}\right) & =\mathcal{\bar{%
K}}_{1}\left( k\right) \boldsymbol{\bar{\hat{\sigma}}}^{local}\left( 
\boldsymbol{k}\right) \quad \text{with\quad }\mathcal{\bar{K}}_{1}\left(
k\right) =\frac{1}{L^{2}}\frac{1}{\frac{1}{L^{2}}+k^{2}},  \label{K1-bar}
\end{align}%
since, according to $\mathcal{F}\left[ \nabla \right] \left( \boldsymbol{k}%
\right) =\mathrm{i}\boldsymbol{k},$ one has $\bar{S}_{1}\left( \boldsymbol{k}%
\right) =-\mathcal{F}\left[ \mathrm{S}_{1}\left( \nabla \right) \right]
\left( \boldsymbol{k}\right) =-\mathcal{F}\left[ \left( \boldsymbol{L}\cdot 
\boldsymbol{L}\right) \left( \nabla \cdot \nabla \right) \right] \left( 
\boldsymbol{k}\right) =L^{2}k^{2},$ see also Table \ref{Modeli}. The inverse
Fourier transform, defined by%
\begin{equation}
f\left( \boldsymbol{r}\right) =\mathcal{F}\left[ \bar{f}\left( \boldsymbol{k}%
\right) \right] \left( \boldsymbol{k}\right) =\frac{1}{\left( 2\pi \right)
^{3}}\int_{%
\mathbb{R}
^{3}}\bar{f}\left( \boldsymbol{k}\right) \mathrm{e}^{\mathrm{i}\boldsymbol{%
k\cdot r}}\mathrm{d}_{\boldsymbol{k}}V,\quad \boldsymbol{r}\in 
\mathbb{R}
^{3},  \label{IFT}
\end{equation}%
transforms the expression (\ref{K1-bar})$_{2}$ into%
\begin{align*}
\mathcal{K}_{1}\left( \boldsymbol{r}\right) & =\mathcal{F}^{-1}\left[ 
\mathcal{\bar{K}}_{1}\left( k\right) \right] \left( \boldsymbol{r}\right) =%
\frac{1}{\left( 2\pi \right) ^{3}L^{2}}\int_{%
\mathbb{R}
^{3}}\frac{1}{\frac{1}{L^{2}}+k^{2}}\mathrm{e}^{\mathrm{i}\boldsymbol{k}%
\cdot \boldsymbol{r}}\mathrm{d}_{\boldsymbol{k}}V \\
& =\frac{1}{\left( 2\pi \right) ^{3}L^{2}}\int_{0}^{2\pi
}\!\!\!\int_{0}^{\pi }\!\!\!\int_{0}^{\infty }\frac{1}{\frac{1}{L^{2}}+k^{2}}%
\mathrm{e}^{\mathrm{i}kr\cos \theta }k^{2}\sin \theta ~\mathrm{d}k~\mathrm{d}%
\theta ~\mathrm{d}\varphi \\
& =\frac{1}{4\pi ^{2}L^{2}}\int_{0}^{\infty }\frac{k^{2}}{\frac{1}{L^{2}}%
+k^{2}}\left( \frac{1}{\mathrm{i}kr}\int_{-\mathrm{i}kr}^{\mathrm{i}kr}%
\mathrm{e}^{u}\mathrm{d}u\right) \mathrm{d}k \\
& =\frac{1}{2\pi ^{2}rL^{2}}\int_{0}^{\infty }\frac{k\sin \left( kr\right) }{%
\frac{1}{L^{2}}+k^{2}}\mathrm{d}k \\
& =-\frac{1}{2\pi ^{2}rL^{2}}\frac{\mathrm{d}}{\mathrm{d}r}\int_{0}^{\infty }%
\frac{\cos \left( kr\right) }{\frac{1}{L^{2}}+k^{2}}\mathrm{d}k \\
& =-\frac{1}{4\pi rL^{2}}\frac{\mathrm{d}}{\mathrm{d}r}\frac{\mathrm{e}^{-%
\frac{r}{L}}}{\frac{1}{L}} \\
& =\frac{1}{4\pi rL^{2}}\mathrm{e}^{-\frac{r}{L}},
\end{align*}%
where the integral%
\begin{equation}
\int_{0}^{\infty }\frac{\cos \left( kx\right) }{k^{2}+\lambda }\mathrm{d}k=%
\frac{\pi }{2}\frac{\mathrm{e}^{-\left\vert x\right\vert \sqrt{\lambda }}}{%
\sqrt{\lambda }},\quad \text{if}\quad \lambda >0,  \label{integral-kosinus}
\end{equation}%
is used. The previous integral is calculated for $x>0$ using the Cauchy
residues theorem as%
\begin{gather*}
\oint\nolimits_{\Gamma }\frac{\mathrm{e}^{\mathrm{i}xz}}{z^{2}+\lambda }%
\mathrm{d}z=2\pi \mathrm{i}~\text{\textrm{Res}}\left[ \frac{\mathrm{e}^{%
\mathrm{i}xz}}{z^{2}+\lambda },\mathrm{i}\sqrt{\lambda }\right] \\
\int_{0}^{\infty }\frac{\mathrm{e}^{\mathrm{i}x\rho }}{\rho ^{2}+\lambda }%
\mathrm{d}\rho +\mathrm{i}\int_{0}^{\pi }\frac{\mathrm{e}^{\mathrm{i}xR%
\mathrm{e}^{\mathrm{i}\varphi }}}{\left( R\mathrm{e}^{\mathrm{i}\varphi
}\right) ^{2}+\lambda }R\mathrm{e}^{\mathrm{i}\varphi }\mathrm{d}\varphi
+\int_{\infty }^{0}\frac{\mathrm{e}^{\mathrm{i}x\rho \mathrm{e}^{\mathrm{i}%
\pi }}}{\left( \rho \mathrm{e}^{\mathrm{i}\pi }\right) ^{2}+\lambda }\mathrm{%
e}^{\mathrm{i}\pi }\mathrm{d}\rho =2\pi \mathrm{i}\frac{\mathrm{e}^{-x\sqrt{%
\lambda }}}{2\mathrm{i}\sqrt{\lambda }} \\
\int_{0}^{\infty }\frac{\mathrm{e}^{\mathrm{i}x\rho }}{\rho ^{2}+\lambda }%
\mathrm{d}\rho +\int_{0}^{\infty }\frac{\mathrm{e}^{-\mathrm{i}x\rho }}{\rho
^{2}+\lambda }\mathrm{d}\rho =\pi \frac{\mathrm{e}^{-x\sqrt{\lambda }}}{%
\sqrt{\lambda }} \\
\int_{0}^{\infty }\frac{\mathrm{\cos }\left( x\rho \right) }{\rho
^{2}+\lambda }\mathrm{d}\rho =\frac{\pi }{2}\frac{\mathrm{e}^{-x\sqrt{%
\lambda }}}{\sqrt{\lambda }},
\end{gather*}%
where $\Gamma $ is the semi-circle located in the upper complex half-plane.
Note, if $\ x<0,$ the integration along the contour $\Gamma $, being a
semi-circle in the lower complex half-plane, ensures the same result for the
integral with variable $x$ replaced with $|x|$.

\subsubsection{Non-locality kernel of model $S_{2}$}

Model $S_{2}$, see the scalar-type stress-gradient Eringen three-dimensional
model (\ref{non-loc-scalar-type}) and Table \ref{Modeli-new}, constituted as
the another three-dimensional generalization of Eringen model (\ref%
{1D-Eringen}), also gives an exponential type of non-locality kernel
represented by%
\begin{equation}
\mathcal{K}_{2}\left( r\right) \mathcal{=}\frac{1}{2L}\delta \left( 
\boldsymbol{r}_{\perp }\right) \mathrm{e}^{-\frac{\left\vert r_{\parallel
}\right\vert }{L}},  \label{K2}
\end{equation}%
where $r_{\parallel }= \boldsymbol{r} \cdot \frac{\boldsymbol{L} }{L}$, $%
\boldsymbol{r}_{\perp }= \boldsymbol{r}-\boldsymbol{r}_{\parallel }$ with $%
\boldsymbol{r}_{\parallel }=r_{\parallel } \frac{\boldsymbol{L} }{L}$, and $%
r_{\perp }= \left\vert \boldsymbol{r}_{\perp } \right\vert $, see also Table %
\ref{Modeli-new}.

Namely, in the case of model $S_{2}$, the expression (\ref{Ki-bar-scalar})
for local stress tensor and non-locality kernel, both in Fourier domain,
becomes%
\begin{align}
\left( 1+\left( \boldsymbol{L}\cdot \boldsymbol{k}\right) ^{2}\right) 
\boldsymbol{\bar{\hat{\sigma}}}\left( \boldsymbol{k}\right) & =\boldsymbol{%
\bar{\hat{\sigma}}}^{local}\left( \boldsymbol{k}\right) ,\quad \text{i.e.,} 
\notag \\
\boldsymbol{\bar{\hat{\sigma}}}\left( \boldsymbol{k}\right) & =\mathcal{\bar{%
K}}_{2}\left( \boldsymbol{k}\right) \boldsymbol{\bar{\hat{\sigma}}}%
^{local}\left( \boldsymbol{k}\right) \quad \text{with\quad }\mathcal{\bar{K}}%
_{2}\left( \boldsymbol{k}\right) =\frac{1}{1+\left( \boldsymbol{L}\cdot 
\boldsymbol{k}\right) ^{2}},  \label{K2-bar}
\end{align}%
since one has $\bar{S}_{2}\left( \boldsymbol{k}\right) =-\mathcal{F}\left[ 
\mathrm{S}_{2}\left( \nabla \right) \right] \left( \boldsymbol{k}\right) =-%
\mathcal{F}\left[ \left( \boldsymbol{L}\cdot \nabla \right) ^{2}\right]
\left( \boldsymbol{k}\right) =\left( \boldsymbol{L}\cdot \boldsymbol{k}%
\right) ^{2},$ see also Table \ref{Modeli}. Assuming $k_{\parallel }=%
\boldsymbol{k}\cdot \frac{\boldsymbol{L}}{L}$ and $\boldsymbol{k}_{\perp }=%
\boldsymbol{k}-k_{\parallel }\frac{\boldsymbol{L}}{L},$ as well as $%
r_{\parallel }=\boldsymbol{r}\cdot \frac{\boldsymbol{L}}{L}$ and $%
\boldsymbol{r}_{\perp }=\boldsymbol{r}-r_{\parallel }\frac{\boldsymbol{L}}{L}%
,$ one has $\boldsymbol{k}\cdot \boldsymbol{r=k}_{\perp }\cdot \boldsymbol{r}%
_{\perp }+k_{\parallel }r_{\parallel }$ and $\mathrm{d}_{\boldsymbol{k}}V=%
\mathrm{d}_{\boldsymbol{k}_{\perp }}S\,\mathrm{d}k_{\parallel },$ so that
the inverse Fourier transform, defined by (\ref{IFT}), transforms the
expression (\ref{K2-bar})$_{2}$ into%
\begin{align}
\mathcal{K}_{2}\left( \boldsymbol{r}\right) & =\mathcal{F}^{-1}\left[ 
\mathcal{\bar{K}}_{2}\left( \boldsymbol{k}\right) \right] \left( \boldsymbol{%
r}\right) =\frac{1}{\left( 2\pi \right) ^{3}}\int_{%
\mathbb{R}
^{3}}\frac{1}{1+\left( \boldsymbol{L}\cdot \boldsymbol{k}\right) ^{2}}%
\mathrm{e}^{\mathrm{i}\boldsymbol{k}\cdot \boldsymbol{r}}\mathrm{d}_{%
\boldsymbol{k}}V  \notag \\
& =\left( \frac{1}{\left( 2\pi \right) ^{2}}\int_{%
\mathbb{R}
^{2}}\mathrm{e}^{\mathrm{i}\boldsymbol{k}_{\perp }\cdot \boldsymbol{r}%
_{\perp }}\mathrm{d}_{\boldsymbol{k}_{\perp }}S\right) \left( \frac{1}{2\pi }%
\int_{%
\mathbb{R}
}\frac{1}{1+L^{2}k_{\parallel }^{2}}\mathrm{e}^{\mathrm{i}k_{\parallel
}r_{\parallel }}\mathrm{d}k_{\parallel }\right)  \notag \\
& =\frac{1}{2\pi L^{2}}\delta \left( \boldsymbol{r}_{\perp }\right)
\int_{-\infty }^{\infty }\frac{1}{\frac{1}{L^{2}}+k_{\parallel }^{2}}\mathrm{%
e}^{\mathrm{i}k_{\parallel }r_{\parallel }}\mathrm{d}k_{\parallel }  \notag
\\
& =\frac{1}{\pi L^{2}}\delta \left( \boldsymbol{r}_{\perp }\right)
\int_{0}^{\infty }\frac{\cos \left( k_{\parallel }r_{\parallel }\right) }{%
\frac{1}{L^{2}}+k_{\parallel }^{2}}\mathrm{d}k_{\parallel }  \label{S-2-K-2}
\\
& =\frac{1}{2L}\delta \left( \boldsymbol{r}_{\perp }\right) \mathrm{e}^{-%
\frac{\left\vert r_{\parallel }\right\vert }{L}}.  \notag
\end{align}%
where $\delta \left( \boldsymbol{r}_{\perp }\right) =\mathcal{F}^{-1}\left[ 1%
\right] \left( \boldsymbol{k}\right) =\frac{1}{\left( 2\pi \right) ^{2}}%
\int_{%
\mathbb{R}
^{2}}\mathrm{e}^{\mathrm{i}\boldsymbol{k}_{\perp }\cdot \boldsymbol{r}%
_{\perp }}\mathrm{d}_{\boldsymbol{k}_{\perp }}S$ is the Dirac delta
distribution in two dimensions, while in (\ref{S-2-K-2}) the integral (\ref%
{integral-kosinus}) is used.

\subsubsection{Non-locality kernel of model $S_{3}$}

Model $S_{3}$, see the scalar-type stress-gradient Eringen three-dimensional
model (\ref{non-loc-scalar-type}) and Table \ref{Modeli-new}, constituted as
the last scalar-type three-dimensional generalization of Eringen model (\ref%
{1D-Eringen}), rather than the exponential type of non-locality kernel,
gives the non-locality kernel in terms of Macdonald function of zeroth order
as%
\begin{equation}
\mathcal{K}_{3}\left( r\right) \mathcal{=}\frac{1}{2\pi L^{2}}\delta \left(
r_{\parallel }\right) K_{0}\left( \frac{r_{\perp }}{L}\right) ,  \label{K3}
\end{equation}%
where $r_{\parallel }=\boldsymbol{r}\cdot \frac{\boldsymbol{L}}{L}$, $%
\boldsymbol{r}_{\perp }=\boldsymbol{r}-\boldsymbol{r}_{\parallel }$ with $%
\boldsymbol{r}_{\parallel }=r_{\parallel }\frac{\boldsymbol{L}}{L}$, and $%
r_{\perp }=\left\vert \boldsymbol{r}_{\perp }\right\vert $, see also Table %
\ref{Modeli-new}.

Namely, in the case of model $S_{3}$, the expression (\ref{Ki-bar-scalar})
for local stress tensor and non-locality kernel, both in Fourier domain,
becomes%
\begin{align}
\left( 1+\left\vert \boldsymbol{L}\times \boldsymbol{k}\right\vert
^{2}\right) \boldsymbol{\bar{\hat{\sigma}}}\left( \boldsymbol{k}\right) & =%
\boldsymbol{\bar{\hat{\sigma}}}^{local}\left( \boldsymbol{k}\right) ,\quad 
\text{i.e.,}  \notag \\
\boldsymbol{\bar{\hat{\sigma}}}\left( \boldsymbol{k}\right) & =\mathcal{\bar{%
K}}_{3}\left( \boldsymbol{k}\right) \boldsymbol{\bar{\hat{\sigma}}}%
^{local}\left( \boldsymbol{k}\right) \quad \text{with\quad }\mathcal{\bar{K}}%
_{3}\left( \boldsymbol{k}\right) =\frac{1}{1+L^{2}k^{2}-\left( \boldsymbol{L}%
\cdot \boldsymbol{k}\right) ^{2}},  \label{K3-bar}
\end{align}%
since one has $\bar{S}_{3}\left( \boldsymbol{k}\right) =-\mathcal{F}\left[ 
\mathrm{S}_{3}\left( \nabla \right) \right] \left( \boldsymbol{k}\right) =-%
\mathcal{F}\left[ \left\vert \boldsymbol{L}\times \boldsymbol{k}\right\vert
^{2}\right] \left( \boldsymbol{k}\right) =L^{2}k^{2}-\left( \boldsymbol{L}%
\cdot \boldsymbol{k}\right) ^{2},$ see also Table \ref{Modeli}, due to%
\begin{align}
\left\vert \boldsymbol{L}\times \boldsymbol{k}\right\vert ^{2}& =\left( 
\boldsymbol{L}\times \boldsymbol{k}\right) \cdot \left( \boldsymbol{L}\times 
\boldsymbol{k}\right)  \notag \\
& =\boldsymbol{L}\cdot \left( \boldsymbol{k}\times \left( \boldsymbol{L}%
\times \boldsymbol{k}\right) \right)  \notag \\
& =\boldsymbol{L}\cdot \left( \boldsymbol{L}\left( \boldsymbol{k}\cdot 
\boldsymbol{k}\right) -\boldsymbol{k}\left( \boldsymbol{L}\cdot \boldsymbol{k%
}\right) \right)  \notag \\
& =L^{2}k^{2}-\left( \boldsymbol{L}\cdot \boldsymbol{k}\right) ^{2}.  \notag
\end{align}%
Assuming $k_{\parallel }=\boldsymbol{k}\cdot \frac{\boldsymbol{L}}{L}$ and $%
\boldsymbol{k}_{\perp }=\boldsymbol{k}-k_{\parallel }\frac{\boldsymbol{L}}{L}
$ with $k_{\perp }^{2}=k^{2}-k_{\parallel }^{2},$ as well as $r_{\parallel }=%
\boldsymbol{r}\cdot \frac{\boldsymbol{L}}{L}$ and $\boldsymbol{r}_{\perp }=%
\boldsymbol{r}-r_{\parallel }\frac{\boldsymbol{L}}{L}$ with $r_{\perp
}=\left\vert \boldsymbol{r}_{\perp }\right\vert ,$ one has $\mathrm{d}_{%
\boldsymbol{k}}V=\mathrm{d}_{\boldsymbol{k}_{\perp }}S\,\mathrm{d}%
k_{\parallel }$ and $\boldsymbol{k}\cdot \boldsymbol{r=k}_{\perp }\cdot 
\boldsymbol{r}_{\perp }+k_{\parallel }r_{\parallel }$ and additionally
assuming $k_{p}=\boldsymbol{k}_{\perp }\cdot \frac{\boldsymbol{r}_{\perp }}{%
r_{\perp }}$ and $\boldsymbol{k}_{n}=\boldsymbol{k}_{\perp }-k_{p}\frac{%
\boldsymbol{r}_{\perp }}{r_{\perp }},$ as well as $\mathrm{d}_{\boldsymbol{k}%
_{\perp }}S=\mathrm{d}k_{p}\,\mathrm{d}k_{n},$ one has that the inverse
Fourier transform, defined by (\ref{IFT}), transforms the expression (\ref%
{K3-bar})$_{2}$ into%
\begin{align}
\mathcal{K}_{3}\left( \boldsymbol{r}\right) & =\mathcal{F}^{-1}\left[ 
\mathcal{\bar{K}}_{3}\left( \boldsymbol{k}\right) \right] \left( \boldsymbol{%
r}\right) =\frac{1}{\left( 2\pi \right) ^{3}}\int_{%
\mathbb{R}
^{3}}\frac{1}{1+L^{2}k^{2}-\left( \boldsymbol{L}\cdot \boldsymbol{k}\right)
^{2}}\mathrm{e}^{\mathrm{i}\boldsymbol{k}\cdot \boldsymbol{r}}\mathrm{d}_{%
\boldsymbol{k}}V  \notag \\
& =\left( \frac{1}{2\pi }\int_{%
\mathbb{R}
}\mathrm{e}^{\mathrm{i}k_{\parallel }r_{\parallel }}\mathrm{d}k_{\parallel
}\right) \left( \frac{1}{\left( 2\pi \right) ^{2}}\int_{%
\mathbb{R}
^{2}}\frac{1}{1+L^{2}k_{\perp }^{2}}\mathrm{e}^{\mathrm{i}\boldsymbol{k}%
_{\perp }\cdot \boldsymbol{r}_{\perp }}\mathrm{d}_{\boldsymbol{k}_{\perp
}}S\right)  \notag \\
& =\frac{1}{4\pi ^{2}L^{2}}\delta \left( r_{\parallel }\right) \int_{-\infty
}^{\infty }\int_{-\infty }^{\infty }\frac{1}{\frac{1}{L^{2}}%
+k_{n}^{2}+k_{p}^{2}}\mathrm{e}^{\mathrm{i}k_{p}r_{\perp }}\mathrm{d}k_{p}\,%
\mathrm{d}k_{n}  \notag \\
& =\frac{1}{2\pi ^{2}L^{2}}\delta \left( r_{\parallel }\right) \int_{-\infty
}^{\infty }\left( \int_{0}^{\infty }\frac{\cos \left( k_{p}r_{\perp }\right) 
}{\frac{1}{L^{2}}+k_{n}^{2}+k_{p}^{2}}\mathrm{d}k_{p}\right) \mathrm{d}k_{n}
\label{S-3-K-3-1} \\
& =\frac{1}{2\pi ^{2}L^{2}}\delta \left( r_{\parallel }\right) \int_{-\infty
}^{\infty }\frac{\pi }{2}\frac{\mathrm{e}^{-\frac{r_{\perp }}{L}\sqrt{%
1+L^{2}k_{n}^{2}}}}{\frac{1}{L}\sqrt{1+L^{2}k_{n}^{2}}}\mathrm{d}k_{n} 
\notag \\
& =\frac{1}{2\pi L}\delta \left( r_{\parallel }\right) \int_{0}^{\infty }%
\frac{\mathrm{e}^{-\frac{r_{\perp }}{L}\sqrt{1+L^{2}k_{n}^{2}}}}{\sqrt{%
1+L^{2}k_{n}^{2}}}\mathrm{d}k_{n}  \label{S-3-K-3-2} \\
& =\frac{1}{2\pi L^{2}}\delta \left( r_{\parallel }\right) \int_{1}^{\infty }%
\frac{\mathrm{e}^{-\frac{r_{\perp }}{L}u}}{\sqrt{u^{2}-1}}\mathrm{d}u  \notag
\\
& =\frac{1}{2\pi L^{2}}\delta \left( r_{\parallel }\right) K_{0}\left( \frac{%
r_{\perp }}{L}\right) ,  \notag
\end{align}%
where $\delta \left( r_{\parallel }\right) =\mathcal{F}^{-1}\left[ 1\right]
\left( k\right) =\frac{1}{2\pi }\int_{%
\mathbb{R}
}\mathrm{e}^{\mathrm{i}kr_{\parallel }}\mathrm{d}k$ is the Dirac delta
distribution in one dimension, while in (\ref{S-3-K-3-1}) the integral (\ref%
{integral-kosinus}) is used and moreover the substitution $u=\sqrt{%
1+L^{2}k_{n}^{2}}$ in (\ref{S-3-K-3-2}) yields the integral representation
of Macdonald function of the zeroth order, since the Macdonald function,
i.e., modified Bessel function of the second kind of order $\nu $ is
represented by%
\begin{equation}
K_{\nu }\left( z\right) =\frac{\sqrt{\pi }\left( \frac{z}{2}\right) ^{\nu }}{%
\Gamma \left( \nu +\frac{1}{2}\right) }\int_{1}^{\infty }\mathrm{e}%
^{-uz}\left( u^{2}-1\right) ^{\nu -\frac{1}{2}}\mathrm{d}u,\quad \text{%
for\quad }\func{Re}z>0\quad \text{and}\quad \func{Re}\nu >-\frac{1}{2}.
\label{Macdonald}
\end{equation}

\subsubsection{Non-locality kernel of model $T_{1}$}

Model $T_{1}$, see the tensor-type stress-gradient Eringen three-dimensional
model (\ref{non-loc-tensor-type}) and Table \ref{Modeli-new}, constituted as
a three-dimensional tensor-type straightforward generalization of Eringen
model (\ref{1D-Eringen}), has a following tensor%
\begin{equation}
\boldsymbol{\hat{\mathcal{K}}}_{1}\left( \boldsymbol{r}\right) =2\delta
\left( \boldsymbol{r}\right) \left( \boldsymbol{\hat{I}}-2\frac{\boldsymbol{r%
}\otimes \boldsymbol{r}}{r^{2}}\right) -\frac{1}{4\pi r^{3}}\mathrm{e}^{-%
\frac{r}{L}}\left( \left( 1+\frac{r}{L}\right) \boldsymbol{\hat{I}}+\left(
3\left( 1+\frac{r}{L}\right) +\frac{r^{2}}{L^{2}}\right) \frac{\boldsymbol{r}%
\otimes \boldsymbol{r}}{r^{2}}\right) ,  \label{kernel-T1}
\end{equation}%
as a non-locality kernel, see also Table \ref{Modeli-new}.

Namely, in the case of model $T_{1}$, the expression (\ref{Ki-bar-tensor})
for local stress tensor and non-locality kernel, both in Fourier domain,
becomes%
\begin{align}
\boldsymbol{\bar{\hat{\mathcal{K}}}}_{1}^{-1}\left( \boldsymbol{k}\right) 
\boldsymbol{\bar{\hat{\sigma}}}\left( \boldsymbol{k}\right) & =\boldsymbol{%
\bar{\hat{\sigma}}}^{local}\left( \boldsymbol{k}\right) ,\quad \text{%
with\quad }\boldsymbol{\bar{\hat{\mathcal{K}}}}_{1}^{-1}\left( \boldsymbol{k}%
\right) =\boldsymbol{\hat{I}}+L^{2}\left( \boldsymbol{k}\otimes \boldsymbol{k%
}\right) ,\quad \text{i.e.,}  \label{K-1-hat-bar-1} \\
\boldsymbol{\bar{\hat{\sigma}}}\left( \boldsymbol{k}\right) & =\boldsymbol{%
\bar{\hat{\mathcal{K}}}}_{1}\left( \boldsymbol{k}\right) \boldsymbol{\bar{%
\hat{\sigma}}}^{local}\left( \boldsymbol{k}\right) \quad \text{with\quad }%
\boldsymbol{\bar{\hat{\mathcal{K}}}}_{1}\left( \boldsymbol{k}\right) =%
\boldsymbol{\hat{I}}-\frac{\boldsymbol{k}\otimes \boldsymbol{k}}{\frac{1}{%
L^{2}}+k^{2}},  \label{K-1-hat-bar-2}
\end{align}%
since one has $\boldsymbol{\bar{\hat{T}}}_{1}\left( \boldsymbol{k}\right) =-%
\mathcal{F}\left[ \mathbf{\hat{T}}_{1}\left( \nabla \right) \right] \left( 
\boldsymbol{k}\right) =-\mathcal{F}\left[ \left( \boldsymbol{L}\cdot 
\boldsymbol{L}\right) \left( \nabla \otimes \nabla \right) \right] \left( 
\boldsymbol{k}\right) =L^{2}\left( \boldsymbol{k}\otimes \boldsymbol{k}%
\right) ,$ see also Table \ref{Modeli}, where the tensor $\boldsymbol{\bar{%
\hat{\mathcal{K}}}}_{1}$ represents an inverse counterpart of tensor $%
\boldsymbol{\bar{\hat{\mathcal{K}}}}_{1}^{-1}$, given by (\ref{K-1-hat-bar-1}%
)$_{2}$, such that $\boldsymbol{\bar{\hat{\mathcal{K}}}}_{1}^{-1}\boldsymbol{%
\bar{\hat{\mathcal{K}}}}_{1}=\boldsymbol{\hat{I}}$, and it is obtained in
the form (\ref{K-1-hat-bar-2})$_{2}$ according to
Sherman--Morrison--Woodbury matrix identity, derived in the middle of the
last century \cite{SMW}, which reduces to the Sherman--Morrison formula when
a matrix operator is represented by two vectors \cite{SM}, as in the case of
the tensors $\boldsymbol{\bar{\hat{\mathcal{K}}}}_{1}^{-1}$ and $\boldsymbol{%
\bar{\hat{\mathcal{K}}}}_{1}$, given by (\ref{K-1-hat-bar-1})$_{2}$ and (\ref%
{K-1-hat-bar-2})$_{2}$. The inverse Fourier transform of tensor $\boldsymbol{%
\bar{\hat{\mathcal{K}}}}_{1}$, see (\ref{K-1-hat-bar-2})$_{2}$, reduces to
the inversion of tensor $\frac{\boldsymbol{k}\otimes \boldsymbol{k}}{\frac{1%
}{L^{2}}+k^{2}}$, which can be written in the matrix form as%
\begin{equation}
\frac{\boldsymbol{k}\otimes \boldsymbol{k}}{\frac{1}{L^{2}}+k^{2}}\!=\!\frac{%
1}{\frac{1}{L^{2}}+k^{2}}\!\left[ 
\begin{array}{ccc}
k_{x}^{2} & k_{x}k_{y} & k_{x}k_{z} \\ 
k_{x}k_{y} & k_{y}^{2} & k_{y}k_{z} \\ 
k_{x}k_{z} & k_{y}k_{z} & k_{z}^{2}%
\end{array}%
\right] \!\!=\frac{k^{2}}{\frac{1}{L^{2}}+k^{2}}\!\left[ 
\begin{array}{ccc}
\sin ^{2}\theta \cos ^{2}\varphi & \sin ^{2}\theta \sin \varphi \cos \varphi
& \sin \theta \cos \theta \cos \varphi \\ 
\sin ^{2}\theta \sin \varphi \cos \varphi & \sin ^{2}\theta \sin ^{2}\varphi
& \sin \theta \cos \theta \sin \varphi \\ 
\sin \theta \cos \theta \cos \varphi & \sin \theta \cos \theta \sin \varphi
& \cos ^{2}\theta%
\end{array}%
\right] ,  \label{matrica-T1}
\end{equation}%
using Cartesian coordinate system for tensor representation and expressing
tensor components in spherical coordinate systems, such that the orientation
of the radius vector $\boldsymbol{r}$ can be chosen to coincide with the $z$%
-axis, i.e., $\boldsymbol{r}=r\boldsymbol{e}_{z},$ while $\boldsymbol{k}%
=\left( k_{x},k_{y},k_{z}\right) =\left( k\sin \theta \cos \varphi ,k\sin
\theta \sin \varphi ,k\cos \theta \right) $. The Fourier transform inversion
of tensor $\frac{\boldsymbol{k}\otimes \boldsymbol{k}}{\frac{1}{L^{2}}+k^{2}}
$ is achieved by inverting each component of the matrix (\ref{matrica-T1}),
according to the definition of the inverse Fourier transform (\ref{IFT}),
implying%
\begin{align}
\mathcal{F}^{-1}\left[ \frac{k_{x}^{2}}{\frac{1}{L^{2}}+k^{2}}\right] \left( 
\boldsymbol{r}\right) & =\frac{1}{\left( 2\pi \right) ^{3}}\int_{%
\mathbb{R}
^{3}}\frac{k_{x}^{2}}{\frac{1}{L^{2}}+k^{2}}\mathrm{e}^{\mathrm{i}%
\boldsymbol{k}\cdot \boldsymbol{r}}\mathrm{d}_{\boldsymbol{k}}V  \notag \\
& =\frac{1}{\left( 2\pi \right) ^{3}}\int_{0}^{2\pi }\!\!\!\int_{0}^{\infty
}\!\!\!\int_{0}^{\pi }\frac{k^{2}}{\frac{1}{L^{2}}+k^{2}}\sin ^{2}\theta
\cos ^{2}\varphi \,\mathrm{e}^{\mathrm{i}kr\cos \theta }k^{2}\sin \theta 
\mathrm{d}\theta \mathrm{d}k\mathrm{d}\varphi  \label{T1-x-0} \\
& =\frac{1}{8\pi ^{2}}\int_{0}^{\infty }\frac{k^{4}}{\frac{1}{L^{2}}+k^{2}}%
\int_{-\mathrm{i}kr}^{\mathrm{i}kr}\left( 1+\frac{u^{2}}{k^{2}r^{2}}\right) 
\mathrm{e}^{u}\frac{\mathrm{d}u}{\mathrm{i}kr}\mathrm{d}k  \notag \\
& =\frac{1}{2\pi ^{2}}\int_{0}^{\infty }\frac{k^{4}}{\frac{1}{L^{2}}+k^{2}}%
\left( -\frac{\cos \left( kr\right) }{k^{2}r^{2}}+\frac{\sin \left(
kr\right) }{k^{3}r^{3}}\right) \mathrm{d}k  \notag \\
& =-\frac{1}{2\pi ^{2}r^{2}}\int_{0}^{\infty }\frac{k^{2}\cos \left(
kr\right) }{\frac{1}{L^{2}}+k^{2}}\mathrm{d}k+\frac{1}{2\pi ^{2}r^{3}}%
\int_{0}^{\infty }\frac{k\sin \left( kr\right) }{\frac{1}{L^{2}}+k^{2}}%
\mathrm{d}k  \notag \\
& =-\frac{1}{2\pi ^{2}r^{2}}\left( -\frac{\partial ^{2}}{\partial r^{2}}%
\int_{0}^{\infty }\frac{\cos \left( kr\right) }{\frac{1}{L^{2}}+k^{2}}%
\mathrm{d}k\right) +\frac{1}{2\pi ^{2}r^{3}}\left( -\frac{\partial }{%
\partial r}\int_{0}^{\infty }\frac{\cos \left( kr\right) }{\frac{1}{L^{2}}%
+k^{2}}\mathrm{d}k\right)  \label{T1-x-1} \\
& =-\frac{1}{2\pi ^{2}r^{2}}\left( \pi \delta \left( r\right) \mathrm{e}^{-%
\frac{\left\vert r\right\vert }{L}}-\frac{\pi }{2}\frac{1}{L}\mathrm{e}^{-%
\frac{\left\vert r\right\vert }{L}}\right) +\frac{1}{2\pi ^{2}r^{3}}\left( 
\frac{\pi }{2}\mathrm{e}^{-\frac{\left\vert r\right\vert }{L}}\right)
\label{T1-x-2} \\
& =-\delta \left( \boldsymbol{r}\right) +\frac{1}{4\pi r^{3}}\left( 1+\frac{r%
}{L}\right) \mathrm{e}^{-\frac{\left\vert r\right\vert }{L}},  \notag
\end{align}%
such that in (\ref{T1-x-0}) the substitution $u=\mathrm{i}kr\cos \theta $ is
used, while in (\ref{T1-x-1}) the following integral%
\begin{equation}
\int_{0}^{\infty }\frac{\cos \left( \alpha x\right) }{\beta ^{2}+x^{2}}%
\mathrm{d}x=\frac{\pi }{2}\frac{\mathrm{e}^{-\beta \left\vert \alpha
\right\vert }}{\beta },  \label{int-sa-kosinusom}
\end{equation}%
is employed, and finally in (\ref{T1-x-2}) properties $\delta \left(
r\right) \mathrm{e}^{-\frac{\left\vert r\right\vert }{L}}=\delta \left(
r\right) $ and $\delta \left( \boldsymbol{r}\right) =\frac{\delta \left(
r\right) }{2\pi r^{2}}$ of Dirac distribution are used. Similarly, it is
obtained that%
\begin{equation*}
\mathcal{F}^{-1}\left[ \frac{k_{y}^{2}}{\frac{1}{L^{2}}+k^{2}}\right] \left( 
\boldsymbol{r}\right) =\delta \left( \boldsymbol{r}\right) -\frac{1}{4\pi
r^{3}}\left( 1+\frac{r}{L}\right) \mathrm{e}^{-\frac{\left\vert r\right\vert 
}{L}}.
\end{equation*}%
Further, one has%
\begin{align}
\mathcal{F}^{-1}& \left[ \frac{k_{z}^{2}}{\frac{1}{L^{2}}+k^{2}}\right]
\left( \boldsymbol{r}\right) =\frac{1}{\left( 2\pi \right) ^{3}}\int_{%
\mathbb{R}
^{3}}\frac{k_{z}^{2}}{\frac{1}{L^{2}}+k^{2}}\mathrm{e}^{\mathrm{i}%
\boldsymbol{k}\cdot \boldsymbol{r}}\mathrm{d}_{\boldsymbol{k}}V  \notag \\
& =\frac{1}{\left( 2\pi \right) ^{3}}\int_{0}^{2\pi }\!\!\!\int_{0}^{\infty
}\!\!\!\int_{0}^{\pi }\frac{k^{2}}{\frac{1}{L^{2}}+k^{2}}\cos ^{2}\theta \,%
\mathrm{e}^{\mathrm{i}kr\cos \theta }k^{2}\sin \theta \mathrm{d}\theta 
\mathrm{d}k\mathrm{d}\varphi  \label{T1-z-0} \\
& =-\frac{1}{4\pi ^{2}}\int_{0}^{\infty }\frac{k^{4}}{\frac{1}{L^{2}}+k^{2}}%
\int_{-\mathrm{i}kr}^{\mathrm{i}kr}\frac{u^{2}}{k^{2}r^{2}}\mathrm{e}^{u}%
\frac{\mathrm{d}u}{\mathrm{i}kr}\mathrm{d}k  \notag \\
& =-\frac{1}{4\pi ^{2}}\int_{0}^{\infty }\frac{k^{4}}{\frac{1}{L^{2}}+k^{2}}%
\left( 4\frac{\sin \left( kr\right) }{k^{3}r^{3}}-4\frac{\cos \left(
kr\right) }{k^{2}r^{2}}-2\frac{\sin \left( kr\right) }{kr}\right) \mathrm{d}k
\notag \\
& =-\frac{1}{\pi ^{2}r^{3}}\int_{0}^{\infty }\frac{k\sin \left( kr\right) }{%
\frac{1}{L^{2}}+k^{2}}\mathrm{d}k+\frac{1}{\pi ^{2}r^{2}}\int_{0}^{\infty }%
\frac{k^{2}\cos \left( kr\right) }{\frac{1}{L^{2}}+k^{2}}\mathrm{d}k+\frac{1%
}{2\pi ^{2}r}\int_{0}^{\infty }\frac{k^{3}\sin \left( kr\right) }{\frac{1}{%
L^{2}}+k^{2}}\mathrm{d}k  \notag \\
& =-\frac{1}{\pi ^{2}r^{3}}\left( -\frac{\partial }{\partial r}%
\int_{0}^{\infty }\frac{\cos \left( kr\right) }{\frac{1}{L^{2}}+k^{2}}%
\mathrm{d}k\right) +\frac{1}{\pi ^{2}r^{2}}\left( -\frac{\partial ^{2}}{%
\partial r^{2}}\int_{0}^{\infty }\frac{\cos \left( kr\right) }{\frac{1}{L^{2}%
}+k^{2}}\mathrm{d}k\right) +\frac{1}{2\pi ^{2}r}\left( \frac{\partial ^{3}}{%
\partial r^{3}}\int_{0}^{\infty }\frac{\cos \left( kr\right) }{\frac{1}{L^{2}%
}+k^{2}}\mathrm{d}k\right)  \label{T1-z-1} \\
& =-\frac{1}{\pi ^{2}r^{3}}\left( \frac{\pi }{2}\mathrm{e}^{-\frac{%
\left\vert r\right\vert }{L}}\right) +\frac{1}{\pi ^{2}r^{2}}\left( \pi
\delta \left( r\right) \mathrm{e}^{-\frac{\left\vert r\right\vert }{L}}-%
\frac{\pi }{2}\frac{1}{L}\mathrm{e}^{-\frac{\left\vert r\right\vert }{L}%
}\right) +\frac{1}{2\pi ^{2}r}\left( -\pi \frac{\mathrm{d}\delta \left(
r\right) }{\mathrm{d}r}-\frac{\pi }{2}\frac{1}{L^{2}}\mathrm{e}^{-\frac{%
\left\vert r\right\vert }{L}}\right)  \notag \\
& =-\frac{1}{2\pi r^{3}}\mathrm{e}^{-\frac{\left\vert r\right\vert }{L}}+%
\frac{1}{\pi r^{2}}\delta \left( r\right) -\frac{1}{2\pi Lr^{2}}\mathrm{e}^{-%
\frac{\left\vert r\right\vert }{L}}-\frac{1}{2\pi r}\frac{\mathrm{d}\delta
\left( r\right) }{\mathrm{d}r}-\frac{1}{4\pi L^{2}r}\mathrm{e}^{-\frac{%
\left\vert r\right\vert }{L}}  \label{T1-z-2} \\
& =3\delta \left( \boldsymbol{r}\right) -\frac{1}{4\pi r^{3}}\left( 2\left(
1+\frac{r}{L}\right) +\frac{r^{2}}{L^{2}}\right) \mathrm{e}^{-\frac{%
\left\vert r\right\vert }{L}},  \notag
\end{align}%
such that in (\ref{T1-z-0}) the substitution $u=\mathrm{i}kr\cos \theta $ is
used, while in (\ref{T1-z-1}) the integral (\ref{int-sa-kosinusom}) is used,
and finally in (\ref{T1-z-2}) the property that connects Dirac delta
distribution with its derivative, namely $-r\frac{\mathrm{d}\delta \left(
r\right) }{\mathrm{d}r}=\delta \left( r\right) $, is used. The inversion of
the off-diagonal terms of the matrix (\ref{matrica-T1}) gives zero for the
result, i.e.,%
\begin{align*}
\mathcal{F}^{-1}\left[ \frac{k_{x}k_{y}}{\frac{1}{L^{2}}+k^{2}}\right]
\left( \boldsymbol{r}\right) & =\frac{1}{\left( 2\pi \right) ^{3}}\int_{%
\mathbb{R}
^{3}}\frac{k_{x}k_{y}}{\frac{1}{L^{2}}+k^{2}}\mathrm{e}^{\mathrm{i}%
\boldsymbol{k}\cdot \boldsymbol{r}}\mathrm{d}_{\boldsymbol{k}}V \\
& =\frac{1}{\left( 2\pi \right) ^{3}}\int_{0}^{2\pi }\!\!\!\int_{0}^{\infty
}\!\!\!\int_{0}^{\pi }\frac{k^{2}}{\frac{1}{L^{2}}+k^{2}}\sin ^{2}\theta
\sin \varphi \cos \varphi \,\mathrm{e}^{\mathrm{i}kr\cos \theta }k^{2}\sin
\theta \mathrm{d}\theta \mathrm{d}k\mathrm{d}\varphi =0, \\
\mathcal{F}^{-1}\left[ \frac{k_{x}k_{z}}{\frac{1}{L^{2}}+k^{2}}\right]
\left( \boldsymbol{r}\right) & =\frac{1}{\left( 2\pi \right) ^{3}}\int_{%
\mathbb{R}
^{3}}\frac{k_{x}k_{z}}{\frac{1}{L^{2}}+k^{2}}\mathrm{e}^{\mathrm{i}%
\boldsymbol{k}\cdot \boldsymbol{r}}\mathrm{d}_{\boldsymbol{k}}V \\
& =\frac{1}{\left( 2\pi \right) ^{3}}\int_{0}^{2\pi }\!\!\!\int_{0}^{\infty
}\!\!\!\int_{0}^{\pi }\frac{k^{2}}{\frac{1}{L^{2}}+k^{2}}\sin \theta \cos
\theta \cos \varphi \,\mathrm{e}^{\mathrm{i}kr\cos \theta }k^{2}\sin \theta 
\mathrm{d}\theta \mathrm{d}k\mathrm{d}\varphi =0, \\
\mathcal{F}^{-1}\left[ \frac{k_{y}k_{z}}{\frac{1}{L^{2}}+k^{2}}\right]
\left( \boldsymbol{r}\right) & =\frac{1}{\left( 2\pi \right) ^{3}}\int_{%
\mathbb{R}
^{3}}\frac{k_{y}k_{z}}{\frac{1}{L^{2}}+k^{2}}\mathrm{e}^{\mathrm{i}%
\boldsymbol{k}\cdot \boldsymbol{r}}\mathrm{d}_{\boldsymbol{k}}V \\
& =\frac{1}{\left( 2\pi \right) ^{3}}\int_{0}^{2\pi }\!\!\!\int_{0}^{\infty
}\!\!\!\int_{0}^{\pi }\frac{k^{2}}{\frac{1}{L^{2}}+k^{2}}\sin \theta \cos
\theta \sin \varphi \,\mathrm{e}^{\mathrm{i}kr\cos \theta }k^{2}\sin \theta 
\mathrm{d}\theta \mathrm{d}k\mathrm{d}\varphi =0,
\end{align*}%
since $\int_{0}^{2\pi }\sin \varphi \cos \varphi \,\mathrm{d}\varphi
=\int_{0}^{2\pi }\cos \varphi \,\mathrm{d}\varphi =\int_{0}^{2\pi }\sin
\varphi \,\mathrm{d}\varphi =0$.

Fourier transform inversion of tensor $\frac{\boldsymbol{k}\otimes 
\boldsymbol{k}}{\frac{1}{L^{2}}+k^{2}},$ given by (\ref{matrica-T1}), after
substitution of previously calculated inversions of its components, yields%
\begin{align*}
\mathcal{F}^{-1} & \left[ \frac{\boldsymbol{k}\otimes \boldsymbol{k}}{\frac{1%
}{L^{2}}+k^{2}}\right] \left( \boldsymbol{r}\right) \\
&=\left[ 
\begin{array}{ccc}
-\delta \left( \boldsymbol{r}\right) +\frac{1}{4\pi r^{3}}\left( 1+\frac{r}{L%
}\right) \mathrm{e}^{-\frac{\left\vert r\right\vert }{L}} & 0 & 0 \\ 
0 & -\delta \left( \boldsymbol{r}\right) +\frac{1}{4\pi r^{3}}\left( 1+\frac{%
r}{L}\right) \mathrm{e}^{-\frac{\left\vert r\right\vert }{L}} & 0 \\ 
0 & 0 & 3\delta \left( \boldsymbol{r}\right) -\frac{1}{4\pi r^{3}}\left(
2\left( 1+\frac{r}{L}\right) +\frac{r^{2}}{L^{2}}\right) \mathrm{e}^{-\frac{%
\left\vert r\right\vert }{L}}%
\end{array}%
\right] \\
& =-\delta \left( \boldsymbol{r}\right) \boldsymbol{\hat{I}}+\frac{1}{4\pi
r^{3}}\left( 1+\frac{r}{L}\right) \mathrm{e}^{-\frac{\left\vert r\right\vert 
}{L}}\boldsymbol{\hat{I}}+4\delta \left( \boldsymbol{r}\right) \frac{%
\boldsymbol{r}\otimes \boldsymbol{r}}{r^{2}}-\frac{1}{4\pi r^{3}}\left(
3\left( 1+\frac{r}{L}\right) +\frac{r^{2}}{L^{2}}\right) \mathrm{e}^{-\frac{%
\left\vert r\right\vert }{L}}\frac{\boldsymbol{r}\otimes \boldsymbol{r}}{%
r^{2}},
\end{align*}%
yielding%
\begin{align*}
\boldsymbol{\hat{\mathcal{K}}}_{1}\left( \boldsymbol{r}\right) & =\mathcal{F}%
^{-1}\left[ \boldsymbol{\bar{\hat{\mathcal{K}}}}_{1}\left( \boldsymbol{k}%
\right) \right] \left( \boldsymbol{r}\right) =\mathcal{\delta }\left( 
\boldsymbol{r}\right) \boldsymbol{\hat{I}}-\mathcal{F}^{-1}\left[ \frac{%
\boldsymbol{k}\otimes \boldsymbol{k}}{\frac{1}{L^{2}}+k^{2}}\right] \left( 
\boldsymbol{r}\right) \\
&=2\delta \left( \boldsymbol{r}\right) \boldsymbol{\hat{I}}-4\delta \left( 
\boldsymbol{r}\right) \frac{\boldsymbol{r}\otimes \boldsymbol{r}}{r^{2}}-%
\frac{1}{4\pi r^{3}}\left( 1+\frac{r}{L}\right) \mathrm{e}^{-\frac{%
\left\vert r\right\vert }{L}}\boldsymbol{\hat{I}}+\frac{1}{4\pi r^{3}}\left(
3\left( 1+\frac{r}{L}\right) +\frac{r^{2}}{L^{2}}\right) \mathrm{e}^{-\frac{%
\left\vert r\right\vert }{L}}\frac{\boldsymbol{r}\otimes \boldsymbol{r}}{%
r^{2}},
\end{align*}%
where $\boldsymbol{\bar{\hat{\mathcal{K}}}}_{1}$ is given by (\ref%
{K-1-hat-bar-2})$_{2},$ and taking the form (\ref{kernel-T1}), where the
absolute values are omitted.

\subsubsection{Non-locality kernel of model $T_{2}$}

Model $T_{2}$, see the tensor-type stress-gradient Eringen three-dimensional
model (\ref{non-loc-tensor-type}) and Table \ref{Modeli-new}, constituted as
another three-dimensional tensor-type generalization of Eringen model (\ref%
{1D-Eringen}), has a following tensor%
\begin{equation}
\boldsymbol{\hat{\mathcal{K}}}_{2}\left( \boldsymbol{r}\right) =\delta
\left( \boldsymbol{r}\right) \left( \boldsymbol{\hat{I}}-\frac{\boldsymbol{L}%
\otimes \boldsymbol{L}}{L^{2}}\right) +\frac{1}{4\pi L^{2}r}\mathrm{e}^{-%
\frac{r}{L}}\frac{\boldsymbol{L}\otimes \boldsymbol{L}}{L^{2}},
\label{kernel-T2}
\end{equation}%
as a non-locality kernel, see also Table \ref{Modeli-new}.

Namely, in the case of model $T_{2}$, the expression (\ref{Ki-bar-tensor})
for local stress tensor and non-locality kernel, both in Fourier domain,
becomes%
\begin{align}
\boldsymbol{\bar{\hat{\mathcal{K}}}}_{2}^{-1}\left( \boldsymbol{k}\right) 
\boldsymbol{\bar{\hat{\sigma}}}\left( \boldsymbol{k}\right) & =\boldsymbol{%
\bar{\hat{\sigma}}}^{local}\left( \boldsymbol{k}\right) ,\quad \text{%
with\quad }\boldsymbol{\bar{\hat{\mathcal{K}}}}_{2}^{-1}\left( \boldsymbol{k}%
\right) =\boldsymbol{\hat{I}}+k^{2}\left( \boldsymbol{L}\otimes \boldsymbol{L%
}\right) ,\quad \text{i.e.,}  \label{K-2-hat-bar-1} \\
\boldsymbol{\bar{\hat{\sigma}}}\left( \boldsymbol{k}\right) & =\boldsymbol{%
\bar{\hat{\mathcal{K}}}}_{2}\left( \boldsymbol{k}\right) \boldsymbol{\bar{%
\hat{\sigma}}}^{local}\left( \boldsymbol{k}\right) \quad \text{with\quad }%
\boldsymbol{\bar{\hat{\mathcal{K}}}}_{2}\left( \boldsymbol{k}\right) =%
\boldsymbol{\hat{I}}-\frac{k^{2}}{\frac{1}{L^{2}}+k^{2}}\frac{\boldsymbol{L}%
\otimes \boldsymbol{L}}{L^{2}},  \label{K-2-hat-bar-2}
\end{align}%
since one has $\boldsymbol{\bar{\hat{T}}}_{2}\left( \boldsymbol{k}\right) =-%
\mathcal{F}\left[ \mathbf{\hat{T}}_{2}\left( \nabla \right) \right] \left( 
\boldsymbol{k}\right) =-\mathcal{F}\left[ \left( \boldsymbol{L}\otimes 
\boldsymbol{L}\right) \left( \nabla \cdot \nabla \right) \right] \left( 
\boldsymbol{k}\right) =k^{2}\left( \boldsymbol{L}\otimes \boldsymbol{L}%
\right) ,$ where the tensor $\boldsymbol{\bar{\hat{\mathcal{K}}}}_{2}$
represents an inverse counterpart of tensor $\boldsymbol{\bar{\hat{\mathcal{K%
}}}}_{2}^{-1}$, given by (\ref{K-2-hat-bar-1})$_{2}$, such that $\boldsymbol{%
\bar{\hat{\mathcal{K}}}}_{2}^{-1}\boldsymbol{\bar{\hat{\mathcal{K}}}}_{2}=%
\boldsymbol{\hat{I}}$, and it is obtained in the form (\ref{K-2-hat-bar-2})$%
_{2}$ according to Sherman--Morrison--Woodbury and Sherman--Morrison matrix
formula, see \cite{SMW} and \cite{SM}. The inverse Fourier transform of
tensor $\boldsymbol{\bar{\hat{\mathcal{K}}}}_{2}$, see (\ref{K-2-hat-bar-2})$%
_{2}$, reduces to the inversion of the scalar function $\frac{k^{2}}{\frac{1%
}{L^{2}}+k^{2}}$, which can be calculated using spherical coordinate system
with fixed radius vector $\boldsymbol{r}=r\boldsymbol{e}_{z}$, implying $%
\boldsymbol{k}\cdot \boldsymbol{r}=kr\cos \theta ,$ so that, the Fourier
transform inversion of scalar function, according to the definition of the
inverse Fourier transform (\ref{IFT}), reads%
\begin{align}
\mathcal{F}^{-1}\left[ \frac{k^{2}}{\frac{1}{L^{2}}+k^{2}}\right] \left( 
\boldsymbol{r}\right) & =\frac{1}{\left( 2\pi \right) ^{3}}\int_{%
\mathbb{R}
^{3}}\frac{k^{2}}{\frac{1}{L^{2}}+k^{2}}\mathrm{e}^{\mathrm{i}\boldsymbol{k}%
\cdot \boldsymbol{r}}\mathrm{d}_{\boldsymbol{k}}V  \notag \\
& =\frac{1}{8\pi ^{3}}\int_{0}^{2\pi }\!\!\!\int_{0}^{\infty
}\!\!\!\int_{0}^{\pi }\frac{k^{2}}{\frac{1}{L^{2}}+k^{2}}\mathrm{e}^{\mathrm{%
i}kr\cos \theta }k^{2}\sin \theta \mathrm{d}\theta \mathrm{d}k\mathrm{d}%
\varphi  \label{T2-0} \\
& =\frac{1}{4\pi ^{2}}\int_{0}^{\infty }\frac{k^{4}}{\frac{1}{L^{2}}+k^{2}}%
\int_{-\mathrm{i}kr}^{\mathrm{i}kr}\mathrm{e}^{u}\frac{\mathrm{d}u}{\mathrm{i%
}kr}\mathrm{d}k  \notag \\
& =\frac{1}{2\pi ^{2}}\int_{0}^{\infty }\frac{k^{4}}{\frac{1}{L^{2}}+k^{2}}%
\frac{\sin \left( kr\right) }{kr}\mathrm{d}k  \notag \\
& =\frac{1}{2\pi ^{2}r}\int_{0}^{\infty }\frac{k^{3}\sin \left( kr\right) }{%
\frac{1}{L^{2}}+k^{2}}\mathrm{d}k  \notag \\
& =\frac{1}{2\pi ^{2}r}\left( \frac{\partial ^{3}}{\partial r^{3}}%
\int_{0}^{\infty }\frac{\cos \left( kr\right) }{\frac{1}{L^{2}}+k^{2}}%
\mathrm{d}k\right)  \label{T2-1} \\
& =\frac{1}{2\pi ^{2}r}\left( -\pi \frac{\mathrm{d}\delta \left( r\right) }{%
\mathrm{d}r}-\frac{\pi }{2}\frac{1}{L^{2}}\mathrm{e}^{-\frac{\left\vert
r\right\vert }{L}}\right)  \label{T2-2} \\
& =\delta \left( \boldsymbol{r}\right) -\frac{1}{4\pi L^{2}r}\mathrm{e}^{-%
\frac{\left\vert r\right\vert }{L}},  \notag
\end{align}%
such that in (\ref{T2-0}) the substitution $u=\mathrm{i}kr\cos \theta $ is
used, while in (\ref{T2-1}) the integral (\ref{int-sa-kosinusom}) is
employed, and finally in (\ref{T2-2}) properties $-r\frac{\mathrm{d}\delta
\left( r\right) }{\mathrm{d}r}=\delta \left( r\right) $ and $\delta \left( 
\boldsymbol{r}\right) =\frac{\delta \left( r\right) }{2\pi r^{2}}$ are used.

Therefore, the expression for non-locality kernel, after Fourier transform
inversion $\mathcal{F}^{-1}\left[ \boldsymbol{\bar{\hat{\mathcal{K}}}}_{2}%
\right] \left( \boldsymbol{r}\right) $, with $\boldsymbol{\bar{\hat{\mathcal{%
K}}}}_{2}$ given by (\ref{K-2-hat-bar-2})$_{2},$ becomes%
\begin{align*}
\boldsymbol{\hat{\mathcal{K}}}_{2}\left( \boldsymbol{r}\right) & =\mathcal{F}%
^{-1}\left[ \boldsymbol{\bar{\hat{\mathcal{K}}}}_{2}\left( \boldsymbol{k}%
\right) \right] \left( \boldsymbol{r}\right) =\mathcal{\delta }\left( 
\boldsymbol{r}\right) \boldsymbol{\hat{I}}-\mathcal{F}^{-1}\left[ \frac{k^{2}%
}{\frac{1}{L^{2}}+k^{2}}\right] \left( \boldsymbol{r}\right) \frac{%
\boldsymbol{L}\otimes \boldsymbol{L}}{L^{2}} \\
& =\delta \left( \boldsymbol{r}\right) \left( \boldsymbol{\hat{I}}-\frac{%
\boldsymbol{L}\otimes \boldsymbol{L}}{L^{2}}\right) +\frac{1}{4\pi L^{2}r}%
\mathrm{e}^{-\frac{\left\vert r\right\vert }{L}}\frac{\boldsymbol{L}\otimes 
\boldsymbol{L}}{L^{2}}
\end{align*}%
and takes the form (\ref{kernel-T2}), where absolute value is omitted.

\subsubsection{Non-locality kernel of model $T_{3}$}

Model $T_{3}$, see the tensor-type stress-gradient Eringen three-dimensional
model (\ref{non-loc-tensor-type}) and Table \ref{Modeli-new}, constituted as
another three-dimensional tensor-type generalization of Eringen model (\ref%
{1D-Eringen}), has a following tensor%
\begin{equation}
\boldsymbol{\hat{\mathcal{K}}}_{3}\left( \boldsymbol{r}\right) =\delta
\left( \boldsymbol{r}\right) \left( \boldsymbol{\hat{I}}-\frac{\boldsymbol{L}%
\otimes \boldsymbol{L}}{L^{2}}\right) +\frac{1}{4L}\delta \left( \boldsymbol{%
r}_{\perp }\right) \mathrm{e}^{-\frac{\left\vert r_{\parallel }\right\vert }{%
L}}\frac{\boldsymbol{L}\otimes \boldsymbol{L}}{L^{2}}+\frac{\delta \left( 
\boldsymbol{r}_{\perp }\right) }{r_{\perp }}\mathrm{e}^{-\frac{\left\vert
r_{\parallel }\right\vert }{L}}\left( \frac{\boldsymbol{L}}{L}\otimes \frac{%
\boldsymbol{r}_{\perp }}{r_{\perp }}\right) ,  \label{kernel-T3}
\end{equation}%
as a non-locality kernel, see also Table \ref{Modeli-new}.

Namely, in the case of model $T_{3}$, the expression (\ref{Ki-bar-tensor})
for local stress tensor and non-locality kernel, both in Fourier domain,
becomes%
\begin{align}
\boldsymbol{\bar{\hat{\mathcal{K}}}}_{3}^{-1}\left( \boldsymbol{k}\right) 
\boldsymbol{\bar{\hat{\sigma}}}\left( \boldsymbol{k}\right) & =\boldsymbol{%
\bar{\hat{\sigma}}}^{local}\left( \boldsymbol{k}\right) ,\quad \text{%
with\quad }\boldsymbol{\bar{\hat{\mathcal{K}}}}_{3}^{-1}\left( \boldsymbol{k}%
\right) =\boldsymbol{\hat{I}}+\left( \boldsymbol{L}\cdot \boldsymbol{k}%
\right) \left( \boldsymbol{L}\otimes \boldsymbol{k}\right) ,\quad \text{i.e.,%
}  \label{K-3-hat-bar-1} \\
\boldsymbol{\bar{\hat{\sigma}}}\left( \boldsymbol{k}\right) & =\boldsymbol{%
\bar{\hat{\mathcal{K}}}}_{3}\left( \boldsymbol{k}\right) \boldsymbol{\bar{%
\hat{\sigma}}}^{local}\left( \boldsymbol{k}\right) \quad \text{with\quad }%
\boldsymbol{\bar{\hat{\mathcal{K}}}}_{3}\left( \boldsymbol{k}\right) =%
\boldsymbol{\hat{I}}-\frac{\left( \boldsymbol{L}\cdot \boldsymbol{k}\right)
\left( \boldsymbol{L}\otimes \boldsymbol{k}\right) }{1+\left( \boldsymbol{L}%
\cdot \boldsymbol{k}\right) ^{2}},  \label{K-3-hat-bar-2}
\end{align}%
since one has $\boldsymbol{\bar{\hat{T}}}_{3}\left( \boldsymbol{k}\right) =-%
\mathcal{F}\left[ \mathbf{\hat{T}}_{3}\left( \nabla \right) \right] \left( 
\boldsymbol{k}\right) =-\mathcal{F}\left[ \left( \boldsymbol{L}\otimes
\nabla \right) \left( \boldsymbol{L}\otimes \nabla \right) \right] \left( 
\boldsymbol{k}\right) =\left( \boldsymbol{k}\cdot \boldsymbol{L}\right)
\left( \boldsymbol{L}\otimes \boldsymbol{k}\right) ,$ where the tensor $%
\boldsymbol{\bar{\hat{\mathcal{K}}}}_{3}$ represents an inverse counterpart
of tensor $\boldsymbol{\bar{\hat{\mathcal{K}}}}_{3}^{-1}$, given by (\ref%
{K-3-hat-bar-1})$_{2}$, such that $\boldsymbol{\bar{\hat{\mathcal{K}}}}%
_{3}^{-1}\boldsymbol{\bar{\hat{\mathcal{K}}}}_{3}=\boldsymbol{\hat{I}}$, and
it is obtained in the form (\ref{K-3-hat-bar-2})$_{2}$ according to
Sherman--Morrison--Woodbury and Sherman--Morrison matrix identity, see \cite%
{SMW} and \cite{SM}. The inverse Fourier transform of tensor $\boldsymbol{%
\bar{\hat{\mathcal{K}}}}_{3}$, see (\ref{K-3-hat-bar-2})$_{2}$, reduces to
the inversion of tensor $\frac{\left( \boldsymbol{L}\cdot \boldsymbol{k}%
\right) \left( \boldsymbol{L}\otimes \boldsymbol{k}\right) }{1+\left( 
\boldsymbol{L}\cdot \boldsymbol{k}\right) ^{2}}$, which can be written in
the matrix form as%
\begin{equation}
\frac{\left( \boldsymbol{L}\cdot \boldsymbol{k}\right) \left( \boldsymbol{L}%
\otimes \boldsymbol{k}\right) }{1+\left( \boldsymbol{L}\cdot \boldsymbol{k}%
\right) ^{2}}=\frac{k_{z}}{\frac{1}{L^{2}}+k_{z}^{2}}\left[ 
\begin{array}{ccc}
0 & 0 & 0 \\ 
0 & 0 & 0 \\ 
k_{x} & k_{y} & k_{z}%
\end{array}%
\right] ,  \label{matrica-T3}
\end{equation}%
using Cartesian coordinate system for tensor representation, since the
orientation of the vector $\boldsymbol{L}$ is chosen to coincide with the $z$%
-axis, i.e., $\boldsymbol{L}=L\boldsymbol{e}_{z},$ while $\boldsymbol{k}%
=\left( k_{x},k_{y},k_{z}\right) $. The Fourier transform inversion of
tensor $\frac{\left( \boldsymbol{L}\cdot \boldsymbol{k}\right) \left( 
\boldsymbol{L}\otimes \boldsymbol{k}\right) }{1+\left( \boldsymbol{L}\cdot 
\boldsymbol{k}\right) ^{2}}$ is achieved by inverting each component of the
matrix (\ref{matrica-T3}), according to the definition of the inverse
Fourier transform (\ref{IFT}), implying%
\begin{align}
\mathcal{F}^{-1}\left[ \frac{k_{x}k_{z}}{\frac{1}{L^{2}}+k_{z}^{2}}\right]
\left( \boldsymbol{r}\right) & =\frac{1}{\left( 2\pi \right) ^{3}}\int_{%
\mathbb{R}
^{3}}\frac{k_{x}k_{z}}{\frac{1}{L^{2}}+k_{z}^{2}}\mathrm{e}^{\mathrm{i}%
\boldsymbol{k}\cdot \boldsymbol{r}}\mathrm{d}_{\boldsymbol{k}}V  \notag \\
& =\frac{1}{\left( 2\pi \right) ^{3}}\int_{-\infty }^{\infty }\int_{-\infty
}^{\infty }\int_{-\infty }^{\infty }\frac{k_{x}k_{z}}{\frac{1}{L^{2}}%
+k_{z}^{2}}\mathrm{e}^{\mathrm{i}\left( xk_{x}+yk_{y}+zk_{z}\right) }\mathrm{%
d}k_{x}\mathrm{d}k_{y}\mathrm{d}k_{z}  \notag \\
& =\frac{1}{\left( 2\pi \right) ^{3}}\int_{-\infty }^{\infty }k_{x}\mathrm{e}%
^{\mathrm{i}xk_{x}}\mathrm{d}k_{x}\int_{-\infty }^{\infty }\mathrm{e}^{%
\mathrm{i}yk_{y}}\mathrm{d}k_{y}\int_{-\infty }^{\infty }\frac{k_{z}}{\frac{1%
}{L^{2}}+k_{z}^{2}}\mathrm{e}^{\mathrm{i}zk_{z}}\mathrm{d}k_{z}  \notag \\
& =\frac{1}{2\pi }\left( -\mathrm{i}\frac{\partial }{\partial x}\left( \frac{%
1}{2\pi }\int_{-\infty }^{\infty }\mathrm{e}^{\mathrm{i}xk_{x}}\mathrm{d}%
k_{x}\right) \right) \left( \frac{1}{2\pi }\int_{-\infty }^{\infty }\mathrm{e%
}^{\mathrm{i}yk_{y}}\mathrm{d}k_{y}\right) \left( 2\mathrm{i}%
\int_{0}^{\infty }\frac{k_{z}\sin \left( zk_{z}\right) }{\frac{1}{L^{2}}%
+k_{z}^{2}}\mathrm{d}k_{z}\right)  \notag \\
& =\frac{1}{\pi }\frac{\mathrm{d}}{\mathrm{d}x}\delta \left( x\right) \delta
\left( y\right) \left( -\frac{\partial }{\partial z}\int_{0}^{\infty }\frac{%
\cos \left( zk_{z}\right) }{\frac{1}{L^{2}}+k_{z}^{2}}\mathrm{d}k_{z}\right)
\label{T3-x-1} \\
& =\frac{1}{\pi }\frac{\mathrm{d}}{\mathrm{d}x}\delta \left( x\right) \delta
\left( y\right) \frac{\pi }{2}\mathrm{e}^{-\frac{\left\vert z\right\vert }{L}%
}  \notag \\
& =\frac{1}{2}\frac{\mathrm{d}}{\mathrm{d}x}\delta \left( x\right) \delta
\left( y\right) \mathrm{e}^{-\frac{\left\vert z\right\vert }{L}},
\label{T3-x-2}
\end{align}%
such that in (\ref{T3-x-1}) the integral (\ref{int-sa-kosinusom}) is used.
Similarly, it is obtained that%
\begin{equation}
\mathcal{F}^{-1}\left[ \frac{k_{y}k_{z}}{\frac{1}{L^{2}}+k_{z}^{2}}\right]
\left( \boldsymbol{r}\right) =\frac{1}{2}\delta \left( x\right) \frac{%
\mathrm{d}}{\mathrm{d}y}\delta \left( y\right) \mathrm{e}^{-\frac{\left\vert
z\right\vert }{L}}.  \label{T3-y-1}
\end{equation}

Inversion of the last non-zero component of the matrix (\ref{matrica-T3})
reads%
\begin{align}
\mathcal{F}^{-1}\left[ \frac{k_{z}^{2}}{\frac{1}{L^{2}}+k_{z}^{2}}\right]
\left( \boldsymbol{r}\right) & =\frac{1}{\left( 2\pi \right) ^{3}}\int_{%
\mathbb{R}
^{3}}\frac{k_{z}^{2}}{\frac{1}{L^{2}}+k_{z}^{2}}\mathrm{e}^{\mathrm{i}%
\boldsymbol{k}\cdot \boldsymbol{r}}\mathrm{d}_{\boldsymbol{k}}V  \notag \\
& =\frac{1}{\left( 2\pi \right) ^{3}}\int_{%
\mathbb{R}
^{3}}\mathrm{e}^{\mathrm{i}\boldsymbol{k}\cdot \boldsymbol{r}}\mathrm{d}_{%
\boldsymbol{k}}V-\frac{1}{L^{2}}\frac{1}{\left( 2\pi \right) ^{3}}\int_{%
\mathbb{R}
^{3}}\frac{1}{\frac{1}{L^{2}}+k_{z}^{2}}\mathrm{e}^{\mathrm{i}\boldsymbol{k}%
\cdot \boldsymbol{r}}\mathrm{d}_{\boldsymbol{k}}V  \notag \\
& =\delta \left( \boldsymbol{r}\right) -\frac{1}{2\pi L^{2}}\left( \frac{1}{%
2\pi }\int_{-\infty }^{\infty }\mathrm{e}^{\mathrm{i}xk_{x}}\mathrm{d}%
k_{x}\right) \left( \frac{1}{2\pi }\int_{-\infty }^{\infty }\mathrm{e}^{%
\mathrm{i}yk_{y}}\mathrm{d}k_{y}\right) \int_{-\infty }^{\infty }\frac{1}{%
\frac{1}{L^{2}}+k_{z}^{2}}\mathrm{e}^{\mathrm{i}zk_{z}}\mathrm{d}k_{z} 
\notag \\
& =\delta \left( \boldsymbol{r}\right) -\frac{1}{2\pi L^{2}}\delta \left(
x\right) \delta \left( y\right) \int_{0}^{\infty }\frac{\cos \left(
zk_{z}\right) }{\frac{1}{L^{2}}+k_{z}^{2}}\mathrm{d}k_{z}  \label{T3-z-1} \\
& =\delta \left( \boldsymbol{r}\right) -\frac{1}{4L}\delta \left( x\right)
\delta \left( y\right) \mathrm{e}^{-\frac{\left\vert z\right\vert }{L}},
\label{T3-z-2}
\end{align}%
such that in (\ref{T3-z-1}) the integral (\ref{int-sa-kosinusom}) is used.

Fourier transform inversion of tensor $\frac{\left( \boldsymbol{L}\cdot 
\boldsymbol{k}\right) \left( \boldsymbol{L}\otimes \boldsymbol{k}\right) }{%
1+\left( \boldsymbol{L}\cdot \boldsymbol{k}\right) ^{2}},$ given by (\ref%
{matrica-T3}), after substitution of previously calculated inversions of its
components, yields%
\begin{align}
\mathcal{F}^{-1}\left[ \frac{\left( \boldsymbol{L}\cdot \boldsymbol{k}%
\right) \left( \boldsymbol{L}\otimes \boldsymbol{k}\right) }{1+\left( 
\boldsymbol{L}\cdot \boldsymbol{k}\right) ^{2}}\right] \left( \boldsymbol{r}%
\right) & =\left[ 
\begin{array}{ccc}
0 & 0 & 0 \\ 
0 & 0 & 0 \\ 
\frac{1}{2}\frac{\mathrm{d}}{\mathrm{d}x}\delta \left( x\right) \delta
\left( y\right) \mathrm{e}^{-\frac{\left\vert z\right\vert }{L}} & \frac{1}{2%
}\delta \left( x\right) \frac{\mathrm{d}}{\mathrm{d}y}\delta \left( y\right) 
\mathrm{e}^{-\frac{\left\vert z\right\vert }{L}} & \delta \left( \boldsymbol{%
r}\right) -\frac{1}{4L}\delta \left( x\right) \delta \left( y\right) \mathrm{%
e}^{-\frac{\left\vert z\right\vert }{L}}%
\end{array}%
\right]  \notag \\
& =\left[ 
\begin{array}{ccc}
0 & 0 & 0 \\ 
0 & 0 & 0 \\ 
0 & 0 & \delta \left( \boldsymbol{r}\right) -\frac{1}{4L}\delta \left(
x\right) \delta \left( y\right) \mathrm{e}^{-\frac{\left\vert z\right\vert }{%
L}}%
\end{array}%
\right] +\frac{1}{2}\left[ 
\begin{array}{ccc}
0 & 0 & 0 \\ 
0 & 0 & 0 \\ 
\frac{\mathrm{d}}{\mathrm{d}x} & \frac{\mathrm{d}}{\mathrm{d}y} & 0%
\end{array}%
\right] \delta \left( x\right) \delta \left( y\right) \mathrm{e}^{-\frac{%
\left\vert z\right\vert }{L}}  \notag \\
& =\delta \left( \boldsymbol{r}\right) \frac{\boldsymbol{L}\otimes 
\boldsymbol{L}}{L^{2}}-\frac{1}{4L}\delta \left( \boldsymbol{r}_{\perp
}\right) \mathrm{e}^{-\frac{\left\vert r_{\parallel }\right\vert }{L}}\frac{%
\boldsymbol{L}\otimes \boldsymbol{L}}{L^{2}}+\frac{1}{2}\frac{\boldsymbol{L}%
}{L}\otimes \left( \nabla _{\perp }\delta \left( \boldsymbol{r}_{\perp
}\right) \right) \mathrm{e}^{-\frac{\left\vert r_{\parallel }\right\vert }{L}%
}  \label{matrica-3} \\
& =\delta \left( \boldsymbol{r}\right) \frac{\boldsymbol{L}\otimes 
\boldsymbol{L}}{L^{2}}-\frac{1}{4L}\delta \left( \boldsymbol{r}_{\perp
}\right) \mathrm{e}^{-\frac{\left\vert r_{\parallel }\right\vert }{L}}\frac{%
\boldsymbol{L}\otimes \boldsymbol{L}}{L^{2}}-\frac{\delta \left( \boldsymbol{%
r}_{\perp }\right) }{r_{\perp }}\mathrm{e}^{-\frac{\left\vert r_{\parallel
}\right\vert }{L}}\left( \frac{\boldsymbol{L}}{L}\otimes \frac{\boldsymbol{r}%
_{\perp }}{r_{\perp }}\right) ,  \label{matrica-3-1}
\end{align}%
where $\boldsymbol{r}_{\parallel }=r_{\parallel }\boldsymbol{e}_{z}=z%
\boldsymbol{e}_{z}$, $\boldsymbol{r}_{\perp }=x\boldsymbol{e}_{x}+y%
\boldsymbol{e}_{y},$ $r_{\perp }=\left\vert \boldsymbol{r}_{\perp
}\right\vert =\sqrt{x^{2}+y^{2}},$ $\delta \left( \boldsymbol{r}_{\perp
}\right) =\delta \left( x\right) \delta \left( y\right) ,$ and $\nabla
_{\perp }=\boldsymbol{e}_{x}\partial _{x}+\boldsymbol{e}_{y}\partial _{y}$,
since $\nabla _{\perp }\delta \left( \boldsymbol{r}_{\perp }\right) $ in the
last term of (\ref{matrica-3}) is calculated as%
\begin{align}
\nabla _{\perp }\delta \left( \boldsymbol{r}_{\perp }\right) & =\nabla
_{\perp }\frac{\delta \left( r_{\perp }\right) }{\pi r_{\perp }}  \notag \\
& =\frac{1}{\pi r_{\perp }}\left( \nabla _{\perp }\delta \left( r_{\perp
}\right) \right) +\frac{1}{\pi }\delta \left( r_{\perp }\right) \left(
\nabla _{\perp }\frac{1}{r_{\perp }}\right)  \notag \\
& =\frac{1}{\pi r_{\perp }}\frac{\mathrm{d}\delta \left( r_{\perp }\right) }{%
\mathrm{d}r_{\perp }}\frac{\boldsymbol{r}_{\perp }}{r_{\perp }}-\frac{1}{\pi 
}\delta \left( r_{\perp }\right) \frac{1}{r_{\perp }^{2}}\frac{\boldsymbol{r}%
_{\perp }}{r_{\perp }}  \notag \\
& =-\frac{1}{\pi r_{\perp }}\frac{1}{r_{\perp }}\delta \left( r_{\perp
}\right) \frac{\boldsymbol{r}_{\perp }}{r_{\perp }}-\frac{1}{\pi }\delta
\left( r_{\perp }\right) \frac{1}{r_{\perp }^{2}}\frac{\boldsymbol{r}_{\perp
}}{r_{\perp }}  \notag \\
& =-2\frac{1}{r_{\perp }}\frac{\delta \left( r_{\perp }\right) }{\pi
r_{\perp }}\frac{\boldsymbol{r}_{\perp }}{r_{\perp }}  \notag \\
& =-2\frac{\delta \left( \boldsymbol{r}_{\perp }\right) }{r_{\perp }}\frac{%
\boldsymbol{r}_{\perp }}{r_{\perp }},  \label{nabla-norm-na-dirak}
\end{align}%
where Dirac delta properties $\delta \left( \boldsymbol{r}_{\perp }\right) =%
\frac{\delta \left( r_{\perp }\right) }{\pi r_{\perp }}$ and $r_{\perp }%
\frac{\mathrm{d}\delta \left( r_{\perp }\right) }{\mathrm{d}r_{\perp }}%
=-\delta \left( r_{\perp }\right) $, as well as the gradient of a
spherically symmetric function $\nabla f\left( r\right) =\frac{\mathrm{d}}{%
\mathrm{d}r}f\left( r\right) \frac{\boldsymbol{r}}{r}$ are used.

Therefore, Fourier transform inversion of tensor $\boldsymbol{\bar{\hat{%
\mathcal{K}}}}_{3}$ becomes%
\begin{align*}
\boldsymbol{\hat{\mathcal{K}}}_{3}\left( \boldsymbol{r}\right) & =\mathcal{F}%
^{-1}\left[ \boldsymbol{\bar{\hat{\mathcal{K}}}}_{3}\left( \boldsymbol{k}%
\right) \right] \left( \boldsymbol{r}\right) =\mathcal{\delta }\left( 
\boldsymbol{r}\right) \boldsymbol{\hat{I}}-\mathcal{F}^{-1}\left[ \frac{%
\left( \boldsymbol{L}\cdot \boldsymbol{k}\right) \left( \boldsymbol{L}%
\otimes \boldsymbol{k}\right) }{1+\left( \boldsymbol{L}\cdot \boldsymbol{k}%
\right) ^{2}}\right] \left( \boldsymbol{r}\right) \\
& =\delta \left( \boldsymbol{r}\right) \left( \boldsymbol{\hat{I}}-\frac{%
\boldsymbol{L}\otimes \boldsymbol{L}}{L^{2}}\right) +\frac{1}{4L}\delta
\left( \boldsymbol{r}_{\perp }\right) \mathrm{e}^{-\frac{\left\vert
r_{\parallel }\right\vert }{L}}\frac{\boldsymbol{L}\otimes \boldsymbol{L}}{%
L^{2}}+\frac{\delta \left( \boldsymbol{r}_{\perp }\right) }{r_{\perp }}%
\mathrm{e}^{-\frac{\left\vert r_{\parallel }\right\vert }{L}}\left( \frac{%
\boldsymbol{L}}{L}\otimes \frac{\boldsymbol{r}_{\perp }}{r_{\perp }}\right) ,
\end{align*}%
where $\boldsymbol{\bar{\hat{\mathcal{K}}}}_{3}$ is given by (\ref%
{K-3-hat-bar-2})$_{2},$ with $\mathcal{F}^{-1}\left[ \frac{\left( 
\boldsymbol{L}\cdot \boldsymbol{k}\right) \left( \boldsymbol{L}\otimes 
\boldsymbol{k}\right) }{1+\left( \boldsymbol{L}\cdot \boldsymbol{k}\right)
^{2}}\right] \left( \boldsymbol{r}\right) $ calculated as in (\ref%
{matrica-3-1}), and taking the form (\ref{kernel-T3}).

\subsubsection{Non-locality kernel of model $T_{4}$}

Model $T_{4}$, see the tensor-type stress-gradient Eringen three-dimensional
model (\ref{non-loc-tensor-type}) and Table \ref{Modeli-new}, is the
transpose counterpart of model $T_{3}$, and represents another
three-dimensional tensor-type generalization of Eringen model (\ref%
{1D-Eringen}). Its non-locality kernel is the tensor 
\begin{equation}
\boldsymbol{\hat{\mathcal{K}}}_{4}\left( \boldsymbol{r}\right) =\delta
\left( \boldsymbol{r}\right) \left( \boldsymbol{\hat{I}}-\frac{\boldsymbol{L}%
\otimes \boldsymbol{L}}{L^{2}}\right) +\frac{1}{4L}\delta \left( \boldsymbol{%
r}_{\perp }\right) \mathrm{e}^{-\frac{\left\vert r_{\parallel }\right\vert }{%
L}}\frac{\boldsymbol{L}\otimes \boldsymbol{L}}{L^{2}}+\frac{\delta \left( 
\boldsymbol{r}_{\perp }\right) }{r_{\perp }}\mathrm{e}^{-\frac{\left\vert
r_{\parallel }\right\vert }{L}}\left( \frac{\boldsymbol{r}_{\perp }}{%
r_{\perp }}\otimes \frac{\boldsymbol{L}}{L}\right) ,  \label{kernel-T4}
\end{equation}%
which is the transpose counterpart of the kernel $\boldsymbol{\hat{\mathcal{K%
}}}_{3}$, given by (\ref{kernel-T3}), see also Table \ref{Modeli-new}.

Namely, in the case of model $T_{4}$, the expression (\ref{Ki-bar-tensor})
for local stress tensor and non-locality kernel, both in Fourier domain,
becomes%
\begin{align}
\boldsymbol{\bar{\hat{\mathcal{K}}}}_{4}^{-1}\left( \boldsymbol{k}\right) 
\boldsymbol{\bar{\hat{\sigma}}}\left( \boldsymbol{k}\right) & =\boldsymbol{%
\bar{\hat{\sigma}}}^{local}\left( \boldsymbol{k}\right) ,\quad \text{%
with\quad }\boldsymbol{\bar{\hat{\mathcal{K}}}}_{4}^{-1}\left( \boldsymbol{k}%
\right) =\boldsymbol{\hat{I}}+\left( \boldsymbol{L}\cdot \boldsymbol{k}%
\right) \left( \boldsymbol{k}\otimes \boldsymbol{L}\right) ,\quad \text{i.e.,%
}  \label{K-4-hat-bar-1} \\
\boldsymbol{\bar{\hat{\sigma}}}\left( \boldsymbol{k}\right) & =\boldsymbol{%
\bar{\hat{\mathcal{K}}}}_{4}\left( \boldsymbol{k}\right) \boldsymbol{\bar{%
\hat{\sigma}}}^{local}\left( \boldsymbol{k}\right) \quad \text{with\quad }%
\boldsymbol{\bar{\hat{\mathcal{K}}}}_{4}\left( \boldsymbol{k}\right) =%
\boldsymbol{\hat{I}}-\frac{\left( \boldsymbol{L}\cdot \boldsymbol{k}\right)
\left( \boldsymbol{k}\otimes \boldsymbol{L}\right) }{1+\left( \boldsymbol{L}%
\cdot \boldsymbol{k}\right) ^{2}},  \label{K-4-hat-bar-2}
\end{align}%
since one has $\boldsymbol{\bar{\hat{T}}}_{4}\left( \boldsymbol{k}\right) =-%
\mathcal{F}\left[ \mathbf{\hat{T}}_{4}\left( \nabla \right) \right] \left( 
\boldsymbol{k}\right) =-\mathcal{F}\left[ \left( \nabla \otimes \boldsymbol{L%
}\right) \left( \nabla \otimes \boldsymbol{L}\right) \right] \left( 
\boldsymbol{k}\right) =\left( \boldsymbol{k}\cdot \boldsymbol{L}\right)
\left( \boldsymbol{k}\otimes \boldsymbol{L}\right) ,$ where the tensor $%
\boldsymbol{\bar{\hat{\mathcal{K}}}}_{4}$ represents an inverse counterpart
of tensor $\boldsymbol{\bar{\hat{\mathcal{K}}}}_{4}^{-1}$, given by (\ref%
{K-4-hat-bar-1})$_{2}$, such that $\boldsymbol{\bar{\hat{\mathcal{K}}}}%
_{4}^{-1}\boldsymbol{\bar{\hat{\mathcal{K}}}}_{4}=\boldsymbol{\hat{I}}$, and
it is obtained in the form (\ref{K-4-hat-bar-2})$_{2}$ according to
Sherman--Morrison--Woodbury and Sherman--Morrison matrix identity, see \cite%
{SMW} and \cite{SM}. The inverse Fourier transform of tensor $\boldsymbol{%
\bar{\hat{\mathcal{K}}}}_{4}$, see (\ref{K-4-hat-bar-2})$_{2}$, reduces to
the inversion of tensor $\frac{\left( \boldsymbol{L}\cdot \boldsymbol{k}%
\right) \left( \boldsymbol{k}\otimes \boldsymbol{L}\right) }{1+\left( 
\boldsymbol{L}\cdot \boldsymbol{k}\right) ^{2}}$, which can be written in
the matrix form as%
\begin{equation}
\frac{\left( \boldsymbol{L}\cdot \boldsymbol{k}\right) \left( \boldsymbol{k}%
\otimes \boldsymbol{L}\right) }{1+\left( \boldsymbol{L}\cdot \boldsymbol{k}%
\right) ^{2}}=\frac{k_{z}}{\frac{1}{L^{2}}+k_{z}^{2}}\left[ 
\begin{array}{ccc}
0 & 0 & k_{x} \\ 
0 & 0 & k_{y} \\ 
0 & 0 & k_{z}%
\end{array}%
\right] ,  \label{matrica-T4}
\end{equation}%
using Cartesian coordinate system for tensor representation, since the
orientation of the vector $\boldsymbol{L}$ is chosen to coincide with the $z$%
-axis, i.e., $\boldsymbol{L}=L\boldsymbol{e}_{z},$ while $\boldsymbol{k}%
=\left( k_{x},k_{y},k_{z}\right) $. The Fourier transform inversion of
tensor $\frac{\left( \boldsymbol{L}\cdot \boldsymbol{k}\right) \left( 
\boldsymbol{k}\otimes \boldsymbol{L}\right) }{1+\left( \boldsymbol{L}\cdot 
\boldsymbol{k}\right) ^{2}}$ is achieved by inverting each component of the
matrix (\ref{matrica-T4}), according to the definition of the inverse
Fourier transform (\ref{IFT}), calculated as (\ref{T3-x-2}), (\ref{T3-y-1}),
and (\ref{T3-z-2}), so that the Fourier transform inversion of tensor $\frac{%
\left( \boldsymbol{L}\cdot \boldsymbol{k}\right) \left( \boldsymbol{k}%
\otimes \boldsymbol{L}\right) }{1+\left( \boldsymbol{L}\cdot \boldsymbol{k}%
\right) ^{2}},$ given by (\ref{matrica-T4}), after substitution of
previously calculated inversions of its components, yields%
\begin{align}
\mathcal{F}^{-1}\left[ \frac{\left( \boldsymbol{L}\cdot \boldsymbol{k}%
\right) \left( \boldsymbol{k}\otimes \boldsymbol{L}\right) }{1+\left( 
\boldsymbol{L}\cdot \boldsymbol{k}\right) ^{2}}\right] \left( \boldsymbol{r}%
\right) & =\left[ 
\begin{array}{ccc}
0 & 0 & \frac{1}{2}\frac{\mathrm{d}}{\mathrm{d}x}\delta \left( x\right)
\delta \left( y\right) \mathrm{e}^{-\frac{\left\vert z\right\vert }{L}} \\ 
0 & 0 & \frac{1}{2}\delta \left( x\right) \frac{\mathrm{d}}{\mathrm{d}y}%
\delta \left( y\right) \mathrm{e}^{-\frac{\left\vert z\right\vert }{L}} \\ 
0 & 0 & \delta \left( \boldsymbol{r}\right) -\frac{1}{4L}\delta \left(
x\right) \delta \left( y\right) \mathrm{e}^{-\frac{\left\vert z\right\vert }{%
L}}%
\end{array}%
\right]  \notag \\
& =\left[ 
\begin{array}{ccc}
0 & 0 & 0 \\ 
0 & 0 & 0 \\ 
0 & 0 & \delta \left( \boldsymbol{r}\right) -\frac{1}{4L}\delta \left(
x\right) \delta \left( y\right) \mathrm{e}^{-\frac{\left\vert z\right\vert }{%
L}}%
\end{array}%
\right] +\frac{1}{2}\left[ 
\begin{array}{ccc}
0 & 0 & \frac{\mathrm{d}}{\mathrm{d}x} \\ 
0 & 0 & \frac{\mathrm{d}}{\mathrm{d}y} \\ 
0 & 0 & 0%
\end{array}%
\right] \delta \left( x\right) \delta \left( y\right) \mathrm{e}^{-\frac{%
\left\vert z\right\vert }{L}}  \notag \\
& =\delta \left( \boldsymbol{r}\right) \frac{\boldsymbol{L}\otimes 
\boldsymbol{L}}{L^{2}}-\frac{1}{4L}\delta \left( \boldsymbol{r}_{\perp
}\right) \mathrm{e}^{-\frac{\left\vert r_{\parallel }\right\vert }{L}}\frac{%
\boldsymbol{L}\otimes \boldsymbol{L}}{L^{2}}+\frac{1}{2}\left( \left( \nabla
_{\perp }\delta \left( \boldsymbol{r}_{\perp }\right) \right) \otimes \frac{%
\boldsymbol{L}}{L}\right) \mathrm{e}^{-\frac{\left\vert r_{\parallel
}\right\vert }{L}}  \label{matrica-4} \\
& =\delta \left( \boldsymbol{r}\right) \frac{\boldsymbol{L}\otimes 
\boldsymbol{L}}{L^{2}}-\frac{1}{4L}\delta \left( \boldsymbol{r}_{\perp
}\right) \mathrm{e}^{-\frac{\left\vert r_{\parallel }\right\vert }{L}}\frac{%
\boldsymbol{L}\otimes \boldsymbol{L}}{L^{2}}-\frac{\delta \left( \boldsymbol{%
r}_{\perp }\right) }{r_{\perp }}\mathrm{e}^{-\frac{\left\vert r_{\parallel
}\right\vert }{L}}\left( \frac{\boldsymbol{r}_{\perp }}{r_{\perp }}\otimes 
\frac{\boldsymbol{L}}{L}\right) ,  \label{matrica-4-1}
\end{align}%
where $\boldsymbol{r}_{\parallel }=r_{\parallel }\boldsymbol{e}_{z}=z%
\boldsymbol{e}_{z}$, $\boldsymbol{r}_{\perp }=x\boldsymbol{e}_{x}+y%
\boldsymbol{e}_{y},$ $r_{\perp }=\left\vert \boldsymbol{r}_{\perp
}\right\vert =\sqrt{x^{2}+y^{2}},$ $\delta \left( \boldsymbol{r}_{\perp
}\right) =\delta \left( x\right) \delta \left( y\right) ,$ and $\nabla
_{\perp }=\boldsymbol{e}_{x}\partial _{x}+\boldsymbol{e}_{y}\partial _{y}$,
since $\nabla _{\perp }\delta \left( \boldsymbol{r}_{\perp }\right) $ in the
last term of (\ref{matrica-4}) is calculated as (\ref{nabla-norm-na-dirak}).

Therefore, Fourier transform inversion of tensor $\boldsymbol{\bar{\hat{%
\mathcal{K}}}}_{4}$ becomes%
\begin{align*}
\boldsymbol{\hat{\mathcal{K}}}_{4}\left( \boldsymbol{r}\right) & =\mathcal{F}%
^{-1}\left[ \boldsymbol{\bar{\hat{\mathcal{K}}}}_{4}\left( \boldsymbol{k}%
\right) \right] \left( \boldsymbol{r}\right) =\mathcal{\delta }\left( 
\boldsymbol{r}\right) \boldsymbol{\hat{I}}-\mathcal{F}^{-1}\left[ \frac{%
\left( \boldsymbol{L}\cdot \boldsymbol{k}\right) \left( \boldsymbol{k}%
\otimes \boldsymbol{L}\right) }{1+\left( \boldsymbol{L}\cdot \boldsymbol{k}%
\right) ^{2}}\right] \left( \boldsymbol{r}\right) \\
& =\delta \left( \boldsymbol{r}\right) \left( \boldsymbol{\hat{I}}-\frac{%
\boldsymbol{L}\otimes \boldsymbol{L}}{L^{2}}\right) +\frac{1}{4L}\delta
\left( \boldsymbol{r}_{\perp }\right) \mathrm{e}^{-\frac{\left\vert
r_{\parallel }\right\vert }{L}}\frac{\boldsymbol{L}\otimes \boldsymbol{L}}{%
L^{2}}+\frac{\delta \left( \boldsymbol{r}_{\perp }\right) }{r_{\perp }}%
\mathrm{e}^{-\frac{\left\vert r_{\parallel }\right\vert }{L}}\left( \frac{%
\boldsymbol{r}_{\perp }}{r_{\perp }}\otimes \frac{\boldsymbol{L}}{L}\right) ,
\end{align*}%
where $\boldsymbol{\bar{\hat{\mathcal{K}}}}_{4}$ is given by (\ref%
{K-4-hat-bar-2})$_{2},$ with $\mathcal{F}^{-1}\left[ \frac{\left( 
\boldsymbol{L}\cdot \boldsymbol{k}\right) \left( \boldsymbol{k}\otimes 
\boldsymbol{L}\right) }{1+\left( \boldsymbol{L}\cdot \boldsymbol{k}\right)
^{2}}\right] \left( \boldsymbol{r}\right) $ calculated as in (\ref%
{matrica-4-1}), and taking the form (\ref{kernel-T4}).

\subsubsection{Non-locality kernel of model $T_{5}$}

Model $T_{5}$, see the tensor-type stress-gradient Eringen three-dimensional
model (\ref{non-loc-tensor-type}) and Table \ref{Modeli-new}, constituted as
the last three-dimensional tensor-type generalization of Eringen model (\ref%
{1D-Eringen}), has a following tensor%
\begin{align}
\boldsymbol{\hat{\mathcal{K}}}_{5}\left( \boldsymbol{r}\right) & =\delta
\left( \boldsymbol{r}\right) \left( \boldsymbol{\hat{I}}-\left( \frac{%
\boldsymbol{L}}{L}\times \frac{\boldsymbol{r}_{\perp }}{r_{\perp }}\right)
\otimes \left( \frac{\boldsymbol{L}}{L}\times \frac{\boldsymbol{r}_{\perp }}{%
r_{\perp }}\right) \right) -\frac{1}{2\pi L}\delta \left( r_{\parallel
}\right) \frac{1}{r_{\perp }}K_{1}\left( \frac{r_{\perp }}{L}\right) \frac{%
\boldsymbol{r}_{\perp }\otimes \boldsymbol{r}_{\perp }}{r_{\perp }^{2}} 
\notag \\
& \quad +\frac{1}{2\pi L}\delta \left( r_{\parallel }\right) \left( \frac{1}{%
L}K_{0}\left( \frac{r_{\perp }}{L}\right) +\frac{1}{r_{\perp }}K_{1}\left( 
\frac{r_{\perp }}{L}\right) \right) \left( \left( \frac{\boldsymbol{L}}{L}%
\times \frac{\boldsymbol{r}_{\perp }}{r_{\perp }}\right) \otimes \left( 
\frac{\boldsymbol{L}}{L}\times \frac{\boldsymbol{r}_{\perp }}{r_{\perp }}%
\right) \right)  \label{kernel-T5}
\end{align}%
as a non-locality kernel, see also Table \ref{Modeli-new}.

Namely, in the case of model $T_{5}$, the expression (\ref{Ki-bar-tensor})
for local stress tensor and non-locality kernel, both in Fourier domain,
becomes%
\begin{align}
\boldsymbol{\bar{\hat{\mathcal{K}}}}_{5}^{-1}\left( \boldsymbol{k}\right) 
\boldsymbol{\bar{\hat{\sigma}}}\left( \boldsymbol{k}\right) & =\boldsymbol{%
\bar{\hat{\sigma}}}^{local}\left( \boldsymbol{k}\right) ,\quad \text{%
with\quad }\boldsymbol{\bar{\hat{\mathcal{K}}}}_{5}^{-1}\left( \boldsymbol{k}%
\right) =\boldsymbol{\hat{I}}+\left( \boldsymbol{L}\times \boldsymbol{k}%
\right) \otimes \left( \boldsymbol{L}\times \boldsymbol{k}\right) ,\quad 
\text{i.e.,}  \label{K-5-hat-bar-1} \\
\boldsymbol{\bar{\hat{\sigma}}}\left( \boldsymbol{k}\right) & =\boldsymbol{%
\bar{\hat{\mathcal{K}}}}_{5}\left( \boldsymbol{k}\right) \boldsymbol{\bar{%
\hat{\sigma}}}^{local}\left( \boldsymbol{k}\right) \quad \text{with\quad }%
\boldsymbol{\bar{\hat{\mathcal{K}}}}_{5}\left( \boldsymbol{k}\right) =%
\boldsymbol{\hat{I}}-\frac{\left( \boldsymbol{L}\times \boldsymbol{k}\right)
\otimes \left( \boldsymbol{L}\times \boldsymbol{k}\right) }{%
1+L^{2}k^{2}-\left( \boldsymbol{L}\cdot \boldsymbol{r}\right) ^{2}},
\label{K-5-hat-bar-2}
\end{align}%
since one has $\boldsymbol{\bar{\hat{T}}}_{5}\left( \boldsymbol{k}\right) =-%
\mathcal{F}\left[ \mathbf{\hat{T}}_{5}\left( \nabla \right) \right] \left( 
\boldsymbol{k}\right) =-\mathcal{F}\left[ \left( \boldsymbol{L}\times \nabla
\right) \otimes \left( \boldsymbol{L}\times \nabla \right) \right] \left( 
\boldsymbol{k}\right) =\left( \boldsymbol{L}\times \boldsymbol{k}\right)
\otimes \left( \boldsymbol{L}\times \boldsymbol{k}\right) ,$ where the
tensor $\boldsymbol{\bar{\hat{\mathcal{K}}}}_{5}$ represents an inverse
counterpart of tensor $\boldsymbol{\bar{\hat{\mathcal{K}}}}_{5}^{-1}$, given
by (\ref{K-5-hat-bar-1})$_{2}$, such that $\boldsymbol{\bar{\hat{\mathcal{K}}%
}}_{5}^{-1}\boldsymbol{\bar{\hat{\mathcal{K}}}}_{5}=\boldsymbol{\hat{I}}$,
and it is obtained in the form (\ref{K-5-hat-bar-2})$_{2}$ according to
Sherman--Morrison--Woodbury matrix identity and the Sherman--Morrison
formula, see \cite{SMW} and \cite{SM}. The inverse Fourier transform of
tensor $\boldsymbol{\bar{\hat{\mathcal{K}}}}_{5}$, see (\ref{K-5-hat-bar-2})$%
_{2}$, reduces to the inversion of tensor $\frac{\left( \boldsymbol{L}\times 
\boldsymbol{k}\right) \otimes \left( \boldsymbol{L}\times \boldsymbol{k}%
\right) }{1+L^{2}k^{2}-\left( \boldsymbol{L}\cdot \boldsymbol{r}\right) ^{2}}
$, which can be written in the matrix form as%
\begin{equation}
\frac{\left( \boldsymbol{L}\times \boldsymbol{k}\right) \otimes \left( 
\boldsymbol{L}\times \boldsymbol{k}\right) }{1+L^{2}k^{2}-\left( \boldsymbol{%
L\cdot r}\right) ^{2}}=\frac{1}{\frac{1}{L^{2}}+k_{x}^{2}+k_{y}^{2}}\left[ 
\begin{array}{ccc}
k_{y}^{2} & -k_{x}k_{y} & 0 \\ 
-k_{x}k_{y} & k_{x}^{2} & 0 \\ 
0 & 0 & 0%
\end{array}%
\right] ,  \label{matrica-T5}
\end{equation}%
using Cartesian coordinate system for tensor representation and expressing
tensor components in Cartesian coordinates, whereby for the sake of a
simpler calculation, it can be chosen that the direction of vector $%
\boldsymbol{L}$ considers with $z$-axis, i.e., $\boldsymbol{L}=L\boldsymbol{e%
}_{z},$ while the radius vector $\boldsymbol{r}$ is taken to lie in the $xz$%
-plane, i.e., $\boldsymbol{r}=x\boldsymbol{e}_{x}+z\boldsymbol{e}_{z},$ so
that its scalar product with vector $\boldsymbol{k}=k_{x}\boldsymbol{e}%
_{x}+k_{y}\boldsymbol{e}_{y}+k_{z}\boldsymbol{e}_{z}$ is $\boldsymbol{k}%
\cdot \boldsymbol{r=}xk_{x}+zk_{z}$. The Fourier transform inversion of
tensor $\frac{\left( \boldsymbol{L}\times \boldsymbol{k}\right) \otimes
\left( \boldsymbol{L}\times \boldsymbol{k}\right) }{1+L^{2}k^{2}-\left( 
\boldsymbol{L}\cdot \boldsymbol{r}\right) ^{2}}$ is achieved by inverting
each component of the matrix (\ref{matrica-T5}), according to the definition
of the inverse Fourier transform (\ref{IFT}), implying%
\begin{align}
\mathcal{F}^{-1}\left[ \frac{k_{y}^{2}}{\frac{1}{L^{2}}+k_{x}^{2}+k_{y}^{2}}%
\right] \left( \boldsymbol{r}\right) & =\frac{1}{\left( 2\pi \right) ^{3}}%
\int_{%
\mathbb{R}
^{3}}\frac{k_{y}^{2}}{\frac{1}{L^{2}}+k_{x}^{2}+k_{y}^{2}}\mathrm{e}^{%
\mathrm{i}\boldsymbol{k}\cdot \boldsymbol{r}}\mathrm{d}_{\boldsymbol{k}}V 
\notag \\
& =\frac{1}{\left( 2\pi \right) ^{3}}\int_{-\infty }^{\infty }\int_{-\infty
}^{\infty }\int_{-\infty }^{\infty }\frac{k_{y}^{2}}{\frac{1}{L^{2}}%
+k_{x}^{2}+k_{y}^{2}}\mathrm{e}^{\mathrm{i}\left( xk_{x}+zk_{z}\right) }%
\mathrm{d}k_{x}\mathrm{d}k_{y}\mathrm{d}k_{z}  \notag \\
& =\frac{1}{4\pi ^{2}}\int_{-\infty }^{\infty }\int_{-\infty }^{\infty }%
\frac{k_{y}^{2}}{\frac{1}{L^{2}}+k_{x}^{2}+k_{y}^{2}}\mathrm{e}^{\mathrm{i}%
xk_{x}}\left( \frac{1}{2\pi }\int_{-\infty }^{\infty }\mathrm{e}^{\mathrm{i}%
zk_{z}}\mathrm{d}k_{z}\right) \mathrm{d}k_{x}\mathrm{d}k_{y}  \notag \\
& =\frac{1}{2\pi }\delta \left( z\right) \int_{-\infty }^{\infty
}k_{y}^{2}\left( \frac{1}{2\pi }\int_{-\infty }^{\infty }\frac{\mathrm{e}^{%
\mathrm{i}xk_{x}}}{\frac{1}{L^{2}}+k_{x}^{2}+k_{y}^{2}}\mathrm{d}%
k_{x}\right) \mathrm{d}k_{y}  \label{T5-y-1} \\
& =\frac{1}{2\pi }\delta \left( z\right) \int_{-\infty }^{\infty }\frac{%
k_{y}^{2}}{2\sqrt{\frac{1}{L^{2}}+k_{y}^{2}}}\mathrm{e}^{\mathrm{-}%
\left\vert x\right\vert \sqrt{\frac{1}{L^{2}}+k_{y}^{2}}}\mathrm{d}k_{y} 
\notag \\
& =\frac{L}{2\pi }\delta \left( z\right) \int_{0}^{\infty }\frac{k_{y}^{2}}{%
\sqrt{1+L^{2}k_{y}^{2}}}\mathrm{e}^{\mathrm{-}\frac{\left\vert x\right\vert 
}{L}\sqrt{1+L^{2}k_{y}^{2}}}\mathrm{d}k_{y}  \label{T5-y-4} \\
& =\frac{1}{2\pi L^{2}}\delta \left( z\right) \int_{1}^{\infty }\sqrt{u^{2}-1%
}\,\mathrm{e}^{\mathrm{-}\frac{\left\vert x\right\vert }{L}u}\,\mathrm{d}u
\label{T5-y-2} \\
& =\frac{1}{2\pi L}\delta \left( z\right) \frac{1}{\left\vert x\right\vert }%
K_{1}\left( \frac{\left\vert x\right\vert }{L}\right) ,  \label{T5-y-3}
\end{align}%
such that in (\ref{T5-y-1}) the integral (\ref{IFT-sa-lam}) is used, while
in (\ref{T5-y-2}) the substitution $u=\sqrt{1+L^{2}k_{y}^{2}}$ is introduced
in order to obtain the integral representation of Macdonald function (\ref%
{T5-y-2}), i.e., modified Bessel function of the second kind, according to (%
\ref{Macdonald}).

For the inversion of another non-zero main-diagonal term of matrix (\ref%
{matrica-T5}) one has%
\begin{align}
\mathcal{F}^{-1}& \left[ \frac{k_{x}^{2}}{\frac{1}{L^{2}}+k_{x}^{2}+k_{y}^{2}%
}\right] \left( \boldsymbol{r}\right) =\frac{1}{\left( 2\pi \right) ^{3}}%
\int_{%
\mathbb{R}
^{3}}\frac{k_{x}^{2}}{\frac{1}{L^{2}}+k_{x}^{2}+k_{y}^{2}}\mathrm{e}^{%
\mathrm{i}\boldsymbol{k}\cdot \boldsymbol{r}}\mathrm{d}_{\boldsymbol{k}}V 
\notag \\
& =\frac{1}{\left( 2\pi \right) ^{3}}\int_{%
\mathbb{R}
^{3}}\mathrm{e}^{\mathrm{i}\boldsymbol{k}\cdot \boldsymbol{r}}\mathrm{d}_{%
\boldsymbol{k}}V-\frac{1}{\left( 2\pi \right) ^{3}}\int_{%
\mathbb{R}
^{3}}\frac{k_{y}^{2}}{\frac{1}{L^{2}}+k_{x}^{2}+k_{y}^{2}}\mathrm{e}^{%
\mathrm{i}\boldsymbol{k}\cdot \boldsymbol{r}}\mathrm{d}_{\boldsymbol{k}}V-%
\frac{1}{\left( 2\pi \right) ^{3}L^{2}}\int_{%
\mathbb{R}
^{3}}\frac{1}{\frac{1}{L^{2}}+k_{x}^{2}+k_{y}^{2}}\mathrm{e}^{\mathrm{i}%
\boldsymbol{k}\cdot \boldsymbol{r}}\mathrm{d}_{\boldsymbol{k}}V  \notag \\
& =\delta \left( \boldsymbol{r}\right) -\mathcal{F}^{-1}\left[ \frac{%
k_{y}^{2}}{\frac{1}{L^{2}}+k_{x}^{2}+k_{y}^{2}}\right] \left( \boldsymbol{r}%
\right) -\frac{1}{L^{2}}\mathcal{F}^{-1}\left[ \frac{1}{\frac{1}{L^{2}}%
+k_{x}^{2}+k_{y}^{2}}\right] \left( \boldsymbol{r}\right) ,  \label{T5-x-1}
\end{align}%
which, with the solution (\ref{T5-y-3}), amounts to inversion%
\begin{align}
\mathcal{F}^{-1}\left[ \frac{1}{\frac{1}{L^{2}}+k_{x}^{2}+k_{y}^{2}}\right]
\left( \boldsymbol{r}\right) & =\frac{1}{\left( 2\pi \right) ^{3}}\int_{%
\mathbb{R}
^{3}}\frac{1}{\frac{1}{L^{2}}+k_{x}^{2}+k_{y}^{2}}\mathrm{e}^{\mathrm{i}%
\boldsymbol{k}\cdot \boldsymbol{r}}\mathrm{d}_{\boldsymbol{k}}V  \notag \\
& =\frac{1}{8\pi ^{3}}\int_{-\infty }^{\infty }\int_{-\infty }^{\infty
}\int_{-\infty }^{\infty }\frac{1}{\frac{1}{L^{2}}+k_{x}^{2}+k_{y}^{2}}%
\mathrm{e}^{\mathrm{i}\left( xk_{x}+zk_{z}\right) }\mathrm{d}k_{x}\mathrm{d}%
k_{y}\mathrm{d}k_{z}  \notag \\
& =\frac{1}{4\pi ^{2}}\int_{-\infty }^{\infty }\int_{-\infty }^{\infty }%
\frac{1}{\frac{1}{L^{2}}+k_{x}^{2}+k_{y}^{2}}\mathrm{e}^{\mathrm{i}%
xk_{x}}\left( \frac{1}{2\pi }\int_{-\infty }^{\infty }\mathrm{e}^{\mathrm{i}%
zk_{z}}\mathrm{d}k_{z}\right) \mathrm{d}k_{x}\mathrm{d}k_{y}  \notag \\
& =\frac{1}{2\pi }\delta \left( z\right) \int_{-\infty }^{\infty }\left( 
\frac{1}{2\pi }\int_{-\infty }^{\infty }\frac{\mathrm{e}^{\mathrm{i}xk_{x}}}{%
\frac{1}{L^{2}}+k_{x}^{2}+k_{y}^{2}}\mathrm{d}k_{x}\right) \mathrm{d}k_{y}
\label{T5-none-1} \\
& =\frac{1}{2\pi }\delta \left( z\right) \int_{-\infty }^{\infty }\frac{1}{2%
\sqrt{\frac{1}{L^{2}}+k_{y}^{2}}}\mathrm{e}^{\mathrm{-}\left\vert
x\right\vert \sqrt{\frac{1}{L^{2}}+k_{y}^{2}}}\mathrm{d}k_{y}  \notag \\
& =\frac{L}{2\pi }\delta \left( z\right) \int_{0}^{\infty }\frac{1}{\sqrt{%
1+L^{2}k_{y}^{2}}}\mathrm{e}^{\mathrm{-}\frac{\left\vert x\right\vert }{L}%
\sqrt{1+L^{2}k_{y}^{2}}}\mathrm{d}k_{y}  \label{T5-none-3} \\
& =\frac{1}{2\pi }\delta \left( z\right) \int_{1}^{\infty }\frac{1}{\sqrt{%
u^{2}-1}}\mathrm{e}^{\mathrm{-}\frac{\left\vert x\right\vert }{L}u}\mathrm{d}%
k_{y}  \label{T5-none-2} \\
& =\frac{1}{2\pi }\delta \left( z\right) K_{0}\left( \frac{\left\vert
x\right\vert }{L}\right) ,  \notag
\end{align}%
such that in (\ref{T5-none-1}) the integral (\ref{IFT-sa-lam}) is used,
while in (\ref{T5-none-3}) the substitution $u=\sqrt{1+L^{2}k_{y}^{2}}$ is
introduced in order to obtain the integral representation of Macdonald
function (\ref{T5-none-2}), see also (\ref{Macdonald}), so that (\ref{T5-x-1}%
) becomes%
\begin{equation*}
\mathcal{F}^{-1}\left[ \frac{k_{x}^{2}}{\frac{1}{L^{2}}+k_{x}^{2}+k_{y}^{2}}%
\right] \left( \boldsymbol{r}\right) =\delta \left( \boldsymbol{r}\right) -%
\frac{1}{2\pi L}\delta \left( z\right) \left( \frac{1}{L}K_{0}\left( \frac{%
\left\vert x\right\vert }{L}\right) +\frac{1}{\left\vert x\right\vert }%
K_{1}\left( \frac{\left\vert x\right\vert }{L}\right) \right) .
\end{equation*}

The remaining two off-diagonal matrix components, see (\ref{matrica-T5}),
have the same Fourier inversion which yields zero contribution, since the
Fourier transform inversion is%
\begin{align}
\mathcal{F}^{-1}\left[ \frac{k_{x}k_{y}}{\frac{1}{L^{2}}+k_{x}^{2}+k_{y}^{2}}%
\right] \left( \boldsymbol{r}\right) & =\frac{1}{\left( 2\pi \right) ^{3}}%
\int_{%
\mathbb{R}
^{3}}\frac{k_{x}k_{y}}{\frac{1}{L^{2}}+k_{x}^{2}+k_{y}^{2}}\mathrm{e}^{%
\mathrm{i}\boldsymbol{k}\cdot \boldsymbol{r}}\mathrm{d}_{\boldsymbol{k}}V 
\notag \\
& =\frac{1}{\left( 2\pi \right) ^{3}}\int_{-\infty }^{\infty }\int_{-\infty
}^{\infty }\int_{-\infty }^{\infty }\frac{k_{x}k_{y}}{\frac{1}{L^{2}}%
+k_{x}^{2}+k_{y}^{2}}\mathrm{e}^{\mathrm{i}\left( xk_{x}+zk_{z}\right) }%
\mathrm{d}k_{x}\mathrm{d}k_{y}\mathrm{d}k_{z}  \notag \\
& =\frac{1}{4\pi ^{2}}\int_{-\infty }^{\infty }k_{y}\int_{-\infty }^{\infty }%
\frac{k_{x}}{\frac{1}{L^{2}}+k_{x}^{2}+k_{y}^{2}}\mathrm{e}^{\mathrm{i}%
xk_{x}}\left( \frac{1}{2\pi }\int_{-\infty }^{\infty }\mathrm{e}^{\mathrm{i}%
zk_{z}}\mathrm{d}k_{z}\right) \mathrm{d}k_{x}\mathrm{d}k_{y}  \notag \\
& =\frac{\mathrm{i}}{2\pi ^{2}}\delta \left( z\right) \int_{-\infty
}^{\infty }k_{y}\int_{0}^{\infty }\frac{k_{x}\sin \left( xk_{x}\right) }{%
\frac{1}{L^{2}}+k_{x}^{2}+k_{y}^{2}}\mathrm{d}k_{x}\mathrm{d}k_{y}  \notag \\
& =\frac{\mathrm{i}}{2\pi ^{2}}\delta \left( z\right) \int_{-\infty
}^{\infty }k_{y}\left( -\frac{\partial }{\partial x}\int_{0}^{\infty }\frac{%
\cos \left( xk_{x}\right) }{\frac{1}{L^{2}}+k_{x}^{2}+k_{y}^{2}}\mathrm{d}%
k_{x}\right) \mathrm{d}k_{y}  \label{T5-xy-1} \\
& =\frac{\mathrm{i}}{2\pi ^{2}}\delta \left( z\right) \int_{-\infty
}^{\infty }k_{y}\left( \frac{\pi }{2}\mathrm{e}^{-\frac{\left\vert
x\right\vert }{L}\sqrt{\frac{1}{L^{2}}+k_{y}^{2}}}\right) \mathrm{d}k_{x}%
\mathrm{d}k_{y}=0,  \notag
\end{align}%
since the integrand is an even function on the interval $k_{y}\in (-\infty
,\infty )$ with respect to $k_{y}=0$, where in (\ref{T5-xy-1}) the integral (%
\ref{int-sa-kosinusom}) is used.

Fourier transform inversion of tensor $\frac{\left( \boldsymbol{L}\times 
\boldsymbol{k}\right) \otimes \left( \boldsymbol{L}\times \boldsymbol{k}%
\right) }{1+L^{2}k^{2}-\left( \boldsymbol{L}\cdot \boldsymbol{k}\right) ^{2}}%
,$ given by (\ref{matrica-T5}), after substitution of previously calculated
inversions of its components, yields%
\begin{equation}
\mathcal{F}^{-1}\left[ \frac{\left( \boldsymbol{L}\times \boldsymbol{k}%
\right) \otimes \left( \boldsymbol{L}\times \boldsymbol{k}\right) }{%
1+L^{2}k^{2}-\left( \boldsymbol{L}\cdot \boldsymbol{k}\right) ^{2}}\right]
\left( \boldsymbol{r}\right) =\left[ 
\begin{array}{ccc}
\frac{1}{2\pi L}\delta \left( r_{\parallel }\right) \frac{1}{r_{\perp }}%
K_{1}\left( \frac{r_{\perp }}{L}\right) & 0 & 0 \\ 
0 & \delta \left( \boldsymbol{r}\right) -\frac{1}{2\pi L}\delta \left(
r_{\parallel }\right) \left( \frac{1}{L}K_{0}\left( \frac{r_{\perp }}{L}%
\right) +\frac{1}{r_{\perp }}K_{1}\left( \frac{r_{\perp }}{L}\right) \right)
& 0 \\ 
0 & 0 & 0%
\end{array}%
\right]  \label{matrica}
\end{equation}%
where $r_{\parallel }=\boldsymbol{r}\cdot \frac{\boldsymbol{L}}{L}$ is used
instead of $z,$ as well as $r_{\perp }=\left\vert \boldsymbol{r}_{\perp
}\right\vert =\left\vert \boldsymbol{r}-r_{\parallel }\frac{\boldsymbol{L}}{L%
}\right\vert $ instead of $\left\vert x\right\vert ,$ since the matrix in
the expression (\ref{matrica}) can be represented by%
\begin{align*}
\mu \frac{\boldsymbol{L}\otimes \boldsymbol{L}}{L^{2}}+\nu \frac{\boldsymbol{%
r}_{\perp }\otimes \boldsymbol{r}_{\perp }}{r_{\perp }^{2}}+\eta \,%
\boldsymbol{\hat{I}}& =\mu \left( \boldsymbol{e}_{z}\otimes \boldsymbol{e}%
_{z}\right) +\nu \left( \boldsymbol{e}_{x}\otimes \boldsymbol{e}_{x}\right)
+\eta \left( \boldsymbol{e}_{x}\otimes \boldsymbol{e}_{x}+\boldsymbol{e}%
_{y}\otimes \boldsymbol{e}_{y}+\boldsymbol{e}_{z}\otimes \boldsymbol{e}%
_{z}\right) \\
& =\left( \nu +\eta \right) \left( \boldsymbol{e}_{x}\otimes \boldsymbol{e}%
_{x}\right) +\eta \,\left( \boldsymbol{e}_{y}\otimes \boldsymbol{e}%
_{y}\right) +\left( \mu +\eta \right) \left( \boldsymbol{e}_{z}\otimes 
\boldsymbol{e}_{z}\right) ,
\end{align*}%
implying%
\begin{equation*}
\left[ 
\begin{array}{ccc}
\nu +\eta & 0 & 0 \\ 
0 & \eta & 0 \\ 
0 & 0 & \mu +\eta%
\end{array}%
\right] =\left[ 
\begin{array}{ccc}
\frac{1}{2\pi L}\delta \left( r_{\parallel }\right) \frac{1}{\left\vert
r_{\perp }\right\vert }K_{1}\left( \frac{\left\vert r_{\perp }\right\vert }{L%
}\right) & 0 & 0 \\ 
0 & \delta \left( \boldsymbol{r}\right) -\frac{1}{2\pi L}\delta \left(
r_{\parallel }\right) \left( \frac{1}{L}K_{0}\left( \frac{r_{\perp }}{L}%
\right) +\frac{1}{r_{\perp }}K_{1}\left( \frac{r_{\perp }}{L}\right) \right)
& 0 \\ 
0 & 0 & 0%
\end{array}%
\right] ,
\end{equation*}%
yielding%
\begin{align*}
\mu & =-\eta =-\delta \left( \boldsymbol{r}\right) +\frac{1}{2\pi L}\delta
\left( r_{\parallel }\right) \left( \frac{1}{L}K_{0}\left( \frac{r_{\perp }}{%
L}\right) +\frac{1}{r_{\perp }}K_{1}\left( \frac{r_{\perp }}{L}\right)
\right) , \\
\nu & =-\delta \left( \boldsymbol{r}\right) +\frac{1}{2\pi L}\delta \left(
r_{\parallel }\right) \left( \frac{1}{L}K_{0}\left( \frac{r_{\perp }}{L}%
\right) +2\frac{1}{r_{\perp }}K_{1}\left( \frac{r_{\perp }}{L}\right)
\right) ,
\end{align*}%
so that%
\begin{align}
\mathcal{F}^{-1}& \left[ \frac{\left( \boldsymbol{L}\times \boldsymbol{k}%
\right) \otimes \left( \boldsymbol{L}\times \boldsymbol{k}\right) }{%
1+L^{2}k^{2}-\left( \boldsymbol{L}\cdot \boldsymbol{k}\right) ^{2}}\right]
\left( \boldsymbol{r}\right) =\mu \frac{\boldsymbol{L}\otimes \boldsymbol{L}%
}{L^{2}}+\nu \frac{\boldsymbol{r}_{\perp }\otimes \boldsymbol{r}_{\perp }}{%
r_{\perp }^{2}}+\eta \,\boldsymbol{\hat{I}}  \notag \\
& =\delta \left( r_{\parallel }\right) \left( \delta \left( \boldsymbol{r}%
_{\perp }\right) -\frac{1}{2\pi L}\left( \frac{1}{L}K_{0}\left( \frac{%
r_{\perp }}{L}\right) +\frac{1}{r_{\perp }}K_{1}\left( \frac{r_{\perp }}{L}%
\right) \right) \right) \left( \boldsymbol{\hat{I}}-\frac{\boldsymbol{L}%
\otimes \boldsymbol{L}}{L^{2}}-\frac{\boldsymbol{r}_{\perp }\otimes 
\boldsymbol{r}_{\perp }}{r_{\perp }^{2}}\right)  \notag \\
& \quad +\frac{1}{2\pi L}\delta \left( r_{\parallel }\right) \frac{1}{%
r_{\perp }}K_{1}\left( \frac{r_{\perp }}{L}\right) \frac{\boldsymbol{r}%
_{\perp }\otimes \boldsymbol{r}_{\perp }}{r_{\perp }^{2}}  \notag \\
& =\delta \left( r_{\parallel }\right) \left( \delta \left( \boldsymbol{r}%
_{\perp }\right) -\frac{1}{2\pi L}\left( \frac{1}{L}K_{0}\left( \frac{%
r_{\perp }}{L}\right) +\frac{1}{r_{\perp }}K_{1}\left( \frac{r_{\perp }}{L}%
\right) \right) \right) \left( \left( \frac{\boldsymbol{L}}{L}\times \frac{%
\boldsymbol{r}_{\perp }}{r_{\perp }}\right) \otimes \left( \frac{\boldsymbol{%
L}}{L}\times \frac{\boldsymbol{r}_{\perp }}{r_{\perp }}\right) \right) 
\notag \\
& \quad +\frac{1}{2\pi L}\delta \left( r_{\parallel }\right) \frac{1}{%
r_{\perp }}K_{1}\left( \frac{r_{\perp }}{L}\right) \frac{\boldsymbol{r}%
_{\perp }\otimes \boldsymbol{r}_{\perp }}{r_{\perp }^{2}},
\label{parce-iz-T5}
\end{align}%
since%
\begin{equation*}
\boldsymbol{\hat{I}}-\frac{\boldsymbol{L}\otimes \boldsymbol{L}}{L^{2}}-%
\frac{\boldsymbol{r}_{\perp }\otimes \boldsymbol{r}_{\perp }}{r_{\perp }^{2}}%
=\boldsymbol{e}_{y}\otimes \boldsymbol{e}_{y}=\left( \boldsymbol{e}%
_{z}\times \boldsymbol{e}_{x}\right) \otimes \left( \boldsymbol{e}_{z}\times 
\boldsymbol{e}_{x}\right) =\left( \frac{\boldsymbol{L}}{L}\times \frac{%
\boldsymbol{r}_{\perp }}{r_{\perp }}\right) \otimes \left( \frac{\boldsymbol{%
L}}{L}\times \frac{\boldsymbol{r}_{\perp }}{r_{\perp }}\right) .
\end{equation*}

Therefore, Fourier transform inversion of tensor $\boldsymbol{\bar{\hat{%
\mathcal{K}}}}_{5}$ becomes%
\begin{align*}
\boldsymbol{\hat{\mathcal{K}}}_{5}\left( \boldsymbol{r}\right) & =\mathcal{F}%
^{-1}\left[ \boldsymbol{\bar{\hat{\mathcal{K}}}}_{5}\left( \boldsymbol{k}%
\right) \right] \left( \boldsymbol{r}\right) =\mathcal{\delta }\left( 
\boldsymbol{r}\right) \boldsymbol{\hat{I}}-\mathcal{F}^{-1}\left[ \frac{%
\left( \boldsymbol{L}\times \boldsymbol{k}\right) \otimes \left( \boldsymbol{%
L}\times \boldsymbol{k}\right) }{1+L^{2}k^{2}-\left( \boldsymbol{L}\cdot 
\boldsymbol{k}\right) ^{2}}\right] \left( \boldsymbol{r}\right) \\
& =\delta \left( \boldsymbol{r}\right) \left( \boldsymbol{\hat{I}}-\left( 
\frac{\boldsymbol{L}}{L}\times \frac{\boldsymbol{r}_{\perp }}{r_{\perp }}%
\right) \otimes \left( \frac{\boldsymbol{L}}{L}\times \frac{\boldsymbol{r}%
_{\perp }}{r_{\perp }}\right) \right) -\frac{1}{2\pi L}\delta \left(
r_{\parallel }\right) \frac{1}{r_{\perp }}K_{1}\left( \frac{r_{\perp }}{L}%
\right) \frac{\boldsymbol{r}_{\perp }\otimes \boldsymbol{r}_{\perp }}{%
r_{\perp }^{2}} \\
& \quad +\frac{1}{2\pi L}\delta \left( r_{\parallel }\right) \left( \frac{1}{%
L}K_{0}\left( \frac{r_{\perp }}{L}\right) +\frac{1}{r_{\perp }}K_{1}\left( 
\frac{r_{\perp }}{L}\right) \right) \left( \left( \frac{\boldsymbol{L}}{L}%
\times \frac{\boldsymbol{r}_{\perp }}{r_{\perp }}\right) \otimes \left( 
\frac{\boldsymbol{L}}{L}\times \frac{\boldsymbol{r}_{\perp }}{r_{\perp }}%
\right) \right)
\end{align*}%
where $\boldsymbol{\bar{\hat{\mathcal{K}}}}_{5}$ is given by (\ref%
{K-5-hat-bar-2})$_{2},$ with $\mathcal{F}^{-1}\left[ \frac{\left( 
\boldsymbol{L}\times \boldsymbol{k}\right) \otimes \left( \boldsymbol{L}%
\times \boldsymbol{k}\right) }{1+L^{2}k^{2}-\left( \boldsymbol{L}\cdot 
\boldsymbol{r}\right) ^{2}}\right] \left( \boldsymbol{r}\right) $ calculated
as in (\ref{parce-iz-T5}), and taking the form (\ref{kernel-T5}).

\section{Non-locality of the Cauchy stress tensor\label{Non-locality of the
Cauchy stress tensor}}

The Cauchy stress tensor, as shown in Section \ref{Three-dimensional-cases},
is expressed as the convolution of local stress tensor with either scalar or
tensor non-locality kernel, taking the form either%
\begin{equation}
\boldsymbol{\hat{\sigma}}\left( \boldsymbol{r}\right) =\mathcal{K}_{i}\left( 
\boldsymbol{r}\right) \ast _{\boldsymbol{r}}\boldsymbol{\hat{\sigma}}%
^{local}\left( \boldsymbol{r}\right) \quad \text{or\quad }\boldsymbol{\hat{%
\sigma}}\left( \boldsymbol{r}\right) =\boldsymbol{\hat{\mathcal{K}}}%
_{i}\left( \boldsymbol{r}\right) \ast _{\boldsymbol{r}}\boldsymbol{\hat{%
\sigma}}^{local}\left( \boldsymbol{r}\right) ,  \label{konvolucija}
\end{equation}%
with non-locality kernels calculated in Section \ref{Three-dimensional-cases}
and listed in Table \ref{Modeli-new}. The convolution in (\ref{konvolucija})
is understood component-wise, so that for scalar-type non-locality kernel
one has the following expression for Cauchy stress tensor 
\begin{equation}
\boldsymbol{\hat{\sigma}}\left( \boldsymbol{r}\right) =\mathcal{K}_{i}\left( 
\boldsymbol{r}\right) \ast _{\boldsymbol{r}}\boldsymbol{\hat{\sigma}}%
^{local}\left( \boldsymbol{r}\right) =\int_{%
\mathbb{R}
^{3}}\mathcal{K}_{i}\left( \boldsymbol{r}-\boldsymbol{r}^{\prime }\right) 
\boldsymbol{\hat{\sigma}}^{local}\left( \boldsymbol{r}^{\prime }\right) 
\mathrm{d}_{\boldsymbol{r}^{\prime }}V,  \label{konvolucija-skalar}
\end{equation}%
which should be understood as%
\begin{equation*}
\sigma _{\alpha \beta }\left( \boldsymbol{r}\right) =\int_{%
\mathbb{R}
^{3}}\mathcal{K}_{i}\left( \boldsymbol{r}-\boldsymbol{r}^{\prime }\right)
\sigma _{\alpha \beta }^{local}\left( \boldsymbol{r}^{\prime }\right) 
\mathrm{d}_{\boldsymbol{r}^{\prime }}V,\quad \alpha ,\beta \in \left\{
1,2,3\right\} ,
\end{equation*}%
where $\sigma _{\alpha \beta }=\left[ \boldsymbol{\hat{\sigma}}\right]
_{\alpha \beta }$ is a tensor component, while for tensor-type non-locality
kernel one has the following expression for Cauchy stress tensor 
\begin{equation}
\boldsymbol{\hat{\sigma}}\left( \boldsymbol{r}\right) =\boldsymbol{\hat{%
\mathcal{K}}}_{i}\left( \boldsymbol{r}\right) \ast _{\boldsymbol{r}}%
\boldsymbol{\hat{\sigma}}^{local}\left( \boldsymbol{r}\right) =\int_{%
\mathbb{R}
^{3}}\boldsymbol{\hat{\mathcal{K}}}_{i}\left( \boldsymbol{r}-\boldsymbol{r}%
^{\prime }\right) \boldsymbol{\hat{\sigma}}^{local}\left( \boldsymbol{r}%
^{\prime }\right) \mathrm{d}_{\boldsymbol{r}^{\prime }}V,
\label{konvolucija-tenzor}
\end{equation}%
which should be understood as%
\begin{equation}
\sigma _{\alpha \beta }\left( \boldsymbol{r}\right) =\int_{%
\mathbb{R}
^{3}}\left[ \boldsymbol{\hat{\mathcal{K}}}_{i}\left( \boldsymbol{r}-%
\boldsymbol{r}^{\prime }\right) \right] _{\alpha \mu }\sigma _{\mu \beta
}^{local}\left( \boldsymbol{r}^{\prime }\right) \mathrm{d}_{\boldsymbol{r}%
^{\prime }}V,\quad \alpha ,\beta \in \left\{ 1,2,3\right\} ,
\label{sigmica-comp}
\end{equation}%
where the summation convention is assumed.

It is well-known that Hooke's law (\ref{Hooke's-law}) represents an
isotropic constitutive equation, which accounts for nonexistence of
shear-extesional coupling, i.e., normal (shear) strains induce normal
(shear) stresses, and which contains only two model parameters, unlike the
constitutive equation for orthotropic material, which also assumes that
shear and extension are decoupled, whereas it contains nine model
parameters. Namely, by forming stress and strain vectors, constituted of
Cauchy stress and strain tensor components, Hooke's law can be written as%
\begin{equation}
\left[ 
\begin{array}{c}
\sigma _{xx}^{local} \\ 
\sigma _{yy}^{local} \\ 
\sigma _{zz}^{local} \\ 
\sigma _{yz}^{local} \\ 
\sigma _{xz}^{local} \\ 
\sigma _{xy}^{local}%
\end{array}%
\right] =\left[ 
\begin{array}{cccccc}
\lambda +2\mu  & \lambda  & \lambda  & 0 & 0 & 0 \\ 
\lambda  & \lambda +2\mu  & \lambda  & 0 & 0 & 0 \\ 
\lambda  & \lambda  & \lambda +2\mu  & 0 & 0 & 0 \\ 
0 & 0 & 0 & 2\mu  & 0 & 0 \\ 
0 & 0 & 0 & 0 & 2\mu  & 0 \\ 
0 & 0 & 0 & 0 & 0 & 2\mu 
\end{array}%
\right] \left[ 
\begin{array}{c}
\varepsilon _{xx} \\ 
\varepsilon _{yy} \\ 
\varepsilon _{zz} \\ 
\varepsilon _{yz} \\ 
\varepsilon _{xz} \\ 
\varepsilon _{xy}%
\end{array}%
\right] ,  \label{isotropic}
\end{equation}%
where $\lambda $ and $%
{\mu}%
$ are Lam\'{e} coefficients.

When examining isotropy properties of non-local elastic models (\ref%
{scalar-non-local-elasticity}) and (\ref{tensor-non-local-elasticity}), the
model is referred as an isotropic, if the Cauchy stress tensor $\boldsymbol{%
\hat{\sigma}}$ is expressed in terms of strain through local stress tensor (%
\ref{isotropic}), such that the constitutive equation retains the form of
Hooke's law (\ref{isotropic}) and retains the number of model parameters, so
that in the case of scalar-type models, by (\ref{konvolucija-skalar}) one has%
\begin{align}
\boldsymbol{\hat{\sigma}}\left( \boldsymbol{r}\right) & =\int_{%
\mathbb{R}
^{3}}\mathcal{K}_{i}\left( \boldsymbol{r}-\boldsymbol{r}^{\prime }\right) 
\boldsymbol{\hat{\sigma}}^{local}\left( \boldsymbol{r}^{\prime }\right) 
\mathrm{d}_{\boldsymbol{r}^{\prime }}V  \notag \\
& =\left[ 
\begin{array}{cccccc}
\lambda +2\mu  & \lambda  & \lambda  & 0 & 0 & 0 \\ 
\lambda  & \lambda +2\mu  & \lambda  & 0 & 0 & 0 \\ 
\lambda  & \lambda  & \lambda +2\mu  & 0 & 0 & 0 \\ 
0 & 0 & 0 & 2\mu  & 0 & 0 \\ 
0 & 0 & 0 & 0 & 2\mu  & 0 \\ 
0 & 0 & 0 & 0 & 0 & 2\mu 
\end{array}%
\right] \left[ 
\begin{array}{c}
\int_{%
\mathbb{R}
^{3}}\mathcal{K}_{i}\left( \boldsymbol{r}-\boldsymbol{r}^{\prime }\right)
\varepsilon _{xx}\left( \boldsymbol{r}^{\prime }\right) \mathrm{d}_{%
\boldsymbol{r}^{\prime }}V \\ 
\int_{%
\mathbb{R}
^{3}}\mathcal{K}_{i}\left( \boldsymbol{r}-\boldsymbol{r}^{\prime }\right)
\varepsilon _{yy}\left( \boldsymbol{r}^{\prime }\right) \mathrm{d}_{%
\boldsymbol{r}^{\prime }}V \\ 
\int_{%
\mathbb{R}
^{3}}\mathcal{K}_{i}\left( \boldsymbol{r}-\boldsymbol{r}^{\prime }\right)
\varepsilon _{zz}\left( \boldsymbol{r}^{\prime }\right) \mathrm{d}_{%
\boldsymbol{r}^{\prime }}V \\ 
\int_{%
\mathbb{R}
^{3}}\mathcal{K}_{i}\left( \boldsymbol{r}-\boldsymbol{r}^{\prime }\right)
\varepsilon _{yz}\left( \boldsymbol{r}^{\prime }\right) \mathrm{d}_{%
\boldsymbol{r}^{\prime }}V \\ 
\int_{%
\mathbb{R}
^{3}}\mathcal{K}_{i}\left( \boldsymbol{r}-\boldsymbol{r}^{\prime }\right)
\varepsilon _{xz}\left( \boldsymbol{r}^{\prime }\right) \mathrm{d}_{%
\boldsymbol{r}^{\prime }}V \\ 
\int_{%
\mathbb{R}
^{3}}\mathcal{K}_{i}\left( \boldsymbol{r}-\boldsymbol{r}^{\prime }\right)
\varepsilon _{xy}\left( \boldsymbol{r}^{\prime }\right) \mathrm{d}_{%
\boldsymbol{r}^{\prime }}V%
\end{array}%
\right] ,  \label{scalar-elastic}
\end{align}%
if Lam\'{e} coefficients are constants, implying the isotropy of the
scalar-type non-local elastic models (\ref{scalar-non-local-elasticity}),
where the non-locality kernels are listed in Table \ref{Modeli-new}. On the
other hand, one may consider non-local anisotropy in the sense that there
might be preferential direction(s), as it is separately discussed for each
scalar-type non-local model.

Contrary to the case of scalar-type non-local elastic models (\ref%
{scalar-non-local-elasticity}), if tensor-type non-local elastic models (\ref%
{tensor-non-local-elasticity}) are considered, from (\ref{konvolucija-tenzor}%
), i.e., (\ref{sigmica-comp}) and Hooke's law (\ref{Hooke's-law}), for the
Cauchy stress-tensor components one has%
\begin{align*}
\sigma _{\alpha \beta }\left( \boldsymbol{r}\right) & =\int_{%
\mathbb{R}
^{3}}\left[ \boldsymbol{\hat{\mathcal{K}}}_{i}\left( \boldsymbol{r}-%
\boldsymbol{r}^{\prime }\right) \right] _{\alpha \eta }\sigma _{\eta \beta
}^{local}\left( \boldsymbol{r}^{\prime }\right) \mathrm{d}_{\boldsymbol{r}%
^{\prime }}V,\quad \alpha ,\beta \in \left\{ 1,2,3\right\} \\
& =\int_{%
\mathbb{R}
^{3}}\left[ \boldsymbol{\hat{\mathcal{K}}}_{i}\left( \boldsymbol{r}-%
\boldsymbol{r}^{\prime }\right) \right] _{\alpha \eta }\left( \lambda
\varepsilon _{\gamma \gamma }\left( \boldsymbol{r}^{\prime }\right) \delta
_{\eta \beta }+2\mu \varepsilon _{\eta \beta }\left( \boldsymbol{r}^{\prime
}\right) \right) \mathrm{d}_{\boldsymbol{r}^{\prime }}V \\
& =\lambda \int_{%
\mathbb{R}
^{3}}\left[ \boldsymbol{\hat{\mathcal{K}}}_{i}\left( \boldsymbol{r}-%
\boldsymbol{r}^{\prime }\right) \right] _{\alpha \beta }\varepsilon _{\gamma
\gamma }\left( \boldsymbol{r}^{\prime }\right) \mathrm{d}_{\boldsymbol{r}%
^{\prime }}V+2\mu \int_{%
\mathbb{R}
^{3}}\left[ \boldsymbol{\hat{\mathcal{K}}}_{i}\left( \boldsymbol{r}-%
\boldsymbol{r}^{\prime }\right) \right] _{\alpha \eta }\varepsilon _{\eta
\beta }\left( \boldsymbol{r}^{\prime }\right) \mathrm{d}_{\boldsymbol{r}%
^{\prime }}V \\
& =\lambda \left[ \boldsymbol{\hat{\mathcal{K}}}_{i}\left( \boldsymbol{r}%
\right) \right] _{\alpha \beta }\ast _{\boldsymbol{r}}\varepsilon _{\gamma
\gamma }\left( \boldsymbol{r}\right) +2\mu \left[ \boldsymbol{\hat{\mathcal{K%
}}}_{i}\left( \boldsymbol{r}\right) \right] _{\alpha \eta }\ast _{%
\boldsymbol{r}}\varepsilon _{\eta \beta }\left( \boldsymbol{r}\right) ,
\end{align*}%
if Lam\'{e} coefficients are constants, implying that diagonal stress
components are%
\begin{align}
\sigma _{11}\left( \boldsymbol{r}\right) &=\left( \lambda +2\mu \right) %
\left[ \boldsymbol{\hat{\mathcal{K}}}_{i}\left( \boldsymbol{r}\right) \right]
_{11}\ast _{\boldsymbol{r}}\varepsilon _{11}\left( \boldsymbol{r}\right)
+\lambda \left[ \boldsymbol{\hat{\mathcal{K}}}_{i}\left( \boldsymbol{r}%
\right) \right] _{11}\ast _{\boldsymbol{r}}\varepsilon _{22}\left( 
\boldsymbol{r}\right) +\lambda \left[ \boldsymbol{\hat{\mathcal{K}}}%
_{i}\left( \boldsymbol{r}\right) \right] _{11}\ast _{\boldsymbol{r}%
}\varepsilon _{33}\left( \boldsymbol{r}\right)  \notag \\
&\quad \quad +2\mu \left[ \boldsymbol{\hat{\mathcal{K}}}_{i}\left( 
\boldsymbol{r}\right) \right] _{12}\ast _{\boldsymbol{r}}\varepsilon
_{12}\left( \boldsymbol{r}\right) +2\mu \left[ \boldsymbol{\hat{\mathcal{K}}}%
_{i}\left( \boldsymbol{r}\right) \right] _{13}\ast _{\boldsymbol{r}%
}\varepsilon _{13}\left( \boldsymbol{r}\right) ,  \label{dijagonalni-1} \\
\sigma _{22}\left( \boldsymbol{r}\right) &=\lambda \left[ \boldsymbol{\hat{%
\mathcal{K}}}_{i}\left( \boldsymbol{r}\right) \right] _{22}\ast _{%
\boldsymbol{r}}\varepsilon _{11}\left( \boldsymbol{r}\right) +\left( \lambda
+2\mu \right) \left[ \boldsymbol{\hat{\mathcal{K}}}_{i}\left( \boldsymbol{r}%
\right) \right] _{22}\ast _{\boldsymbol{r}}\varepsilon _{22}\left( 
\boldsymbol{r}\right) +\lambda \left[ \boldsymbol{\hat{\mathcal{K}}}%
_{i}\left( \boldsymbol{r}\right) \right] _{22}\ast _{\boldsymbol{r}%
}\varepsilon _{33}\left( \boldsymbol{r}\right)  \notag \\
&\quad \quad +2\mu \left[ \boldsymbol{\hat{\mathcal{K}}}_{i}\left( 
\boldsymbol{r}\right) \right] _{21}\ast _{\boldsymbol{r}}\varepsilon
_{12}\left( \boldsymbol{r}\right) +2\mu \left[ \boldsymbol{\hat{\mathcal{K}}}%
_{i}\left( \boldsymbol{r}\right) \right] _{23}\ast _{\boldsymbol{r}%
}\varepsilon _{23}\left( \boldsymbol{r}\right) ,  \label{dijagonalni-2} \\
\sigma _{33}\left( \boldsymbol{r}\right) &=\lambda \left[ \boldsymbol{\hat{%
\mathcal{K}}}_{i}\left( \boldsymbol{r}\right) \right] _{33}\ast _{%
\boldsymbol{r}}\varepsilon _{11}\left( \boldsymbol{r}\right) +\lambda \left[ 
\boldsymbol{\hat{\mathcal{K}}}_{i}\left( \boldsymbol{r}\right) \right]
_{33}\ast _{\boldsymbol{r}}\varepsilon _{22}\left( \boldsymbol{r}\right)
+\left( \lambda +2\mu \right) \left[ \boldsymbol{\hat{\mathcal{K}}}%
_{i}\left( \boldsymbol{r}\right) \right] _{33}\ast _{\boldsymbol{r}%
}\varepsilon _{33}\left( \boldsymbol{r}\right)  \notag \\
&\quad \quad +2\mu \left[ \boldsymbol{\hat{\mathcal{K}}}_{i}\left( 
\boldsymbol{r}\right) \right] _{31}\ast _{\boldsymbol{r}}\varepsilon
_{13}\left( \boldsymbol{r}\right) +2\mu \left[ \boldsymbol{\hat{\mathcal{K}}}%
_{i}\left( \boldsymbol{r}\right) \right] _{32}\ast _{\boldsymbol{r}%
}\varepsilon _{23}\left( \boldsymbol{r}\right) ,  \label{dijagonalni-3}
\end{align}%
while off-diagonal stress components are%
\begin{align}
\sigma _{31}\left( \boldsymbol{r}\right) &=\left( \lambda +2\mu \right) %
\left[ \boldsymbol{\hat{\mathcal{K}}}_{i}\left( \boldsymbol{r}\right) \right]
_{31}\ast _{\boldsymbol{r}}\varepsilon _{11}\left( \boldsymbol{r}\right)
+\lambda \left[ \boldsymbol{\hat{\mathcal{K}}}_{i}\left( \boldsymbol{r}%
\right) \right] _{31}\ast _{\boldsymbol{r}}\varepsilon _{22}\left( 
\boldsymbol{r}\right) +\lambda \left[ \boldsymbol{\hat{\mathcal{K}}}%
_{i}\left( \boldsymbol{r}\right) \right] _{31}\ast _{\boldsymbol{r}%
}\varepsilon _{33}\left( \boldsymbol{r}\right)  \notag \\
& \quad \quad +2\mu \left[ \boldsymbol{\hat{\mathcal{K}}}_{i}\left( 
\boldsymbol{r}\right) \right] _{32}\ast _{\boldsymbol{r}}\varepsilon
_{12}\left( \boldsymbol{r}\right) +2\mu \left[ \boldsymbol{\hat{\mathcal{K}}}%
_{i}\left( \boldsymbol{r}\right) \right] _{33}\ast _{\boldsymbol{r}%
}\varepsilon _{13}\left( \boldsymbol{r}\right) ,  \label{off-dijagonalni-1}
\\
\sigma _{12}\left( \boldsymbol{r}\right) &=\lambda \left[ \boldsymbol{\hat{%
\mathcal{K}}}_{i}\left( \boldsymbol{r}\right) \right] _{12}\ast _{%
\boldsymbol{r}}\varepsilon _{11}\left( \boldsymbol{r}\right) +\left( \lambda
+2\mu \right) \left[ \boldsymbol{\hat{\mathcal{K}}}_{i}\left( \boldsymbol{r}%
\right) \right] _{12}\ast _{\boldsymbol{r}}\varepsilon _{22}\left( 
\boldsymbol{r}\right) +\lambda \left[ \boldsymbol{\hat{\mathcal{K}}}%
_{i}\left( \boldsymbol{r}\right) \right] _{12}\ast _{\boldsymbol{r}%
}\varepsilon _{33}\left( \boldsymbol{r}\right)  \notag \\
& \quad \quad +2\mu \left[ \boldsymbol{\hat{\mathcal{K}}}_{i}\left( 
\boldsymbol{r}\right) \right] _{11}\ast _{\boldsymbol{r}}\varepsilon
_{12}\left( \boldsymbol{r}\right) +2\mu \left[ \boldsymbol{\hat{\mathcal{K}}}%
_{i}\left( \boldsymbol{r}\right) \right] _{13}\ast _{\boldsymbol{r}%
}\varepsilon _{23}\left( \boldsymbol{r}\right) ,  \label{off-dijagonalni-2}
\\
\sigma _{23}\left( \boldsymbol{r}\right) &=\lambda \left[ \boldsymbol{\hat{%
\mathcal{K}}}_{i}\left( \boldsymbol{r}\right) \right] _{23}\ast _{%
\boldsymbol{r}}\varepsilon _{11}\left( \boldsymbol{r}\right) +\lambda \left[ 
\boldsymbol{\hat{\mathcal{K}}}_{i}\left( \boldsymbol{r}\right) \right]
_{23}\ast _{\boldsymbol{r}}\varepsilon _{22}\left( \boldsymbol{r}\right)
+\left( \lambda +2\mu \right) \left[ \boldsymbol{\hat{\mathcal{K}}}%
_{i}\left( \boldsymbol{r}\right) \right] _{23}\ast _{\boldsymbol{r}%
}\varepsilon _{33}\left( \boldsymbol{r}\right)  \notag \\
& \quad \quad +2\mu \left[ \boldsymbol{\hat{\mathcal{K}}}_{i}\left( 
\boldsymbol{r}\right) \right] _{21}\ast _{\boldsymbol{r}}\varepsilon
_{13}\left( \boldsymbol{r}\right) +2\mu \left[ \boldsymbol{\hat{\mathcal{K}}}%
_{i}\left( \boldsymbol{r}\right) \right] _{22}\ast _{\boldsymbol{r}%
}\varepsilon _{23}\left( \boldsymbol{r}\right) .  \label{off-dijagonalni-3}
\end{align}%
Obviously, shear-extesional coupling does not exist if tensor non-locality
kernel is diagonal, see (\ref{dijagonalni-1}) - (\ref{dijagonalni-3}) and (%
\ref{off-dijagonalni-1}) - (\ref{off-dijagonalni-3}), and moreover if $\left[
\boldsymbol{\hat{\mathcal{K}}}_{i}\right] _{11}=\left[ \boldsymbol{\hat{%
\mathcal{K}}}_{i}\right] _{22}=\left[ \boldsymbol{\hat{\mathcal{K}}}_{i}%
\right] _{33}=\mathcal{K}_{i},$ i.e., $\boldsymbol{\hat{\mathcal{K}}}_{i}=%
\mathcal{K}_{i}\boldsymbol{\hat{I}},$ then tensor-type non-local elastic
model is isotropic, however the assumption $\boldsymbol{\hat{\mathcal{K}}}%
_{i}=\mathcal{K}_{i}\boldsymbol{\hat{I}}$ implies that non-locality kernel
is of scalar-type, i.e., all tensor-type non-local elastic models are
anisotropic. On the other hand, similarly as in the case of scalar-type
non-local elastic models, one may consider non-local anisotropy in the sense
that there might be preferential direction(s), as it is discussed for each
tensor-type non-local model.

Furthermore, it is discussed whether the Cauchy stress tensor is non-locally
isotropic or anisotropic, depending on whether the non-locality is uniformly
represented in all directions or only along specific directions.

\subsection{Model $S_{1}$}

\label{Model-S-1}

In the case of model $S_{1},$ the expression (\ref{konvolucija-skalar}),
connecting the non-local and local Cauchy stress tensors, along with the
non-locality kernel (\ref{K1}) being of the scalar-type, see also Table \ref%
{Modeli-new}, yields%
\begin{align*}
\boldsymbol{\hat{\sigma}}\left( \boldsymbol{r}\right) & =\mathcal{K}%
_{1}\left( \boldsymbol{r}\right) \ast _{\boldsymbol{r}}\boldsymbol{\hat{%
\sigma}}^{local}\left( \boldsymbol{r}\right) =\int_{%
\mathbb{R}
^{3}}\mathcal{K}_{1}\left( \boldsymbol{r}-\boldsymbol{r}^{\prime }\right) 
\boldsymbol{\hat{\sigma}}^{local}\left( \boldsymbol{r}^{\prime }\right) 
\mathrm{d}_{\boldsymbol{r}^{\prime }}V \\
& =\frac{1}{4\pi L^{2}}\int_{%
\mathbb{R}
^{3}}\frac{1}{\left\vert \boldsymbol{r}-\boldsymbol{r}^{\prime }\right\vert }%
\mathrm{e}^{-\frac{\left\vert \boldsymbol{r}-\boldsymbol{r}^{\prime
}\right\vert }{L}}\boldsymbol{\hat{\sigma}}^{local}\left( \boldsymbol{r}%
^{\prime }\right) \mathrm{d}_{\boldsymbol{r}^{\prime }}V.
\end{align*}%
The non-locality kernel in the case of model $S_{1}$ is a scalar function of
Yukawa type, therefore on one hand implying an isotropic character of the
Cauchy stress tensor, and on the other hand implying an isotropic
non-locality of the Cauchy stress tensor, since the Yukawa type function is
isotropic and therefore exhibits no directional preference.

\subsection{Model $S_{2}$}

In the case of model $S_{2},$ the expression (\ref{konvolucija-skalar}),
connecting the non-local and local Cauchy stress tensors, along with the
non-locality kernel (\ref{K2}) being of the scalar-type, see also Table \ref%
{Modeli-new}, yields 
\begin{align*}
\boldsymbol{\hat{\sigma}}\left( \boldsymbol{r}\right) & =\mathcal{K}%
_{2}\left( \boldsymbol{r}\right) \ast _{\boldsymbol{r}}\boldsymbol{\hat{%
\sigma}}^{local}\left( \boldsymbol{r}\right) =\int_{%
\mathbb{R}
^{3}}\mathcal{K}_{2}\left( \boldsymbol{r}-\boldsymbol{r}^{\prime }\right) 
\boldsymbol{\hat{\sigma}}^{local}\left( \boldsymbol{r}^{\prime }\right) 
\mathrm{d}_{\boldsymbol{r}^{\prime }}V \\
& =\frac{1}{2L}\int_{%
\mathbb{R}
}\mathrm{e}^{-\frac{\left\vert r_{\parallel }-r_{\parallel }^{\prime
}\right\vert }{L}}\int_{%
\mathbb{R}
^{2}}\delta \left( \boldsymbol{r}_{\perp }-\boldsymbol{r}_{\perp }^{\prime
}\right) \boldsymbol{\hat{\sigma}}^{local}\left( \boldsymbol{r}_{\perp
}^{\prime },r_{\parallel }^{\prime }\right) \mathrm{d}_{\boldsymbol{r}%
_{\perp }^{\prime }}S\,\mathrm{d}r_{\parallel }^{\prime } \\
& =\frac{1}{2L}\int_{%
\mathbb{R}
}\mathrm{e}^{-\frac{\left\vert r_{\parallel }-r_{\parallel }^{\prime
}\right\vert }{L}}\boldsymbol{\hat{\sigma}}^{local}\left( \boldsymbol{r}%
_{\perp },r_{\parallel }^{\prime }\right) \mathrm{d}r_{\parallel }^{\prime }.
\end{align*}%
As in the case of model $S_{1}$, model $S_{2}$ assumes Cauchy stress tensor
that exhibits isotropic character, however being non-localy anisotropic,
such that non-locality of exponential type occurs only along the direction
determined by the non-locality vector $\boldsymbol{L}$, while the Cauchy
stress tensor does not exhibit non-locality in the plane perpendicular to
this direction.

\subsection{Model $S_{3}$}

\label{Model-S-3}

In the case of model $S_{3},$ the expression (\ref{konvolucija-skalar}),
connecting the non-local and local Cauchy stress tensors, along with the
non-locality kernel (\ref{K3}) being of the scalar-type, see also Table \ref%
{Modeli-new}, yields 
\begin{align*}
\boldsymbol{\hat{\sigma}}\left( \boldsymbol{r}\right) & =\mathcal{K}%
_{3}\left( \boldsymbol{r}\right) \ast _{\boldsymbol{r}}\boldsymbol{\hat{%
\sigma}}^{local}\left( \boldsymbol{r}\right) =\int_{%
\mathbb{R}
^{3}}\mathcal{K}_{3}\left( \boldsymbol{r}-\boldsymbol{r}^{\prime }\right) 
\boldsymbol{\hat{\sigma}}^{local}\left( \boldsymbol{r}^{\prime }\right) 
\mathrm{d}_{\boldsymbol{r}^{\prime }}V \\
& =\frac{1}{2\pi L^{2}}\int_{%
\mathbb{R}
^{2}}K_{0}\left( \frac{\left\vert \boldsymbol{r}_{\perp }-\boldsymbol{r}%
_{\perp }^{\prime }\right\vert }{L}\right) \int_{%
\mathbb{R}
}\delta \left( r_{\parallel }-r_{\parallel }^{\prime }\right) \boldsymbol{%
\hat{\sigma}}^{local}\left( \boldsymbol{r}_{\perp }^{\prime },r_{\parallel
}^{\prime }\right) \mathrm{d}_{\boldsymbol{r}_{\perp }^{\prime }}S\,\mathrm{d%
}r_{\parallel }^{\prime } \\
& =\frac{1}{2\pi L^{2}}\int_{%
\mathbb{R}
^{2}}K_{0}\left( \frac{\left\vert \boldsymbol{r}_{\perp }-\boldsymbol{r}%
_{\perp }^{\prime }\right\vert }{L}\right) \boldsymbol{\hat{\sigma}}%
^{local}\left( \boldsymbol{r}_{\perp }^{\prime },r_{\parallel }\right) 
\mathrm{d}_{\boldsymbol{r}_{\perp }^{\prime }}S.
\end{align*}%
As in the case of models $S_{1}$ and $S_{2}$, model $S_{3}$ assumes
isotropic Cauchy stress tensor having anisotropic non-local character so as
the model $S_{2}$. However in the case of model $S_{3},$ the Cauchy stress
tensor does not exhibit a non-local character along the direction determined
by the non-locality vector $\boldsymbol{L}$, but in the plane perpendicular
to this direction.

\subsection{Model $T_{1}$}

In the case of model $T_{1},$ the expression (\ref{konvolucija-tenzor}),
connecting the non-local and local Cauchy stress tensors, along with the
non-locality kernel (\ref{kernel-T1}) being of the tensor-type, see also
Table \ref{Modeli-new}, yields 
\begin{align*}
\boldsymbol{\hat{\sigma}}\left( \boldsymbol{r}\right) & =\boldsymbol{\hat{%
\mathcal{K}}}_{1}\left( \boldsymbol{r}\right) \ast _{\boldsymbol{r}}%
\boldsymbol{\hat{\sigma}}^{local}\left( \boldsymbol{r}\right) =\int_{%
\mathbb{R}
^{3}}\boldsymbol{\hat{\mathcal{K}}}_{1}\left( \boldsymbol{r}-\boldsymbol{r}%
^{\prime }\right) \boldsymbol{\hat{\sigma}}^{local}\left( \boldsymbol{r}%
^{\prime }\right) \mathrm{d}_{\boldsymbol{r}^{\prime }}V \\
& =2\int_{%
\mathbb{R}
^{3}}\delta \left( \boldsymbol{r}-\boldsymbol{r}^{\prime }\right) \left( 
\boldsymbol{\hat{I}}-2\frac{\left( \boldsymbol{r}-\boldsymbol{r}^{\prime
}\right) \otimes \left( \boldsymbol{r}-\boldsymbol{r}^{\prime }\right) }{%
\left\vert \boldsymbol{r}-\boldsymbol{r}^{\prime }\right\vert ^{2}}\right) 
\boldsymbol{\hat{\sigma}}^{local}\left( \boldsymbol{r}^{\prime }\right) 
\mathrm{d}_{\boldsymbol{r}^{\prime }}V-\frac{1}{4\pi }\int_{%
\mathbb{R}
^{3}}\frac{1}{\left\vert \boldsymbol{r}-\boldsymbol{r}^{\prime }\right\vert
^{3}}\mathrm{e}^{-\frac{\left\vert \boldsymbol{r}-\boldsymbol{r}^{\prime
}\right\vert }{L}} \\
& \quad \quad \quad \quad \quad \times \left( \left( 1+\frac{\left\vert 
\boldsymbol{r}-\boldsymbol{r}^{\prime }\right\vert }{L}\right) \boldsymbol{%
\hat{I}}+\left( 3+3\frac{\left\vert \boldsymbol{r}-\boldsymbol{r}^{\prime
}\right\vert }{L}+\frac{\left\vert \boldsymbol{r}-\boldsymbol{r}^{\prime
}\right\vert ^{2}}{L^{2}}\right) \frac{\left( \boldsymbol{r}-\boldsymbol{r}%
^{\prime }\right) \otimes \left( \boldsymbol{r}-\boldsymbol{r}^{\prime
}\right) }{\left\vert \boldsymbol{r}-\boldsymbol{r}^{\prime }\right\vert ^{2}%
}\right) \boldsymbol{\hat{\sigma}}^{local}\left( \boldsymbol{r}^{\prime
}\right) \mathrm{d}_{\boldsymbol{r}^{\prime }}V \\
& =2\boldsymbol{\hat{\sigma}}^{local}\left( \boldsymbol{r}\right) -4\frac{%
\boldsymbol{r}\otimes \boldsymbol{r}}{r^{2}}\boldsymbol{\hat{\sigma}}%
^{local}\left( \boldsymbol{r}\right) -\frac{1}{4\pi }\int_{%
\mathbb{R}
^{3}}\frac{1}{\left\vert \boldsymbol{r}-\boldsymbol{r}^{\prime }\right\vert
^{3}}\left( 1+\frac{\left\vert \boldsymbol{r}-\boldsymbol{r}^{\prime
}\right\vert }{L}\right) \mathrm{e}^{-\frac{\left\vert \boldsymbol{r}-%
\boldsymbol{r}^{\prime }\right\vert }{L}}\boldsymbol{\hat{\sigma}}%
^{local}\left( \boldsymbol{r}^{\prime }\right) \mathrm{d}_{\boldsymbol{r}%
^{\prime }}V \\
& \quad \quad \quad \quad \quad -\frac{1}{4\pi }\int_{%
\mathbb{R}
^{3}}\frac{1}{\left\vert \boldsymbol{r}-\boldsymbol{r}^{\prime }\right\vert
^{3}}\left( 3+3\frac{\left\vert \boldsymbol{r}-\boldsymbol{r}^{\prime
}\right\vert }{L}+\frac{\left\vert \boldsymbol{r}-\boldsymbol{r}^{\prime
}\right\vert ^{2}}{L^{2}}\right) \mathrm{e}^{-\frac{\left\vert \boldsymbol{r}%
-\boldsymbol{r}^{\prime }\right\vert }{L}}\frac{\left( \boldsymbol{r}-%
\boldsymbol{r}^{\prime }\right) \otimes \left( \boldsymbol{r}-\boldsymbol{r}%
^{\prime }\right) }{\left\vert \boldsymbol{r}-\boldsymbol{r}^{\prime
}\right\vert ^{2}}\boldsymbol{\hat{\sigma}}^{local}\left( \boldsymbol{r}%
^{\prime }\right) \mathrm{d}_{\boldsymbol{r}^{\prime }}V
\end{align*}%
Concerning model $T_{1}$, the Cauchy stress tensor consists of four terms:
two local ones of which one is spherically symmetric, since tensor%
\begin{equation*}
\frac{\boldsymbol{r}\otimes \boldsymbol{r}}{r^{2}}=\left[ 
\begin{array}{c}
\boldsymbol{e}_{r} \\ 
0\boldsymbol{e}_{\theta } \\ 
0\boldsymbol{e}_{\varphi }%
\end{array}%
\right] =\left[ 
\begin{array}{c}
\boldsymbol{e}_{r} \\ 
\boldsymbol{0} \\ 
\boldsymbol{0}%
\end{array}%
\right] ,
\end{equation*}%
where $\boldsymbol{e}_{r},$ $\boldsymbol{e}_{\theta },$ and $\boldsymbol{e}%
_{\varphi }$ are unit vectors of the spherical coordinate system, as well as
of two non-local terms displaying exponetial-type isotropic non-locality,
since there is no preferred direction, differing in existence of spherically
symmetric tensor $\frac{\boldsymbol{r}\otimes \boldsymbol{r}}{r^{2}}$ acting
on local Cauchy stress tensor.

\subsection{Model $T_{2}$}

In the case of model $T_{2},$ the expression (\ref{konvolucija-tenzor}),
connecting the non-local and local Cauchy stress tensors, along with the
non-locality kernel (\ref{kernel-T2}) being of the tensor-type, see also
Table \ref{Modeli-new}, yields 
\begin{align*}
\boldsymbol{\hat{\sigma}}\left( \boldsymbol{r}\right) & =\boldsymbol{\hat{%
\mathcal{K}}}_{2}\left( \boldsymbol{r}\right) \ast _{\boldsymbol{r}}%
\boldsymbol{\hat{\sigma}}^{local}\left( \boldsymbol{r}\right) =\int_{%
\mathbb{R}
^{3}}\boldsymbol{\hat{\mathcal{K}}}_{2}\left( \boldsymbol{r}-\boldsymbol{r}%
^{\prime }\right) \boldsymbol{\hat{\sigma}}^{local}\left( \boldsymbol{r}%
^{\prime }\right) \mathrm{d}_{\boldsymbol{r}^{\prime }}V \\
& =\int_{%
\mathbb{R}
^{3}}\delta \left( \boldsymbol{r}-\boldsymbol{r}^{\prime }\right) \left( 
\boldsymbol{\hat{I}}-\frac{\boldsymbol{L}\otimes \boldsymbol{L}}{L^{2}}%
\right) \boldsymbol{\hat{\sigma}}^{local}\left( \boldsymbol{r}^{\prime
}\right) \mathrm{d}_{\boldsymbol{r}^{\prime }}V+\frac{1}{4\pi L^{2}}\int_{%
\mathbb{R}
^{3}}\frac{1}{\left\vert \boldsymbol{r}-\boldsymbol{r}^{\prime }\right\vert }%
\mathrm{e}^{-\frac{\left\vert \boldsymbol{r}-\boldsymbol{r}^{\prime
}\right\vert }{L}}\frac{\boldsymbol{L}\otimes \boldsymbol{L}}{L^{2}}%
\boldsymbol{\hat{\sigma}}^{local}\left( \boldsymbol{r}^{\prime }\right) 
\mathrm{d}_{\boldsymbol{r}^{\prime }}V \\
& =\boldsymbol{\hat{\sigma}}^{local}\left( \boldsymbol{r}\right) -\frac{%
\boldsymbol{L}\otimes \boldsymbol{L}}{L^{2}}\boldsymbol{\hat{\sigma}}%
^{local}\left( \boldsymbol{r}\right) +\frac{1}{4\pi L^{2}}\frac{\boldsymbol{L%
}\otimes \boldsymbol{L}}{L^{2}}\int_{%
\mathbb{R}
^{3}}\frac{1}{\left\vert \boldsymbol{r}-\boldsymbol{r}^{\prime }\right\vert }%
\mathrm{e}^{-\frac{\left\vert \boldsymbol{r}-\boldsymbol{r}^{\prime
}\right\vert }{L}}\boldsymbol{\hat{\sigma}}^{local}\left( \boldsymbol{r}%
^{\prime }\right) \mathrm{d}_{\boldsymbol{r}^{\prime }}V.
\end{align*}%
As in the case of model $T_{1}$, the Cauchy stress tensor corresponding to
model $T_{2}$ also consists of local and non-local terms, where the second
term assumes directional dependence on non-locality vector $\boldsymbol{L},$
since tensor $\frac{\boldsymbol{L}\otimes \boldsymbol{L}}{L^{2}}$ favors a
particular direction and acts as a projector onto the direction of $%
\boldsymbol{L}$, while in the third term the direction of non-locality
vector $\boldsymbol{L}$ is also favored through tensor $\frac{\boldsymbol{L}%
\otimes \boldsymbol{L}}{L^{2}},$ while being non-localy isotropic of
Yukawa-type.

\subsection{Model $T_{3}$}

In the case of model $T_{3},$ the expression (\ref{konvolucija-tenzor}),
connecting the non-local and local Cauchy stress tensors, along with the
non-locality kernel (\ref{kernel-T3}) being of the tensor-type, see also
Table \ref{Modeli-new}, yields 
\begin{align*}
\boldsymbol{\hat{\sigma}}\left( \boldsymbol{r}\right) & =\boldsymbol{\hat{%
\mathcal{K}}}_{3}\left( \boldsymbol{r}\right) \ast _{\boldsymbol{r}}%
\boldsymbol{\hat{\sigma}}^{local}\left( \boldsymbol{r}\right) =\int_{%
\mathbb{R}
^{3}}\boldsymbol{\hat{\mathcal{K}}}_{3}\left( \boldsymbol{r}-\boldsymbol{r}%
^{\prime }\right) \boldsymbol{\hat{\sigma}}^{local}\left( \boldsymbol{r}%
^{\prime }\right) \mathrm{d}_{\boldsymbol{r}^{\prime }}V \\
& =\int_{%
\mathbb{R}
^{3}}\!\delta \left( \boldsymbol{r}\right) \left( \boldsymbol{\hat{I}}-\frac{%
\boldsymbol{L}\otimes \boldsymbol{L}}{L^{2}}\right) \boldsymbol{\hat{\sigma}}%
^{local}\left( \boldsymbol{r}^{\prime }\right) \mathrm{d}_{\boldsymbol{r}%
^{\prime }}V+\frac{1}{4L}\int_{%
\mathbb{R}
}\!\mathrm{e}^{-\frac{\left\vert r_{\parallel }-r_{\parallel }^{\prime
}\right\vert }{L}}\int_{%
\mathbb{R}
^{2}}\!\delta \left( \boldsymbol{r}_{\perp }-\boldsymbol{r}_{\perp }^{\prime
}\right) \frac{\boldsymbol{L}\otimes \boldsymbol{L}}{L^{2}}\boldsymbol{\hat{%
\sigma}}^{local}\left( \boldsymbol{r}_{\perp }^{\prime },r_{\parallel
}^{\prime }\right) \mathrm{d}_{\boldsymbol{r}_{\perp }^{\prime }}S\,\mathrm{d%
}r_{\parallel }^{\prime } \\
& \quad \quad \quad +\int_{%
\mathbb{R}
}\mathrm{e}^{-\frac{\left\vert r_{\parallel }-r_{\parallel }^{\prime
}\right\vert }{L}}\int_{%
\mathbb{R}
^{2}}\frac{\delta \left( \boldsymbol{r}_{\perp }-\boldsymbol{r}_{\perp
}^{\prime }\right) }{\left\vert \boldsymbol{r}_{\perp }-\boldsymbol{r}%
_{\perp }^{\prime }\right\vert }\left( \frac{\boldsymbol{L}}{L}\otimes \frac{%
\boldsymbol{r}_{\perp }-\boldsymbol{r}_{\perp }^{\prime }}{\left\vert 
\boldsymbol{r}_{\perp }-\boldsymbol{r}_{\perp }^{\prime }\right\vert }%
\right) \boldsymbol{\hat{\sigma}}^{local}\left( \boldsymbol{r}_{\perp
}^{\prime },r_{\parallel }^{\prime }\right) \mathrm{d}_{\boldsymbol{r}%
_{\perp }^{\prime }}S\,\mathrm{d}r_{\parallel }^{\prime } \\
& =\boldsymbol{\hat{\sigma}}^{local}\left( \boldsymbol{r}\right) -\frac{%
\boldsymbol{L}\otimes \boldsymbol{L}}{L^{2}}\boldsymbol{\hat{\sigma}}%
^{local}\left( \boldsymbol{r}\right) +\frac{1}{4L}\frac{\boldsymbol{L}%
\otimes \boldsymbol{L}}{L^{2}}\int_{%
\mathbb{R}
}\mathrm{e}^{-\frac{\left\vert r_{\parallel }-r_{\parallel }^{\prime
}\right\vert }{L}}\boldsymbol{\hat{\sigma}}^{local}\left( \boldsymbol{r}%
_{\perp },r_{\parallel }^{\prime }\right) \mathrm{d}r_{\parallel }^{\prime }
\\
& \quad \quad \quad +\frac{1}{r_{\perp }}\left( \frac{\boldsymbol{L}}{L}%
\otimes \frac{\boldsymbol{r}_{\perp }}{r_{\perp }}\right) \int_{%
\mathbb{R}
}\mathrm{e}^{-\frac{\left\vert r_{\parallel }-r_{\parallel }^{\prime
}\right\vert }{L}}\boldsymbol{\hat{\sigma}}^{local}\left( \boldsymbol{r}%
_{\perp },r_{\parallel }^{\prime }\right) \mathrm{d}r_{\parallel }^{\prime }.
\end{align*}%
Similarly as in the case of model $T_{2},$ the Cauchy stress tensor in the
case of model $T_{3}$ also contains two local-type terms, of which one also
favours the direction of non-locality vector $\boldsymbol{L},$ while both
non-local terms are non-localy anisotropic, such that non-locality, taken
into account through the exponential function, occurs along direction
determined by the non-locality vector $\boldsymbol{L}$, while non-locality
is not displayed in the plane perpendicular to this direction. The first
non-local term arises from the projector $\frac{\boldsymbol{L}\otimes 
\boldsymbol{L}}{L^{2}}$, while the second non-local term is generated by the
tensor $\frac{\boldsymbol{L}}{L}\otimes \frac{\boldsymbol{r}_{\perp }}{%
r_{\perp }}$, which contracts the Cauchy stress tensor with the unit vector $%
\frac{\boldsymbol{r}_{\perp }}{r_{\perp }}$ and subsequently reconstructs
the tensor along the direction of $\frac{\boldsymbol{L}}{L}$ via dyadic
product, and therefore anisotropy of the Cauchy stress tensor manifests
along the directions determined by the non-locality vector $\boldsymbol{L}$
and a vector $\boldsymbol{r}_{\perp },$ lying in the plane perpendicular to
the vector $\boldsymbol{L}.$

\subsection{Model $T_{4}$}

In the case of model $T_{4},$ the expression (\ref{konvolucija-tenzor}),
connecting the non-local and local Cauchy stress tensors, along with the
non-locality kernel (\ref{kernel-T4}) being of the tensor-type, see also
Table \ref{Modeli-new}, yields 
\begin{align*}
\boldsymbol{\hat{\sigma}}\left( \boldsymbol{r}\right) & =\boldsymbol{\hat{%
\mathcal{K}}}_{4}\left( \boldsymbol{r}\right) \ast _{\boldsymbol{r}}%
\boldsymbol{\hat{\sigma}}^{local}\left( \boldsymbol{r}\right) =\int_{%
\mathbb{R}
^{3}}\boldsymbol{\hat{\mathcal{K}}}_{4}\left( \boldsymbol{r}-\boldsymbol{r}%
^{\prime }\right) \boldsymbol{\hat{\sigma}}^{local}\left( \boldsymbol{r}%
^{\prime }\right) \mathrm{d}_{\boldsymbol{r}^{\prime }}V \\
& =\int_{%
\mathbb{R}
^{3}}\!\delta \left( \boldsymbol{r}\right) \left( \boldsymbol{\hat{I}}-\frac{%
\boldsymbol{L}\otimes \boldsymbol{L}}{L^{2}}\right) \boldsymbol{\hat{\sigma}}%
^{local}\left( \boldsymbol{r}^{\prime }\right) \mathrm{d}_{\boldsymbol{r}%
^{\prime }}V+\frac{1}{4L}\int_{%
\mathbb{R}
}\!\mathrm{e}^{-\frac{\left\vert r_{\parallel }-r_{\parallel }^{\prime
}\right\vert }{L}}\int_{%
\mathbb{R}
^{2}}\!\delta \left( \boldsymbol{r}_{\perp }-\boldsymbol{r}_{\perp }^{\prime
}\right) \frac{\boldsymbol{L}\otimes \boldsymbol{L}}{L^{2}}\boldsymbol{\hat{%
\sigma}}^{local}\left( \boldsymbol{r}_{\perp }^{\prime },r_{\parallel
}^{\prime }\right) \mathrm{d}_{\boldsymbol{r}_{\perp }^{\prime }}S\,\mathrm{d%
}r_{\parallel }^{\prime } \\
& \quad \quad \quad +\int_{%
\mathbb{R}
}\mathrm{e}^{-\frac{\left\vert r_{\parallel }-r_{\parallel }^{\prime
}\right\vert }{L}}\int_{%
\mathbb{R}
^{2}}\frac{\delta \left( \boldsymbol{r}_{\perp }-\boldsymbol{r}_{\perp
}^{\prime }\right) }{\left\vert \boldsymbol{r}_{\perp }-\boldsymbol{r}%
_{\perp }^{\prime }\right\vert }\left( \frac{\boldsymbol{r}_{\perp }-%
\boldsymbol{r}_{\perp }^{\prime }}{\left\vert \boldsymbol{r}_{\perp }-%
\boldsymbol{r}_{\perp }^{\prime }\right\vert }\otimes \frac{\boldsymbol{L}}{L%
}\right) \boldsymbol{\hat{\sigma}}^{local}\left( \boldsymbol{r}_{\perp
}^{\prime },r_{\parallel }^{\prime }\right) \mathrm{d}_{\boldsymbol{r}%
_{\perp }^{\prime }}S\,\mathrm{d}r_{\parallel }^{\prime } \\
& =\boldsymbol{\hat{\sigma}}^{local}\left( \boldsymbol{r}\right) -\frac{%
\boldsymbol{L}\otimes \boldsymbol{L}}{L^{2}}\boldsymbol{\hat{\sigma}}%
^{local}\left( \boldsymbol{r}\right) +\frac{1}{4L}\frac{\boldsymbol{L}%
\otimes \boldsymbol{L}}{L^{2}}\int_{%
\mathbb{R}
}\mathrm{e}^{-\frac{\left\vert r_{\parallel }-r_{\parallel }^{\prime
}\right\vert }{L}}\boldsymbol{\hat{\sigma}}^{local}\left( \boldsymbol{r}%
_{\perp },r_{\parallel }^{\prime }\right) \mathrm{d}r_{\parallel }^{\prime }
\\
& \quad \quad \quad +\frac{1}{r_{\perp }}\left( \frac{\boldsymbol{r}_{\perp }%
}{r_{\perp }}\otimes \frac{\boldsymbol{L}}{L}\right) \int_{%
\mathbb{R}
}\mathrm{e}^{-\frac{\left\vert r_{\parallel }-r_{\parallel }^{\prime
}\right\vert }{L}}\boldsymbol{\hat{\sigma}}^{local}\left( \boldsymbol{r}%
_{\perp },r_{\parallel }^{\prime }\right) \mathrm{d}r_{\parallel }^{\prime }.
\end{align*}%
The difference between models $T_{3}$ and $T_{4}$ lies in the order of
vectors in dyadic product in the last term of the Cauchy stress tensor,
generated by the tensor $\frac{\boldsymbol{r}_{\perp }}{r_{\perp }}\otimes 
\frac{\boldsymbol{L}}{L}$, which contracts the Cauchy stress tensor with the
unit vector $\frac{\boldsymbol{L}}{L}$ and subsequently reconstructs the
tensor along the direction of vector $\frac{\boldsymbol{r}_{\perp }}{%
r_{\perp }}$ via dyadic product.

\subsection{Model $T_{5}$}

In the case of model $T_{5},$ the expression (\ref{konvolucija-tenzor}),
connecting the non-local and local Cauchy stress tensors, along with the
non-locality kernel (\ref{kernel-T5}) being of the tensor-type, see also
Table \ref{Modeli-new}, yields 
\begin{align*}
\boldsymbol{\hat{\sigma}}\left( \boldsymbol{r}\right) & =\boldsymbol{\hat{%
\mathcal{K}}}_{5}\left( \boldsymbol{r}\right) \ast _{\boldsymbol{r}}%
\boldsymbol{\hat{\sigma}}^{local}\left( \boldsymbol{r}\right) =\int_{%
\mathbb{R}
^{3}}\boldsymbol{\hat{\mathcal{K}}}_{5}\left( \boldsymbol{r}-\boldsymbol{r}%
^{\prime }\right) \boldsymbol{\hat{\sigma}}^{local}\left( \boldsymbol{r}%
^{\prime }\right) \mathrm{d}_{\boldsymbol{r}^{\prime }}V \\
& =\int_{%
\mathbb{R}
}\delta \left( r_{\parallel }-r_{\parallel }^{\prime }\right) \int_{%
\mathbb{R}
^{2}}\delta \left( \boldsymbol{r}_{\perp }-\boldsymbol{r}_{\perp }^{\prime
}\right) \left( \boldsymbol{\hat{I}}-\left( \frac{\boldsymbol{L}}{L}\times 
\frac{\boldsymbol{r}_{\perp }-\boldsymbol{r}_{\perp }^{\prime }}{\left\vert 
\boldsymbol{r}_{\perp }-\boldsymbol{r}_{\perp }^{\prime }\right\vert }%
\right) \otimes \left( \frac{\boldsymbol{L}}{L}\times \frac{\boldsymbol{r}%
_{\perp }-\boldsymbol{r}_{\perp }^{\prime }}{\left\vert \boldsymbol{r}%
_{\perp }-\boldsymbol{r}_{\perp }^{\prime }\right\vert }\right) \right) 
\boldsymbol{\hat{\sigma}}^{local}\left( \boldsymbol{r}_{\perp }^{\prime
},r_{\parallel }^{\prime }\right) \mathrm{d}_{\boldsymbol{r}_{\perp
}^{\prime }}S\,\mathrm{d}r_{\parallel }^{\prime } \\
& \quad -\frac{1}{2\pi L}\int_{%
\mathbb{R}
}\delta \left( r_{\parallel }-r_{\parallel }^{\prime }\right) \int_{%
\mathbb{R}
^{2}}\frac{1}{\left\vert \boldsymbol{r}_{\perp }-\boldsymbol{r}_{\perp
}^{\prime }\right\vert }K_{1}\left( \frac{\left\vert \boldsymbol{r}_{\perp }-%
\boldsymbol{r}_{\perp }^{\prime }\right\vert }{L}\right) \frac{\left( 
\boldsymbol{r}_{\perp }-\boldsymbol{r}_{\perp }^{\prime }\right) \otimes
\left( \boldsymbol{r}_{\perp }-\boldsymbol{r}_{\perp }^{\prime }\right) }{%
\left\vert \boldsymbol{r}_{\perp }-\boldsymbol{r}_{\perp }^{\prime
}\right\vert ^{2}}\boldsymbol{\hat{\sigma}}^{local}\left( \boldsymbol{r}%
_{\perp }^{\prime },r_{\parallel }^{\prime }\right) \mathrm{d}_{\boldsymbol{r%
}_{\perp }^{\prime }}S\,\mathrm{d}r_{\parallel }^{\prime } \\
& \quad +\frac{1}{2\pi L}\int_{%
\mathbb{R}
}\delta \left( r_{\parallel }-r_{\parallel }^{\prime }\right) \int_{%
\mathbb{R}
^{2}}\left( \frac{1}{L}K_{0}\left( \frac{\left\vert \boldsymbol{r}_{\perp }-%
\boldsymbol{r}_{\perp }^{\prime }\right\vert }{L}\right) +\frac{1}{%
\left\vert \boldsymbol{r}_{\perp }-\boldsymbol{r}_{\perp }^{\prime
}\right\vert }K_{1}\left( \frac{\left\vert \boldsymbol{r}_{\perp }-%
\boldsymbol{r}_{\perp }^{\prime }\right\vert }{L}\right) \right)  \\
& \quad \quad \quad \quad \quad \quad \quad \quad \quad \quad \quad \quad
\quad \cdot \left( \left( \frac{\boldsymbol{L}}{L}\times \frac{\boldsymbol{r}%
_{\perp }-\boldsymbol{r}_{\perp }^{\prime }}{\left\vert \boldsymbol{r}%
_{\perp }-\boldsymbol{r}_{\perp }^{\prime }\right\vert }\right) \otimes
\left( \frac{\boldsymbol{L}}{L}\times \frac{\boldsymbol{r}_{\perp }-%
\boldsymbol{r}_{\perp }^{\prime }}{\left\vert \boldsymbol{r}_{\perp }-%
\boldsymbol{r}_{\perp }^{\prime }\right\vert }\right) \right) \boldsymbol{%
\hat{\sigma}}^{local}\left( \boldsymbol{r}_{\perp }^{\prime },r_{\parallel
}^{\prime }\right) \mathrm{d}_{\boldsymbol{r}_{\perp }^{\prime }}S\,\mathrm{d%
}r_{\parallel }^{\prime } \\
& =\boldsymbol{\hat{\sigma}}^{local}\left( \boldsymbol{r}\right) -\left( 
\frac{\boldsymbol{L}}{L}\times \frac{\boldsymbol{r}_{\perp }}{r_{\perp }}%
\right) \otimes \left( \frac{\boldsymbol{L}}{L}\times \frac{\boldsymbol{r}%
_{\perp }}{r_{\perp }}\right) \boldsymbol{\hat{\sigma}}^{local}\left( 
\boldsymbol{r}\right)  \\
& \quad -\frac{1}{2\pi L}\int_{%
\mathbb{R}
^{2}}\frac{1}{\left\vert \boldsymbol{r}_{\perp }-\boldsymbol{r}_{\perp
}^{\prime }\right\vert }K_{1}\left( \frac{\left\vert \boldsymbol{r}_{\perp }-%
\boldsymbol{r}_{\perp }^{\prime }\right\vert }{L}\right) \frac{\left( 
\boldsymbol{r}_{\perp }-\boldsymbol{r}_{\perp }^{\prime }\right) \otimes
\left( \boldsymbol{r}_{\perp }-\boldsymbol{r}_{\perp }^{\prime }\right) }{%
\left\vert \boldsymbol{r}_{\perp }-\boldsymbol{r}_{\perp }^{\prime
}\right\vert ^{2}}\boldsymbol{\hat{\sigma}}^{local}\left( \boldsymbol{r}%
_{\perp }^{\prime },r_{\parallel }\right) \mathrm{d}_{\boldsymbol{r}_{\perp
}^{\prime }}S \\
& \quad +\frac{1}{2\pi L}\int_{%
\mathbb{R}
^{2}}\left( \frac{1}{L}K_{0}\left( \frac{\left\vert \boldsymbol{r}_{\perp }-%
\boldsymbol{r}_{\perp }^{\prime }\right\vert }{L}\right) +\frac{1}{%
\left\vert \boldsymbol{r}_{\perp }-\boldsymbol{r}_{\perp }^{\prime
}\right\vert }K_{1}\left( \frac{\left\vert \boldsymbol{r}_{\perp }-%
\boldsymbol{r}_{\perp }^{\prime }\right\vert }{L}\right) \right)  \\
& \quad \quad \quad \quad \quad \quad \quad \cdot \left( \left( \frac{%
\boldsymbol{L}}{L}\times \frac{\boldsymbol{r}_{\perp }-\boldsymbol{r}_{\perp
}^{\prime }}{\left\vert \boldsymbol{r}_{\perp }-\boldsymbol{r}_{\perp
}^{\prime }\right\vert }\right) \otimes \left( \frac{\boldsymbol{L}}{L}%
\times \frac{\boldsymbol{r}_{\perp }-\boldsymbol{r}_{\perp }^{\prime }}{%
\left\vert \boldsymbol{r}_{\perp }-\boldsymbol{r}_{\perp }^{\prime
}\right\vert }\right) \right) \boldsymbol{\hat{\sigma}}^{local}\left( 
\boldsymbol{r}_{\perp }^{\prime },r_{\parallel }\right) \mathrm{d}_{%
\boldsymbol{r}_{\perp }^{\prime }}S,
\end{align*}%
where $\cdot $ is used for multiplication of scalars. The second local-type
term of the Cauchy stress tensor in the case of model $T_{5}$ is obtained as
a contraction of local Cauchy stress tensor with the vector in the plane
perpendicular to non-locality vector $\boldsymbol{L},$ namely $\frac{%
\boldsymbol{L}}{L}\times \frac{\boldsymbol{r}_{\perp }}{r_{\perp }},$ and
the tensor is subsequently reconstructed by the same vector via dyadic
product, thus having anisotropic character. The non-local terms are
non-localy anisotropic, such that non-locality, taken into account through
the modified Bessel functions of the second kind, occurs in the plane
perpendicular to the direction determined by the non-locality vector $%
\boldsymbol{L}$, while non-locality is not displayed in the direction of
vector $\boldsymbol{L}$. The anisotropy of the first non-local term arises
from the projector $\frac{\boldsymbol{r}_{\perp }\otimes \boldsymbol{r}%
_{\perp }}{r_{\perp }^{2}}$, while the anisotropy of the second non-local
term is generated by the tensor $\left( \frac{\boldsymbol{L}}{L}\times \frac{%
\boldsymbol{r}_{\perp }}{r_{\perp }}\right) \otimes \left( \frac{\boldsymbol{%
L}}{L}\times \frac{\boldsymbol{r}_{\perp }}{r_{\perp }}\right) =\left( \frac{%
\boldsymbol{L}}{L}\times \frac{\boldsymbol{r}}{r}\right) \otimes \left( 
\frac{\boldsymbol{L}}{L}\times \frac{\boldsymbol{r}}{r}\right) $. Both
non-local terms exhibit anisotropy manifesting in the plane perpendicular to
the direction of the non-locality vector $\boldsymbol{L}$.

\section{Conclusion}

Following the example of Eringen stress-gradient model (\ref{1D-Eringen}),
containing the non-locality operator $l^{2}\frac{\mathrm{d}^{2}}{\mathrm{d}%
x^{2}},$ and corresponding to an one-dimensional non-local elastic body, two
groups of models, namely scalar-type (\ref{non-loc-scalar-type}) and
tensor-type (\ref{non-loc-tensor-type}), are formulated in order to
generalize the Eringen model for the case of a three-dimensional non-local
body by defining non-locality operators $\mathrm{S}_{i}$, $i\in \left\{
1,2,3\right\} $, and $\mathbf{\hat{T}}_{i}$, $i\in \left\{ 1,2,3,4,5\right\}
,$ see Table \ref{Modeli-new}, as the three dimensional scalar- and
tensor-type non-local counterparts of the operator $l^{2}\frac{\mathrm{d}^{2}%
}{\mathrm{d}x^{2}}$, combining the non-locality vector $\boldsymbol{L}$,
instead of the scalar parameter $l$, with the nabla operator $\nabla $
instead of the derivative $\frac{\mathrm{d}}{\mathrm{d}x}$, using the scalar
product $\cdot $, vector product $\times $, dyadic product $\otimes $, and
tensor contraction $:$. Some of scalar- and tensor-type non-local operators $%
\mathrm{S}_{i}$ and $\mathbf{\hat{T}}_{i}$ can be written in several
additional equivalent forms, as discussed in Section \ref{konstraksn}.
Moreover, in the case of the scalar-type model (\ref{non-loc-scalar-type})
both non-local $\boldsymbol{\hat{\sigma}}$ and local $\boldsymbol{\hat{\sigma%
}}^{local}$ Cauchy stress tensors are symmetric, while in the case of the
tensor-type model (\ref{non-loc-tensor-type}) the compatibility conditions
are necessary to guarantee the symmetric form of the local Cauchy stress
tensor $\boldsymbol{\hat{\sigma}}^{local}$, since non-local Cauchy stress
tensor $\boldsymbol{\hat{\sigma}}$ is assumed to be symmetric.

Using the Fourier integral transform with respect to spatial coordinates,
non-locality kernels $\mathcal{K}_{i}$ and $\boldsymbol{\hat{\mathcal{K}}}%
_{i},$ reflecting the non-locality character of the material, are derived in
Section \ref{Three-dimensional-cases} for each of the proposed models and
presented in Table \ref{Modeli-new}. Unlike the first scalar-type model,
denoted by $S_{1}$, in the case of all other models the non-locality kernel
contains, in addition to the functional part, a distributional part given by
the Dirac delta distribution $\delta $, thus accounting for both local and
non-local contributions to Cauchy stress tensor.

It is examined in Section \ref{Non-locality of the Cauchy stress tensor}
whether the constitutive equations, i.e., the Cauchy stress tensors, in the
case of scalar- and tensor-type non-local elastic models (\ref%
{scalar-non-local-elasticity}) and (\ref{tensor-non-local-elasticity}) are
isotropic, as well as whether the non-locality in the case of these models
is isotropic. Recall, the model is referred as isotropic if there exists no
shear-extensional coupling and if there is only two independent model
parameters, while model is considered non-locally isotropic if there is no
preferential direction(s). Namely, in the case of the scalar-type models (%
\ref{scalar-non-local-elasticity}), the constitutive equations, i.e., the
Cauchy stress tensors, represented by (\ref{konvolucija-skalar}), are
isotropic when local Cauchy stress tensor $\boldsymbol{\hat{\sigma}}^{local}$
is connected to the strain tensor via Hooke's law (\ref{isotropic}), since
the non-locality kernels $\mathcal{K}_{i}$ determine only the non-local
character of the Cauchy stress tensors $\boldsymbol{\hat{\sigma}}$ and do
not change isotropy of the constitutive equation, see (\ref{scalar-elastic}%
). On the other hand, for the local Cauchy stress tensor $\boldsymbol{\hat{%
\sigma}}^{local}$, connected to the strain tensor via Hooke's law (\ref%
{Hooke's-law}), in the case of the tenor-type models (\ref%
{tensor-non-local-elasticity}), the constitutive equations, i.e., the Cauchy
stress tensors $\boldsymbol{\hat{\sigma}}$, represented by (\ref%
{konvolucija-tenzor}), are anisotropic since the non-locality kernels $%
\boldsymbol{\hat{\mathcal{K}}}_{i}$ are given as tensors, and therefore
shear-extesional coupling occurs in the case of all proposed tensor-type
non-local models, see (\ref{dijagonalni-1}) - (\ref{dijagonalni-3}) and (\ref%
{off-dijagonalni-1}) - (\ref{off-dijagonalni-3}).

Unlike the case of tensor-type models, in the case of scalar-type models,
the Cauchy stress tensor (\ref{konvolucija-skalar}) consists of a single
term, see Sections \ref{Model-S-1} - \ref{Model-S-3}, such that only in the
case of model $S_{1}$, the preferred direction of non-locality is not
distinguished, contrary to the remaining two scalar models $S_{2}$ and $S_{3}
$ in which the Cauchy stress tensor exhibits non-locality along a certain
direction(s). Therefore, model $S_{1}$ describes non-locally isotropic three
dimensional body, while models $S_{2}$ and $S_{3}$ describe non-locally
anisotropic bodies. Namely, non-local (un)isotropy is determined by the
character of non-locality kernel $\mathcal{K}_{i}$, which may not
distinguish a particular direction(s) along which the body is non-local, as
assumed by model $S_{1}$, while it also may imply that body is non-local
along a certain direction(s), such as along the direction determined by the
non-locality vector $\boldsymbol{L}$ as in the case of model $S_{2}$, or in
the plane perpendicular to direction determined by the non-locality vector $%
\boldsymbol{L},$ as predicted by the model $S_{3}$.

In the case of tensor-type models, the Cauchy stress tensor (\ref%
{konvolucija-tenzor}) consists of multiple terms, so only in the case of
models $T_{1}$ and $T_{2}$ there is no preferred direction of non-locality,
thus these models describe a non-locally isotropic body, in contrast to the
remaining three tensor-type models $T_{3}$, $T_{4}$, and $T_{5}$ in which
the Cauchy stress tensor exhibits non-locality along a specific direction(s)
and therefore these models describe non-locally anisotropic
three-dimensional bodies. Namely, in the case of tensor-type models,
non-local (un)isotropy is determined by the character of non-locality kernel 
$\boldsymbol{\hat{\mathcal{K}}}_{i}$, which may not distinguish a particular
direction(s) along which the body is non-local, as assumed by models $T_{1}$
and $T_{2}$, while it also may imply that body is non-local along a certain
direction(s), such as along the direction determined by the non-locality
vector $\boldsymbol{L}$ as in the case of models $T_{3}$ and $T_{4}$, or in
the plane perpendicular to direction determined by the non-locality vector $%
\boldsymbol{L},$ as predicted by the model $T_{5}$.

\bigskip

\noindent \textbf{CRediT authorship contribution statement}\smallskip

\noindent \textbf{Sla\dj an Jeli\'{c}}: Conceptualization, Formal analysis,
Investigation, Methodology, Writing original draft, Writing - review \&
editing.

\noindent \textbf{Du\v{s}an Zorica}: Conceptualization, Formal analysis,
Investigation, Methodology, Writing original draft, Writing - review \&
editing. \smallskip

\noindent \textbf{Declaration of Competing Interest} \smallskip

\noindent The authors declare that they have no conflict of
interest.\smallskip

\noindent \textbf{Data availability }\smallskip

\noindent No data was used for the research described in the article.
\smallskip

\noindent \textbf{Acknowledgements} \smallskip

\noindent The work is supported by the Ministry of Science, Technological
Development and Innovation of the Republic of Serbia under grants
451-03-33/2026-03/200125 \& 451-03-34/2026-03/200125 (SJ and DZ).

\bibliographystyle{plain}
\bibliography{dz}

\end{document}